\documentclass[iop,twocolappendix,numberedappendix,appendixfloats]{emulateapj}

\usepackage{natbib}
\usepackage[bookmarks,breaklinks]{hyperref}
   \hypersetup{
     colorlinks,
     citecolor=blue,
     filecolor=magenta,
     linkcolor=blue,
     urlcolor=cyan
}
\usepackage{multirow}
\usepackage{longtable}
\usepackage{threeparttablex}
\usepackage{csquotes}

\newcommand{\OIII}{\mbox{[O\,\textsc{iii}]}}

\graphicspath{{J0918/}{J1135/}{J1404/}{J1606/}{J1622/}{J1720/}}

\shorttitle{Structure of ionized gas outflows in Type 2 AGN - I}
\shortauthors{Karouzos et al.}

\begin{document}


\title{Unravelling the complex structure of AGN-driven outflows:\\ I. Kinematics and sizes}


\author{Marios Karouzos$^{1}$, Jong-Hak Woo$^{1}$, Hyun-Jin Bae$^{2}$}
\affil{$^{1}$Astronomy Program, Department of Physics and Astronomy, Seoul National University, Seoul 151-742, Republic of Korea}
\affil{$^{2}$Department of Astronomy and Center for Galaxy EVolution Research, Yonsei University, Seoul 120-749, Republic of Korea}
\email{email: woo@astro.snu.ac.kr}




\begin{abstract}
Outflows driven by active galactic nuclei (AGNs) are often invoked as agents of the long-sought AGN feedback. Yet, characterizing and quantifying the impact on their host galaxies has been challenging. We present Gemini Multi-Object Spectrograph integral field unit data of 6 local (z$<0.1$) and luminous (L$_{\mathrm{[OIII]}}>10^{42}$ erg s$^{-1}$) Type 2 AGNs. In the first of a series of papers, we investigate the kinematics and constrain the size of the outflows. The ionized gas kinematics can be described as a superposition of a gravitational component that follows the stellar motion and an outflow-driven component that shows large velocity (up to 600 km s$^{-1}$) and large velocity dispersion (up to 800 km s$^{-1}$). Using the spatially resolved measurements of the gas, we kinematically measure the size of the outflow, which is found to be between 1.3 and 2.1 kpc. Due to the lack of a detailed kinematic analysis, previous outflow studies likely overestimate their size by up to more than a factor of 2, depending on how the size is estimated and whether the [OIII] or H$\alpha$ emission line is used. The relatively small size of the outflows for all 6 of our objects casts doubts on their potency as a mechanism for negative AGN feedback.
\end{abstract}



\section{Introduction}
\label{sec:intro}

Astrophysical outflows can be found in a number of diverse environments and have been associated with different processes that span several orders of magnitude in spatial and temporal scale and energetics. Beyond those associated with stellar processes (e.g., \citealt{Herbig1960,Crowther2007,Humphreys1994,Woosley2006}), some of the most energetic outflows are linked to supermassive black holes (SMBHs) during the accretion phase.
Such outflows have been observed in different phases of the interstellar medium (ISM): hot plasma 
(radio jets; \citealt{Akiyama2015}), ionized gas 
(e.g., \citealt{Konigl1994,Murray1995,Crenshaw2003}), and molecular gas (e.g., \citealt{Cicone2014}), although the connection to 
active galactic nuclei (AGNs) is still under debate. Galaxy-wide star formation or compact starburst regions are alternative drivers of such outflows (e.g., \citealt{Chevalier1985,Veilleux2005}). AGN-driven outflows are considered to be 
channels of the AGN feedback, which appears as necessary to explain the present-day scaling relations between black hole mass and host galaxy properties (e.g., \citealt{Springel2005,DiMatteo2005,Croton2006,Somerville2008,Schaye2010}, but also see \citealt{Hopkins2014}).

Outflows in AGN host galaxies are prevalent, as suggested by extensive observational studies of the ionized gas kinematics
using large samples
(e.g., \citealt{Heckman1981,Whittle1985a,Nelson1995,Greene2005b,Crenshaw2010,Villar2011,Mullaney2013,Bae2014,Woo2015b}). {Corroborating evidence for these outflows also come from studies of absorption lines in the rest-frame UV spectrum of AGN in the local Universe (e.g., \citealt{Arav2012,Borguet2012,Crenshaw2012,Liu2015}) and at higher redshift (e.g., \citealt{Moe2009,Lucy2014,Chamberlain2015}).}
While these studies
provide convincing evidence for strong non-gravitational motions affecting the ionized gas of luminous AGNs, they were unable to properly constrain either the detailed physical properties of these outflows or their impact on the host galaxies due to the lack of spatially resolved information. 

The recent development of powerful integral field unit (IFU) spectrographs has allowed us an unprecedented view of the spatially resolved properties of these outflows. A number of detailed investigations on the kinematics of AGN ionized gas in both the local Universe (e.g., \citealt{Storchi2010,Sharp2010,Riffel2013,Rupke2013,Riffel2014,McElroy2015}) and at higher redshifts (e.g., \citealt{Nesvadba2006,Nesvadba2008,Harrison2012,Carniani2015}) have elucidated the geometry and the kinematics of AGN-driven outflows. 

Despite the significant progress in understanding AGN outflows, studies of gas kinematics based on IFU data have often suffered from
the lack of information about the underlying gravitational potential as manifested by stellar kinematics. Without characterizing the gravitational potential within which the ionized gas is moving, precise determination of any non-gravitational kinematics, i.e. an outflow, is impossible. In fact, \citet{Woo2015b} showed the presence of such a non-gravitational component in an increasing fraction of Type 2 AGNs with increasing [OIII] luminosity and Eddington ratio. This non-gravitational kinematic component would be otherwise watered down 
without removing (or correcting for) the gravitational component as traced by stellar motions. 
Thus, proper spatial {\it and} kinematic decomposition of the ionized gas emission is the key to constraining both the size and the kinetic energy of the outflows, which in turn allows the fundamental understanding of the AGN feedback.

Here we present the analysis of Gemini Multi-Object Spectrograph (GMOS) optical IFU data for 6 luminous Type 2 AGNs, drawn from our previous study of a sample of $\sim$39,000 SDSS Type 2 AGNs (\citealt{Bae2014,Woo2015b}). We characterize the kinematics and size of the outflows in these objects, by taking into account the gravitational and non-gravitational kinematics. 

Section \ref{sec:sample} describes the sample selection, observations and data reduction. In Section \ref{sec:method} we discuss our method of measuring kinematics, including continuum and emission line fitting as well as measurement uncertainties.
In Sections \ref{sec:results} and \ref{sec:physical} we present the main results. 
Section \ref{sec:discussion} provides discussion and Section \ref{sec:conclusions} presents the summary and conclusions. 
Finally, in the Appendix we provide comments on individual sources.
We use the following cosmological parameters for a $\Lambda$-CDM Universe: h$_{0}$=0.71, $\Omega_{m}$=0.27, and $\Omega_{\Lambda}$=0.73.

\section{Sample and observations}
\label{sec:sample}

\subsection{Sample selection}
\label{sec:selection}

\begin{figure}[tb]
\begin{center}
\includegraphics[width=0.4\textwidth,angle=0]{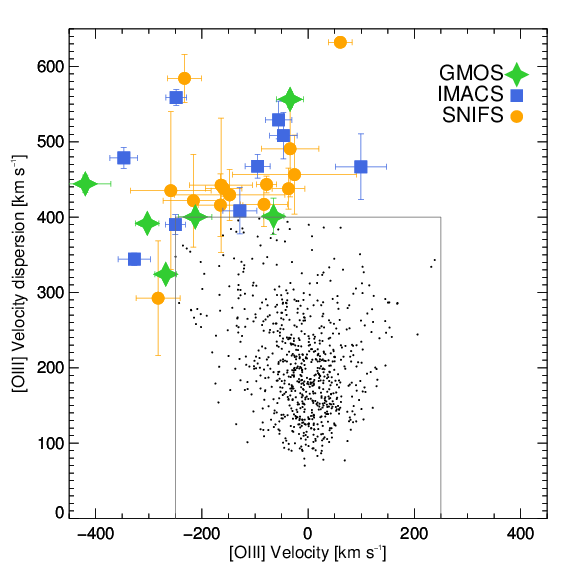}
\caption{\mbox{[O\,\textsc{iii}]} velocity versus velocity dispersion (VVD) diagram of the luminosity-limited sample of
AGNs with $L_{\mathrm{\mbox{[O\,\textsc{iii}]};cor}}>10^{42}$ erg s$^{-1}$ and z $\leqslant0.1$ from \citet{Woo2015b}. 
{The} \OIII\ {velocity dispersion is calculated from the second moment of the emission line profile based on Eq. \ref{eq:mom}.}
 The sample of 6 AGNs presented in this paper (green stars), together with the 23 objects (blue squares and orange circles) 
 under investigation with other facilities, are selected based on their high velocity and velocity dispersion (solid box)
 as the best candidates for follow-up IFU studies.} 
\label{fig:sample}
\end{center}
\end{figure}

In our previous statistical study using $\sim$39,000 Type 2 AGNs at z$<$0.3 based on the archival spectra of the Sloan Digital Sky Survey (\citealt{Woo2015b}), 
we identified AGNs with extreme kinematic signatures by analysing the velocity and velocity dispersion of the \mbox{[O\,\textsc{iii}]} $\lambda$ 5007\AA\ emission line. Among them, we selected 6 AGNs for a pilot study using the following criteria.
First, we limited the redshift range to z$<$ 0.1 to obtain sub-kpc spatial resolution.
Second, since the outflow kinematics is closely linked to AGN luminosity, we focused on high luminosity AGNs by selecting 902 Type 2 AGNs with extinction-corrected \mbox{[O\,\textsc{iii}]} luminosity L$_{\mathrm{\mbox{[O\,\textsc{iii}]};cor}}$$>$10$^{42}$ erg s$^{-1}$.
Third, out of 902 objects, we selected 29 AGNs with an \mbox{[O\,\textsc{iii}]} velocity shift $|v_{\mbox{[O\,\textsc{iii}]}}|>250$ km s$^{-1}$ (with respect to the systemic velocity measured from stellar absorption lines), and an \mbox{[O\,\textsc{iii}]} velocity dispersion $\sigma_{\mbox{[O\,\textsc{iii}]}}>400$ km s$^{-1}$, based on measurements using SDSS spectra (\citealt{Woo2015b}). We argue that these AGNs with strong outflow signatures are the best candidates for detecting spatially-resolved outflows and to understand the impact on their host galaxies.

In Fig. \ref{fig:sample}, we present the \mbox{[O\,\textsc{iii}]} velocity-velocity dispersion (VVD) diagram for the luminosity-limited
sample of 902 AGNs with $L_{\mathrm{\mbox{[O\,\textsc{iii}]};cor}}>10^{42}$ erg s$^{-1}$ along with the kinematically selected 29 AGNs.
The selection criteria of the \mbox{[O\,\textsc{iii}]} kinematics (gray box) indicate that the kinematically selected 29 AGNs have extreme velocities compared to the majority of AGNs with similar luminosities. The fraction of the selected AGNs based on the kinematic criteria is 3.2\%. If we do not apply the luminosity limit, a total of 95 objects (0.4\%) out of the $\sim$23,000 AGNs at z$<$0.1 satisfies the \mbox{[O\,\textsc{iii}]} kinematic criteria. Given the sample completeness in terms of the \mbox{[O\,\textsc{iii}]} luminosity and kinematics, we will use this 
sample to understand AGN-driven outflows, including kinematics, energy budgets, and the impact on the ISM for high luminosity AGNs in the local Universe. 

We are performing an integral field spectroscopy survey of these 29 AGNs using three different instruments: the GMOS IFU (\citealt{Allington2002}) on Gemini-North (6 sources, presented here), the Inamori MAgellan Cassegrain Spectrograph (IMACS) IFU (\citealt{Dressler2011}) on the 6m Magellan-Baade telescope (9 sources), and the SuperNova Integral Field Spectrograph (SNIFS) IFU (\citealt{Aldering2002,Lantz2004}) on the 2.2m University of Hawaii telescope (14 sources). Given the significant differences in spatial and spectral resolution and the achieved S/N among the three data sets, we present first the results using the highest quality Gemini GMOS-N data. We provide the properties of the Gemini/GMOS sample in Table \ref{tab:sample}.

\begin{deluxetable*}{c c c c c c c c c c c c c}
\tabletypesize{\footnotesize}
\tablecolumns{10}
\tablewidth{0pt}
\tablecaption{Our sample of 6 Type 2 AGNs with extreme \mbox{[O\,\textsc{iii}]} kinematics. \label{tab:sample}}
\tablehead{\colhead{ID}	&	\colhead{RA}	&	\colhead{DEC}	&	\colhead{z}	&	\colhead{v$_{\mbox{[O\,\textsc{iii}]}}$}	&	\colhead{$\sigma_{\mbox{[O\,\textsc{iii}]}}$}	&	\colhead{$\log{L_{\mathrm{\mbox{[O\,\textsc{iii}]};cor}}}$}	&	\colhead{m$_{r}$}	& \colhead{b/a}	&	\colhead{Date}	&	\colhead{t$_{\mathrm{exp}}$}	&	\colhead{Seeing}	&	\colhead{AM}	\\
\colhead{ }	&	\colhead{[hh:mm:ss] } 	&	\colhead{[dd:mm:ss] } 	& \colhead{ }	&	\multicolumn{2}{c}{[km s$^{-1}$]}		&	\colhead{[erg s$^{-1}$]}	&	\colhead{[AB]}	&	\colhead{ } & 	\colhead{ }	&	\colhead{[min]}	&	\colhead{[\arcsec]}	&	\colhead{ }}
\startdata
J091808+343946 	&	09:18:08	&	+34:39:46		&	0.0973	&	-419	&	444	&	42.9	&	16.45	&	0.94 &	03/18/15	&	133	&	0.5	&	1.25\\
J113549+565708 	&	11:35:49 	&	+56:57:08 	&	0.0515	&	-212	&	400	&	43.1	&	14.72	&	0.82 &	03/17/15	&	45	&	0.8	&	1.13\\
J140453+532332	&	14:04:53	&	+53:23:32		&	0.0813	&	-303	&	392	&	42.6	&	16.11	&	0.38 &	03/17/15	&	45	&	0.6	&	1.24\\
J160652+275539	&	16:06:52	&	+27:55:39		&	0.0461	&	-268	&	324	&	42.2	&	15.22	&	0.88 &	03/17/15	&	45	&	0.5	&	1.16\\
J162233+395650	&	16:22:33	&	+39:56:50 	&	0.0631	&	-34	&	556	& 	42.4	&	15.61	&	0.76 &	03/17/15	&	45	&	0.5	&	1.11\\
J172038+294112	&	17:20:38 	&	+29:41:12 	&	0.0995	&	-65	&	401	&	42.3	&	16.73	&	0.95 &	03/24/15	&	142	&	0.6	&	1.27\\
\enddata
\tablecomments{
Col. 1: target ID, Col. 2: right ascension (J2000), Col. 3: declination (J2000), Col. 4: the redshift based on the stellar absorption lines 
using the SDSS spectra \citep{Bae2014}, Col. 5: the \mbox{[O\,\textsc{iii}]} velocity shift with respect to the systemic velocity,
Col. 6: the \mbox{[O\,\textsc{iii}]} velocity dispersion measured from SDSS spectra, Col. 7: the dust-corrected \mbox{[O\,\textsc{iii}]}  luminosity (see \citealt{Bae2014}), Col. 8: the \textit{r}-band SDSS magnitude, Col. 9: the b/a minor-to-major axis ratio, Col. 10: the date of observations, Col. 11: the exposure time, Col. 12: seeing, Col. 13: the average airmass.}
\end{deluxetable*}%

\subsection{Observations with the GMOS IFU on Gemini-North}
\label{sec:observations}

We observed 6 AGNs using the GMOS-N IFU in 1-slit mode in the 2015A semester (ID: GN-2015A-Q-204, PI: Woo).
The field of view (FoV) was $5\farcs0\times3\farcs5$, which corresponds to 3-9 kpc at the redshift of our targets,
with a spatial pixel (spaxel) scale $\sim0\farcs07$.
We used the B600 grating, which delivers a spectral resolution of R$\sim1700$ over the spectral range $\sim$4500-6800\AA. Given the configuration of the lenslits, we used a 2 pixel spectral binning, degrading the nominal resolution to R $\sim1400$, corresponding to a full width at half-maximum (FWHM) velocity resolution of $\sim215$ km s$^{-1}$. 
 Exposure times were determined based on surface brightness measurements in B band calculated from SDSS photometry, ranging between 45 and 140 minutes (see Table 1).

During our observations, which were performed as part of the K-GMT Science Program in priority visitor mode, weather conditions were stable with low wind and moderate humidity values. 
Seeing values ranged between 0\farcs5 and 0\farcs8, corresponding to sub-kpc spatial resolutions for most of our targets (e.g.,  $\sim$1 kpc for the farthest target J172038+294112 at z=0.0995).
In Table \ref{tab:sample} we provide the log of the observations.

\subsection{Data reduction}
We used the Gemini \textit{IRAF} package for the data reduction, following the standard procedures for the 1-slit mode IFU on the Gemini webpage\footnote{http://www.gemini.edu/sciops/data/IRAFdoc/}. First, we subtracted the CCD bias from all frames by using the standard bias frame and removed cosmic rays by using the PyCosmic routine \citep{2012A&A...545A.137H}, which shows robust results for cosmic-ray removal in fiber-fed IFU data. Second, we obtained the extraction solution based on the flat-field images from the afternoon calibration and corrected for the fiber-to-fiber variation by using the twilight flat-field images. The obtained extraction solution was used to extract spectra from the science and calibration frames. The arc spectra were used to determine the wavelength solution for the science frames. Third, the science spectra were calibrated using the wavelength solution and corrected for the sky background using the mean sky spectra from the dedicated sky fibers. Fourth, flux calibration was performed using the spectrophotometric standard stars. Finally, we obtained a 3-D cube for each target by resampling the spaxel size to 0\farcs1$\times$0\farcs1. 

\section{Methodology}
\label{sec:method}

\subsection{Emission line fitting}
{To measure the kinematic properties of AGN emission lines, we follow the procedure we adopted for the SDSS AGNs studies
(e.g., \citealt{Bae2014, Woo2015b}). Here, we briefly describe the measurement steps. 
First, we model the stellar component by fitting the stellar continuum with the best stellar population models using the pPXF code \citep{Cappellari2004}. For the pPXF modeling, we adopted 47 MILES simple-stellar population models with different ages, but with solar metallicity \citep{2011A&A...532A..95F}. The best-fit pPXF stellar emission model is then subtracted from the original spectrum to produce a pure emission line spectrum.} The fitting solution provides the line-of-sight velocity of the stellar component in each spaxel. The systemic velocity of each host galaxy is determined from the velocity of the stellar component using the spectrum extracted from the central 3\arcsec\ spaxels, while the stellar velocity dispersion, $\sigma_{*}$, of each galaxy was adopted from the SDSS measurements based on the spatially integrated 3\arcsec -fiber SDSS spectrum. 

Using the emission line spectra, we fit the \mbox{[O\,\textsc{iii}]} and H$\alpha$ regions, employing the Levenberg-Marquardt least-squares algorithm (\citealt{Marquardt1963,More1978}) as implemented in the IDL procedure \textit{MPFIT} (\citealt{Markwardt2009}). Each individual line is fit with a double Gaussian since for most emission lines two Gaussian components are required except for the outer spaxels in the FoV.  We employ an iterative fitting method to decide whether a secondary broad component is needed for fitting each individual line. The secondary component is adopted if its peak flux is at least equal to the noise measured at the continuum near the respective emission line. A more stringent criterion for the retention of broad components ($>3$ times the noise) does not affect the results presented here. The double Gaussian profile can be understood as that gas kinematics is characterized by a narrow emission component representing the gravitational potential while a broad emission component reflects the non-gravitational potential such as an AGN outflow, as demonstrated by \citet{Woo2015b} based on the detailed analysis of the integrated SDSS spectra
(see Section 4.2 for details). 

\begin{figure*}[tb]
\begin{center}
\includegraphics[width=0.48\textwidth,angle=0]{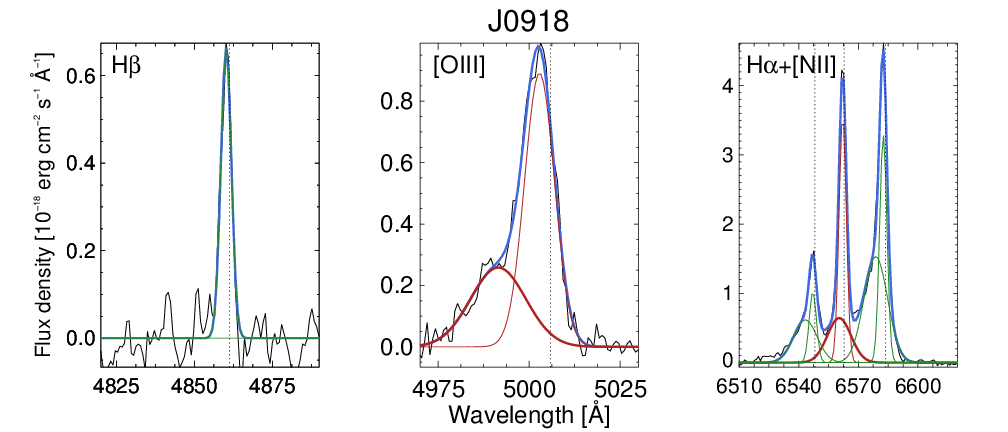}
\includegraphics[width=0.48\textwidth,angle=0]{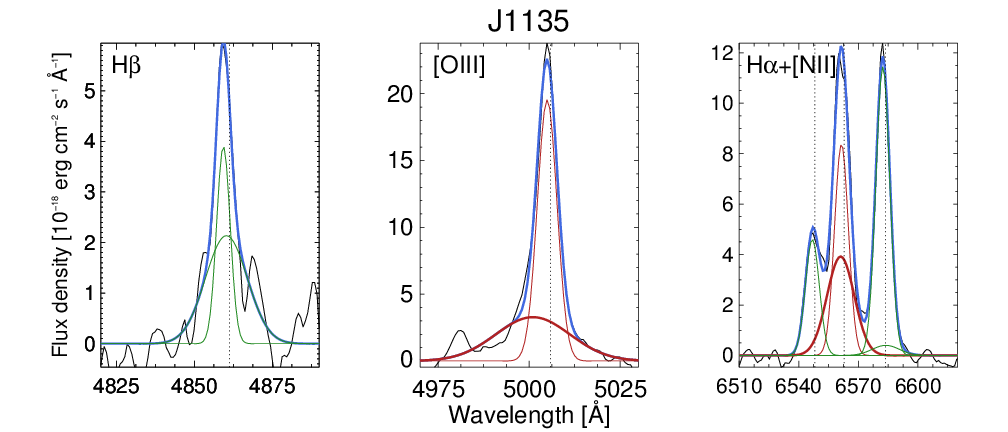}\\
\includegraphics[width=0.48\textwidth,angle=0]{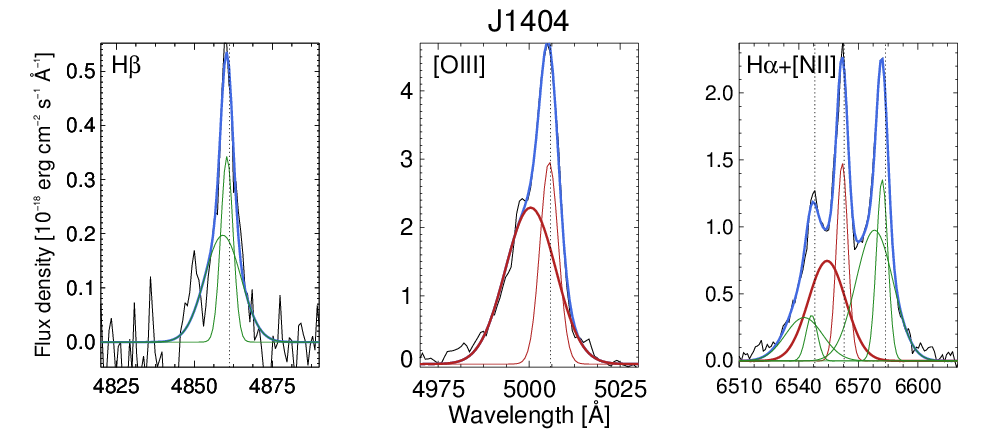}
\includegraphics[width=0.48\textwidth,angle=0,trim={20 0 0 0},clip]{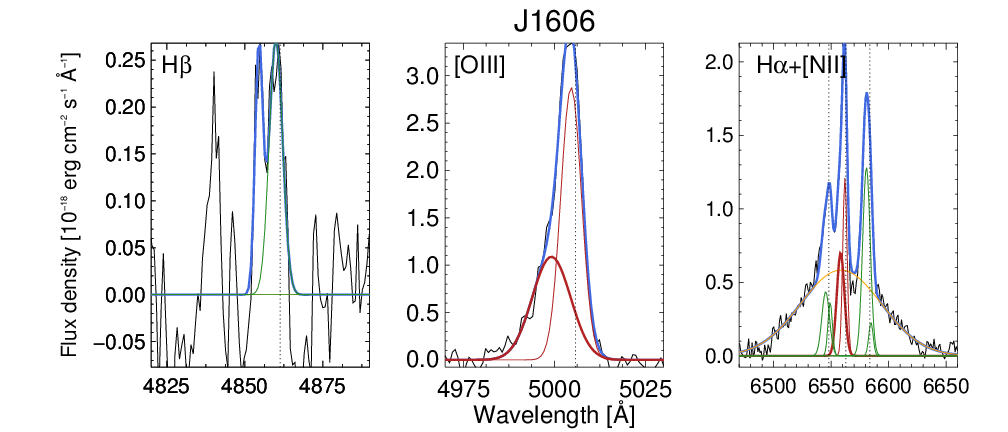}\\
\includegraphics[width=0.48\textwidth,angle=0]{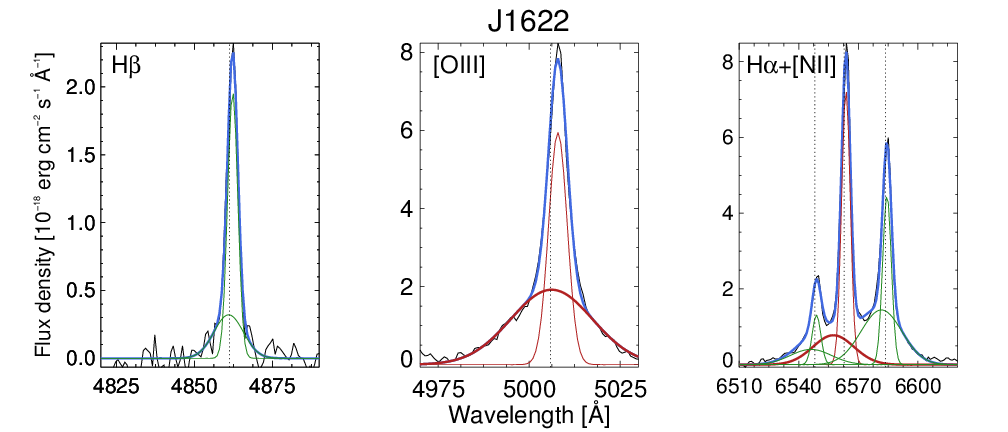}
\includegraphics[width=0.48\textwidth,angle=0]{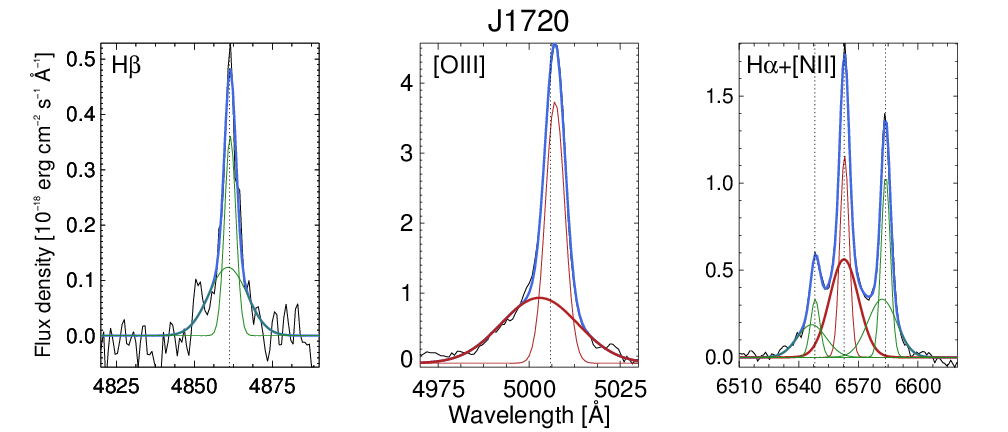}
\caption{Examples of emission line fits of H$\beta$, \mbox{[O\,\textsc{iii}]}, and H$\alpha$ in the central spaxel of each source. The best total fit (blue) is shown together with the individual Gaussian components: red for \mbox{[O\,\textsc{iii}]} and H$\alpha$, and green for H$\beta$, [NII]. 
The expected center of each line based on the systemic velocity is denoted with vertical dotted lines. 
For J1606, the very broad (Type 1) component is shown in orange.}
\label{fig:examplefit}
\end{center}
\end{figure*}

For the H$\alpha$ region, we simultaneously fit the H$\alpha$ line and the [NII] doublet, while the redward [SII] doublet is fitted separately. All lines are fitted with a single or double Gaussian profiles as done for \mbox{[O\,\textsc{iii}]}. To reduce the degrees of freedom of the fitting, we assume the same dispersion and velocity shift (with respect to the systemic velocity) values for the doublets ([NII] and [SII]) while velocity dispersion is in turn tied to the dispersion of the individual H$\alpha$ components. For \mbox{[O\,\textsc{iii}]} at $\lambda$5007\AA\ and H$\beta$ we also used a double Gaussian profile if required, fitting the two lines separately. In Fig. \ref{fig:examplefit} we demonstrate the emission line fitting results for the central spaxel of each galaxy.

The only exception to the above is source J160652+275539, which upon closer inspection was found to show a very broad H$\alpha$ component ($\sigma>2000$ km s$^{-1}$ ), which seems to originate from the broad-line region. A very broad component of similar width was also seen in the H$\beta$ emission for some spaxels\footnote{In general, H$\beta$ emission is very faint for this object, with reliable H$\beta$ measurements for only a few tens of spaxels. There are indications of heavy dust extinction as implied by the Balmer decrement of the spaxels where both lines could be measured reliably.}. As a result we had to allow for one more Gaussian component for this very broad component in the fitting of H$\alpha$ and H$\beta$ of this target. This component was left to have free central wavelength and dispersion. As a result, for the core area of the IFU ($\sim2\arcsec$ diameter) both H$\alpha$ and H$\beta$ are fitted with three components each, while for the outer regions two components were sufficient. The very broad component is excluded from our analysis since we assume that it represents the broad-line region kinematics. 
For some spaxels, a very broad component and a narrow component were sufficient to fit the observed emission profile. For these cases the intermediately broad component, seen in other spaxels and sources may be either truly missing or undetectable due to the powerful broad emission from the core of the AGN. For these cases, we choose to conservatively assume that for these spaxels the narrow emission can be sufficiently described with a single narrow component. 

For each emission line in each spaxel, we measure the first moment, $\lambda_{0}$, and second moment, $\sigma$, of the line profile that are defined as
\begin{equation}
\label{eq:mom}
\mathrm{\lambda_{0}}=\frac{\int\lambda f_{\lambda}d\lambda}{\int f_{\lambda}d\lambda},\;  \mathrm{\sigma}=\frac{\int\lambda^{2} f_{\lambda}d\lambda}{\int f_{\lambda}d\lambda}-\mathrm{\lambda_{0}}^{2}.
\end{equation}
We then calculate the velocity with respect to the systemic velocity (measured from the stellar absorption lines extracted from the central 3\arcsec\ spaxels)
and velocity dispersion after correcting for the instrumental resolution ($\sigma_{inst}$$\sim91$ km s$^{-1}$). For some spaxels, emission lines have velocity dispersion consistent with zero after the instrumental resolution correction. These lines are considered unresolved and set to a velocity dispersion value of zero. In addition to the total best-fit model, we also derive the line flux, velocity, and velocity dispersion
of the narrow and broad components separately.

The flux-weighted kinematic measures used here present distinct advantages over the peak velocity and FWHM velocity, typically used to study the shape of emission lines, because they better represent the emission line wings and asymmetric features. Also, velocity dispersion (i.e., second moment of the line profile)  is more robust against low S/N data than measures such as W$_{80}$, velocity width containing 80\% of the total emission-line flux (e.g., \citealt{Liu2013,Zakamska2014}), and $u_{02}$, the 2nd percentile of the total emission-line flux (e.g., \citealt{Rupke2013,Harrison2014}), which are also often used in the literature.

\subsection{Simulating uncertainties}
\label{sec:mc}

We use the noise spectra for each spaxel to derive the uncertainties of the values calculated from our fits, using Monte Carlo (MC) simulations. {This is necessary because the lines we want to model are highly asymmetric and require two Gaussian components to be described. The measurement uncertainties of the first and second moment therefore are not straightforward to directly derive from the best-fit model. Instead,} for each spaxel we produce a 100 new spectra by randomizing the flux using the flux error. We then fit each of these 100 spectra and take the standard deviation of the resulting distributions as the uncertainty for each parameter. We employ an iterative 4$\sigma$ clipping algorithm to exclude catastrophic fits for determining the final uncertainty.
The clipping is stopped once the change in the values of the distribution is less than 10\%.

Beyond the uncertainties calculated from the MC simulation, we also employ the estimated peak S/N to exclude spaxels with weak or spurious lines from our analysis. Throughout the paper we use an S/N limit of 3 in an effort to maximize the amount of information extracted from the spectra. Given the high quality of the GMOS data, the choice of S/N limit does not affect our results significantly.

\subsection{Analysis}
\label{sec:analysis}

The power of IFU data lies in the combination of spatial and spectral information, effectively producing a three dimensional parameter space within which both morphological and kinematical features can be traced. Here we briefly lay out the analysis steps 
and the particular information that we extract from them.

\begin{enumerate}
\item Flux maps: We investigate the spatial distributions of the continuum and emission lines flux. Line fluxes are measured by integrating the flux over (a) the whole emission line profile, (b) the broad component, and (c) the narrow component. Flux maps provide information both in terms of the extent of each component as well as a measure of trustworthiness of subsequent maps, as all kinematic measurements are flux-weighted and S/N dependent.
\item Kinematics maps: We produce stellar velocity maps based on the line-of-sight velocities measured from stellar absorption lines. We also obtain gas velocity and velocity dispersion maps, which are divided into individual emission components (i.e., broad and narrow). Gas velocity dispersion maps are compared to the flux-weighted $\sigma_{*}$ measured from the integrated SDSS spectra\footnote{The quality of the GMOS data, does not allow us to calculate a $\sigma_{*}$ at each spaxel.}).
\item Kinematic and spatial decomposition: We use the velocity-velocity dispersion (VVD) diagram (e.g., \citealt{Woo2015b}) to explore the kinematics of the gas as a function of its spatial distribution.
We study the distribution of spaxels with respect to the systemic velocity and $\sigma_{*}$, as well as the spaxel distance from the center of the galaxy.
\item Outflow structure: We project the 2D IFU maps on the radial axis and investigate both the \mbox{[O\,\textsc{iii}]} velocity and velocity dispersion as a function of distance from the center of the galaxy. We define the effective radius, r$_{\mathrm{eff}}$, of the outflow based on either the broad \mbox{[O\,\textsc{iii}]} or broad H$\alpha$ emission flux distribution and investigate consistent kinematic patterns as a function of this relative distance unit. 
\item Integrated outflow kinematics: Based on the previous steps, we calculate the outflow sizes and the spatially integrated kinematic properties for \mbox{[O\,\textsc{iii}]} and H$\alpha$ for total emission profiles, as well as broad and narrow emission components separately.
\end{enumerate}

\section{Results}
\label{sec:results}


\begin{figure*}[bpt]
\begin{center}
\raisebox{-0.5\height}{\includegraphics[width=0.16\textwidth,angle=0]{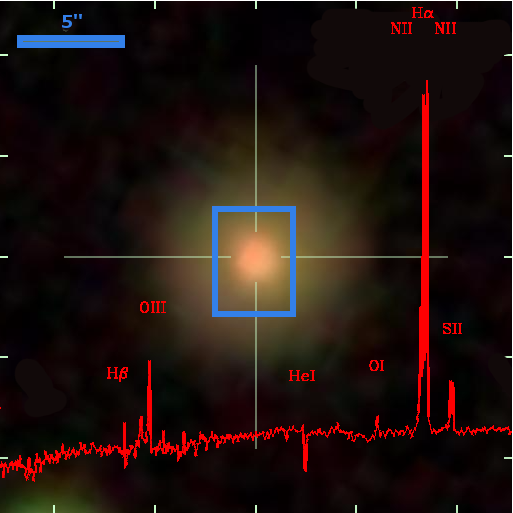}}
\raisebox{-0.5\height}{\includegraphics[width=0.16\textwidth,angle=0]{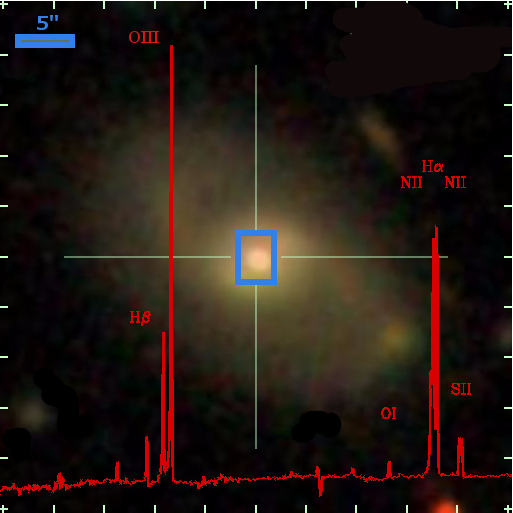}}
\raisebox{-0.5\height}{\includegraphics[width=0.16\textwidth,angle=0]{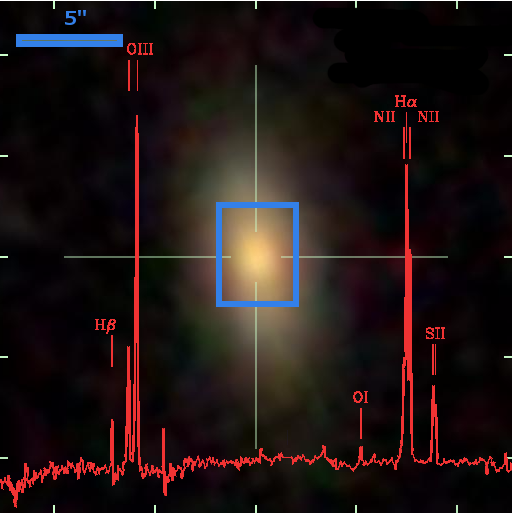}}
\raisebox{-0.5\height}{\includegraphics[width=0.16\textwidth,angle=0]{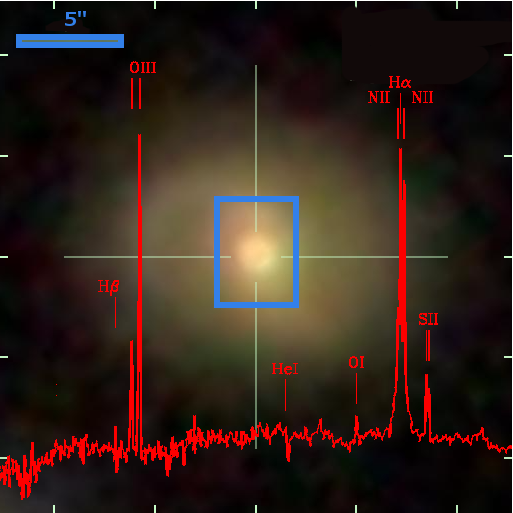}}
\raisebox{-0.5\height}{\includegraphics[width=0.16\textwidth,angle=0]{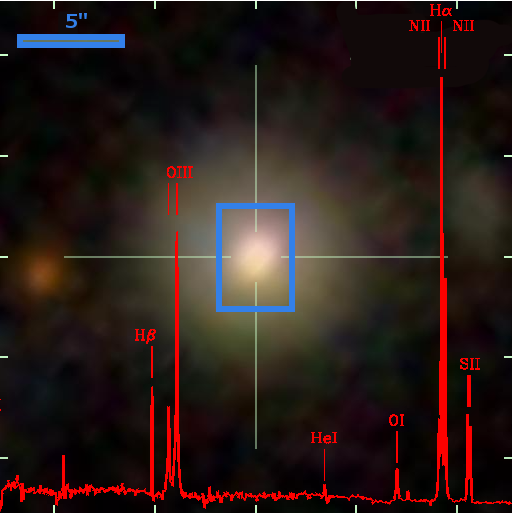}}
\raisebox{-0.5\height}{\includegraphics[width=0.16\textwidth,angle=0]{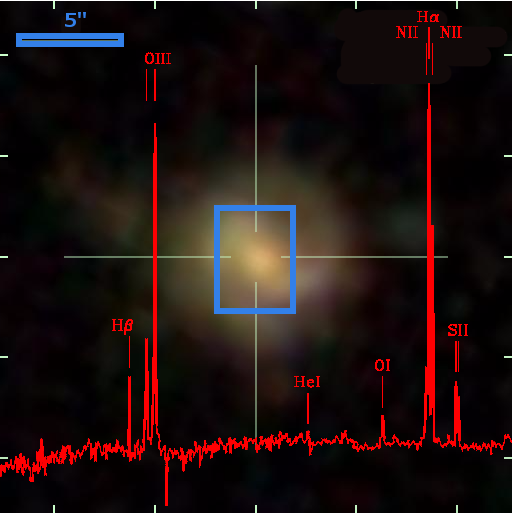}}\\
\raisebox{-0.5\height}{\includegraphics[width=0.16\textwidth,angle=0,trim={50 60 20 50},clip]{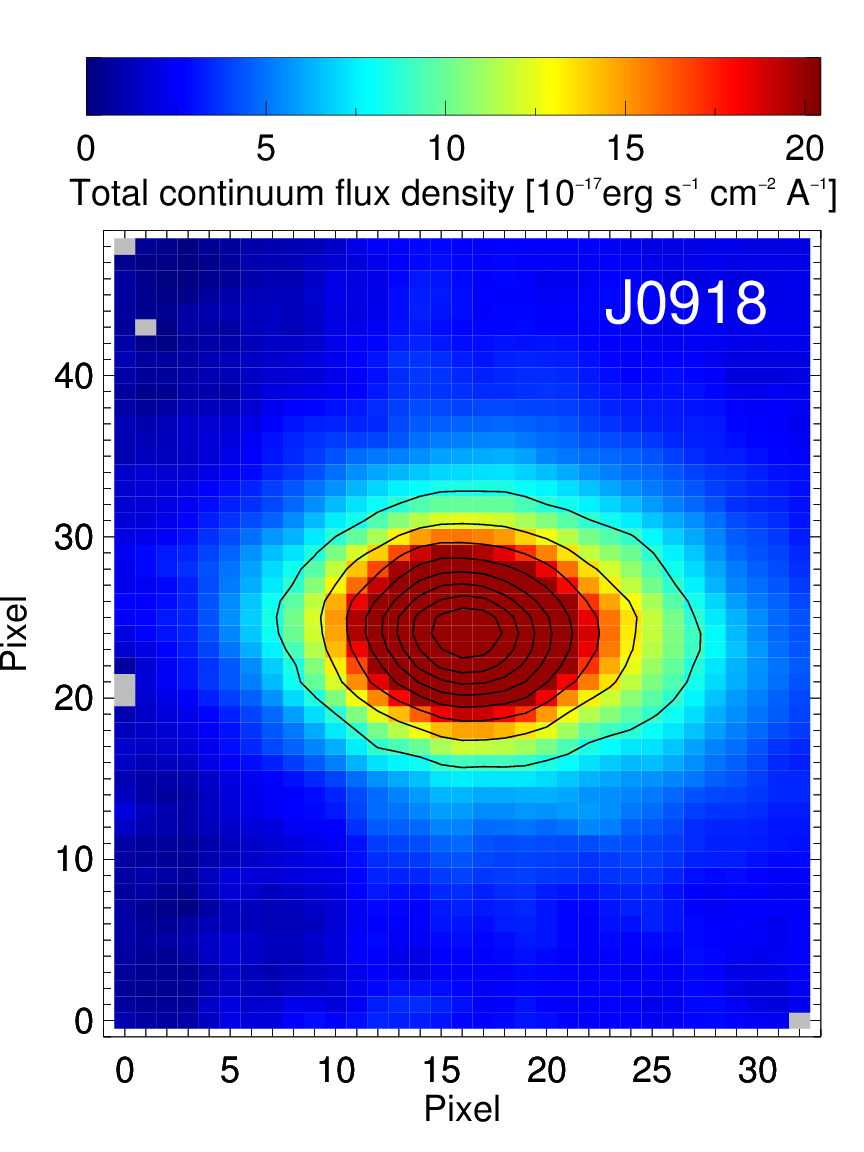}}
\raisebox{-0.5\height}{\includegraphics[width=0.16\textwidth,angle=0,trim={50 60 20 50},clip]{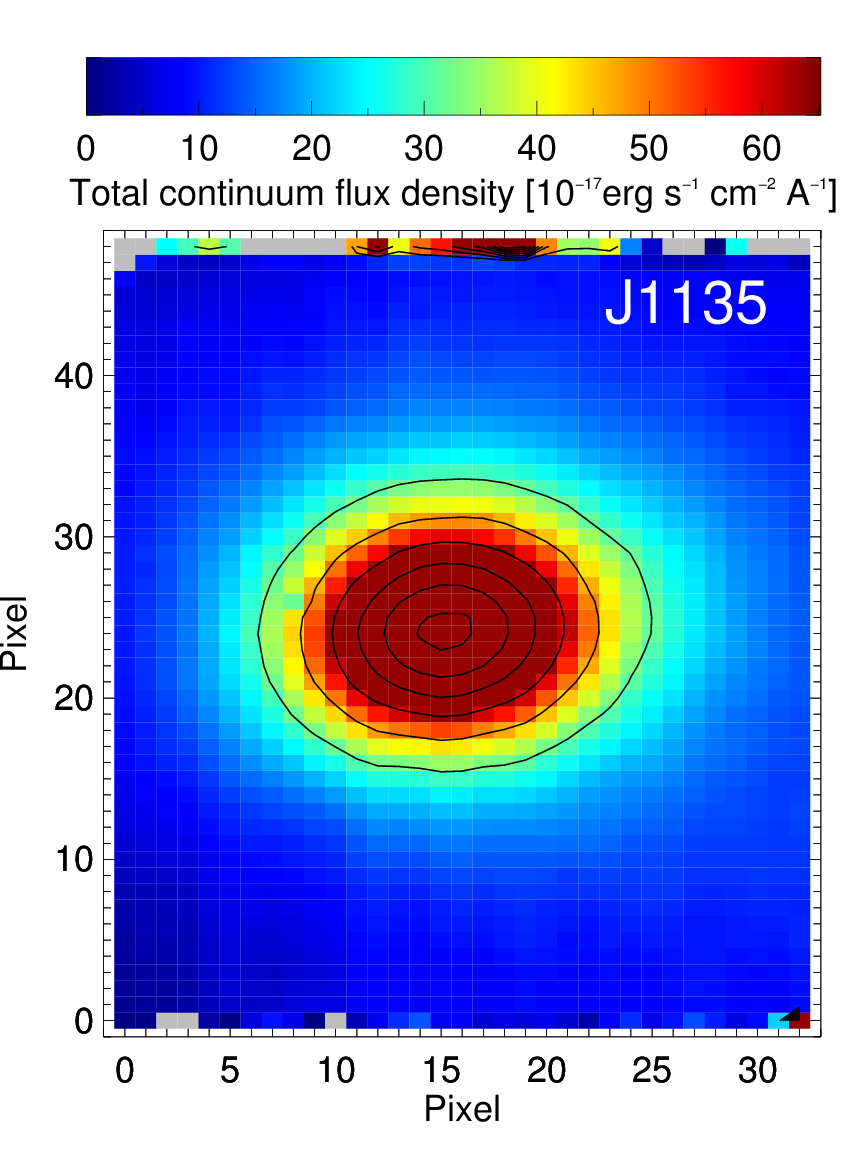}}
\raisebox{-0.5\height}{\includegraphics[width=0.16\textwidth,angle=0,trim={50 60 20 50},clip]{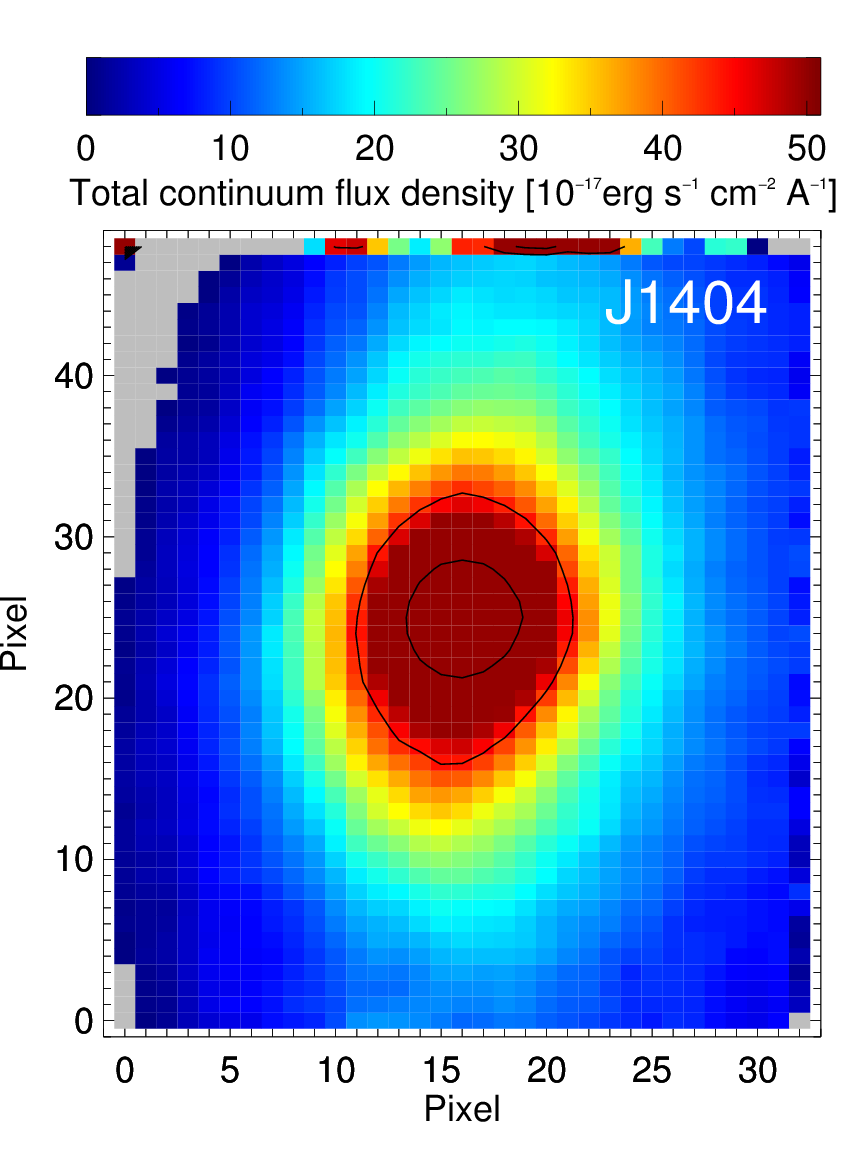}}
\raisebox{-0.5\height}{\includegraphics[width=0.16\textwidth,angle=0,trim={50 60 20 50},clip]{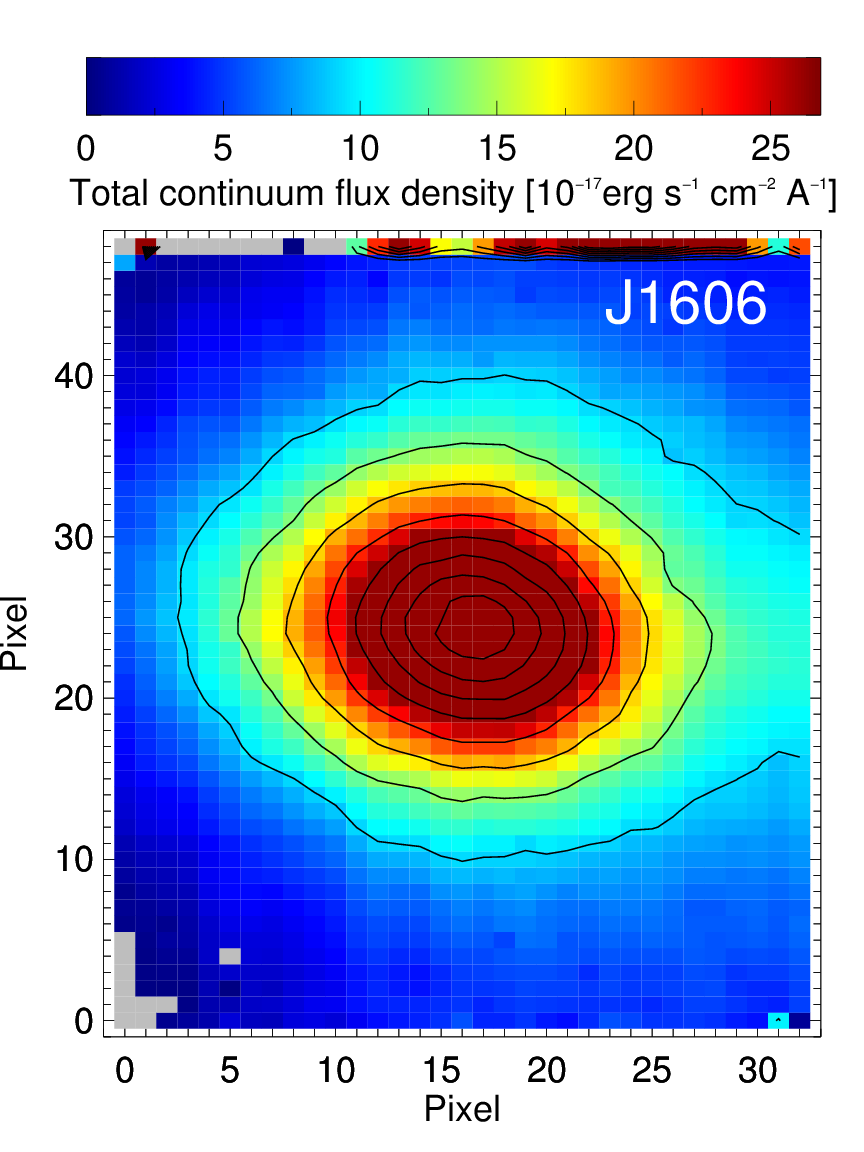}}
\raisebox{-0.5\height}{\includegraphics[width=0.16\textwidth,angle=0,trim={50 60 20 50},clip]{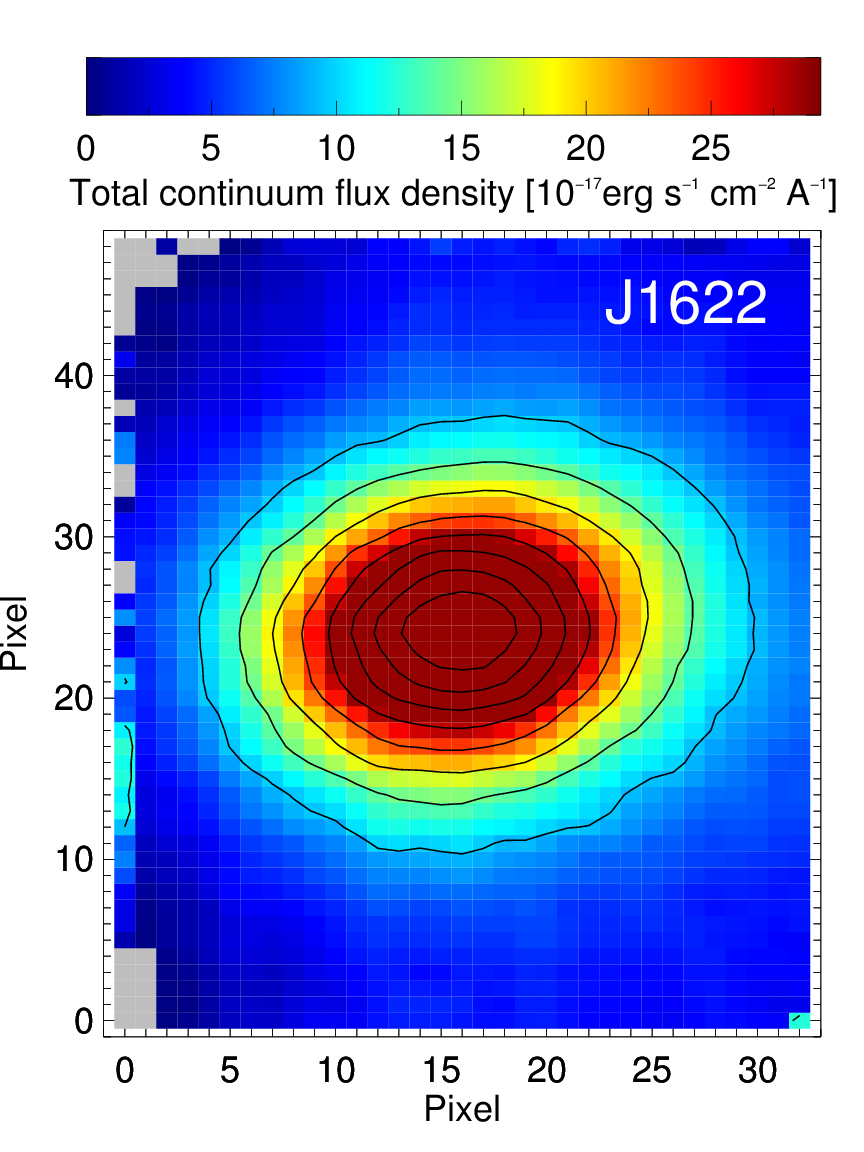}}
\raisebox{-0.5\height}{\includegraphics[width=0.16\textwidth,angle=0,trim={50 60 20 50},clip]{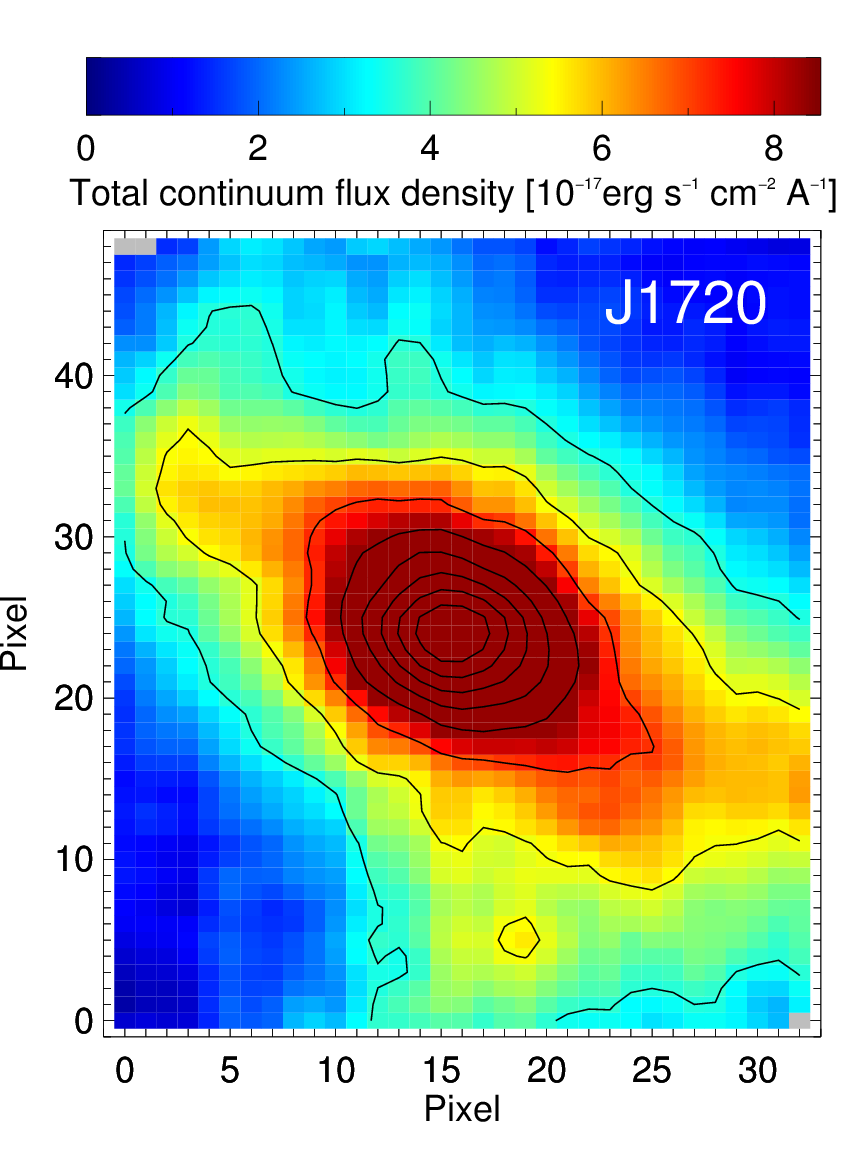}}\\
\raisebox{-0.5\height}{\includegraphics[width=0.16\textwidth,angle=0,trim={50 60 30 50},clip]{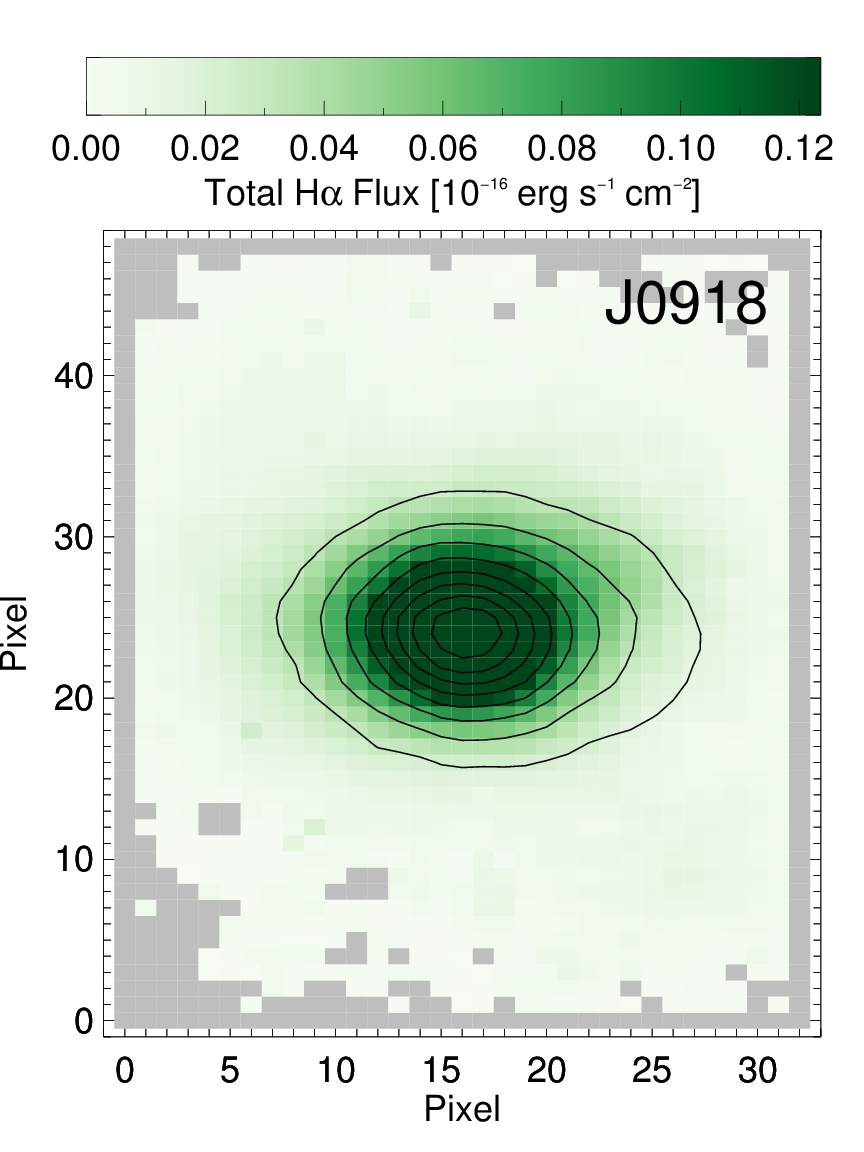}}
\raisebox{-0.5\height}{\includegraphics[width=0.16\textwidth,angle=0,trim={50 60 30 50},clip]{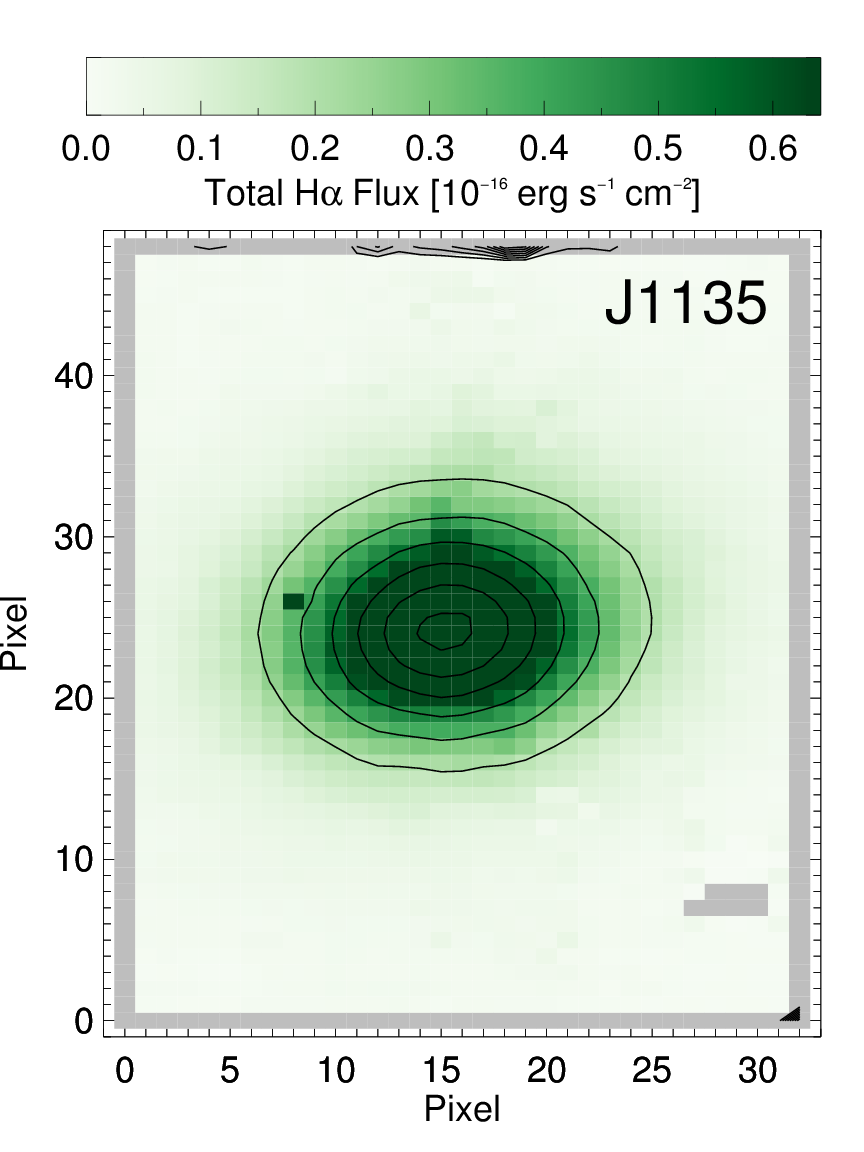}}
\raisebox{-0.5\height}{\includegraphics[width=0.16\textwidth,angle=0,trim={50 60 30 50},clip]{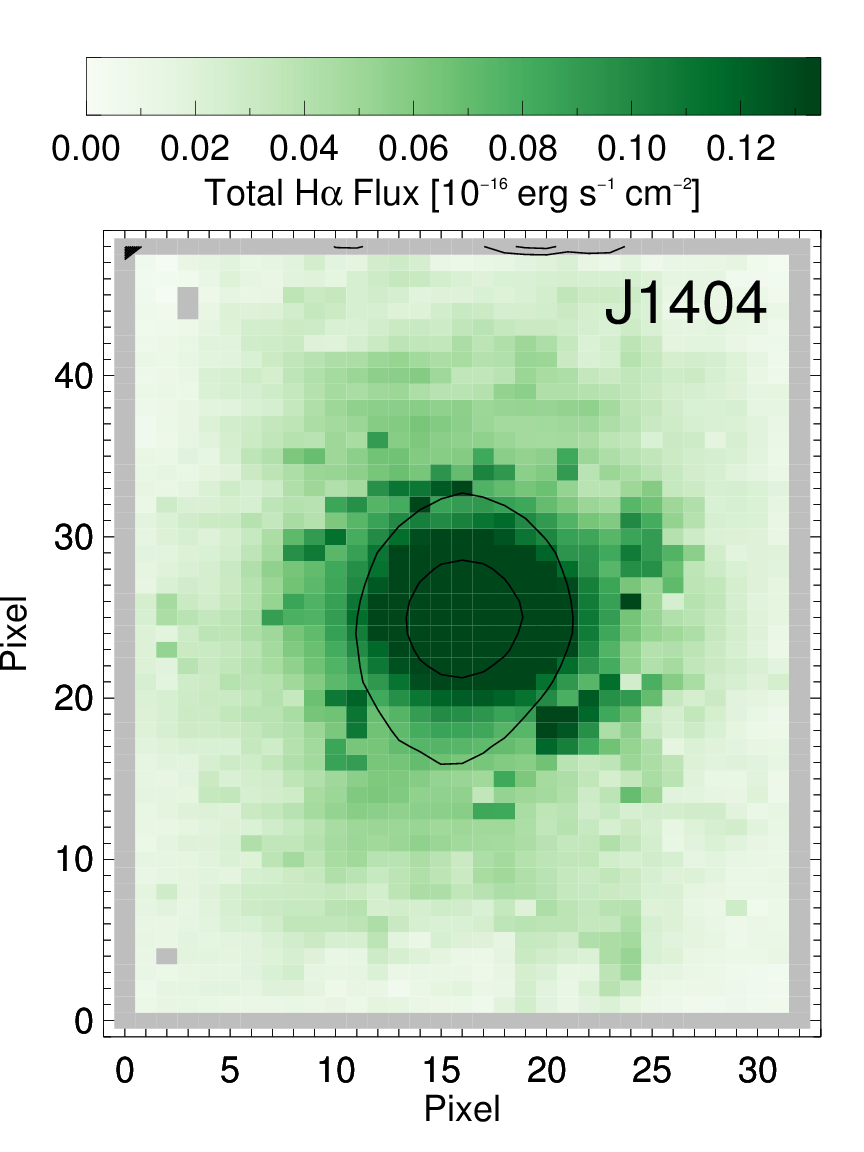}}
\raisebox{-0.5\height}{\includegraphics[width=0.16\textwidth,angle=0,trim={50 60 30 50},clip]{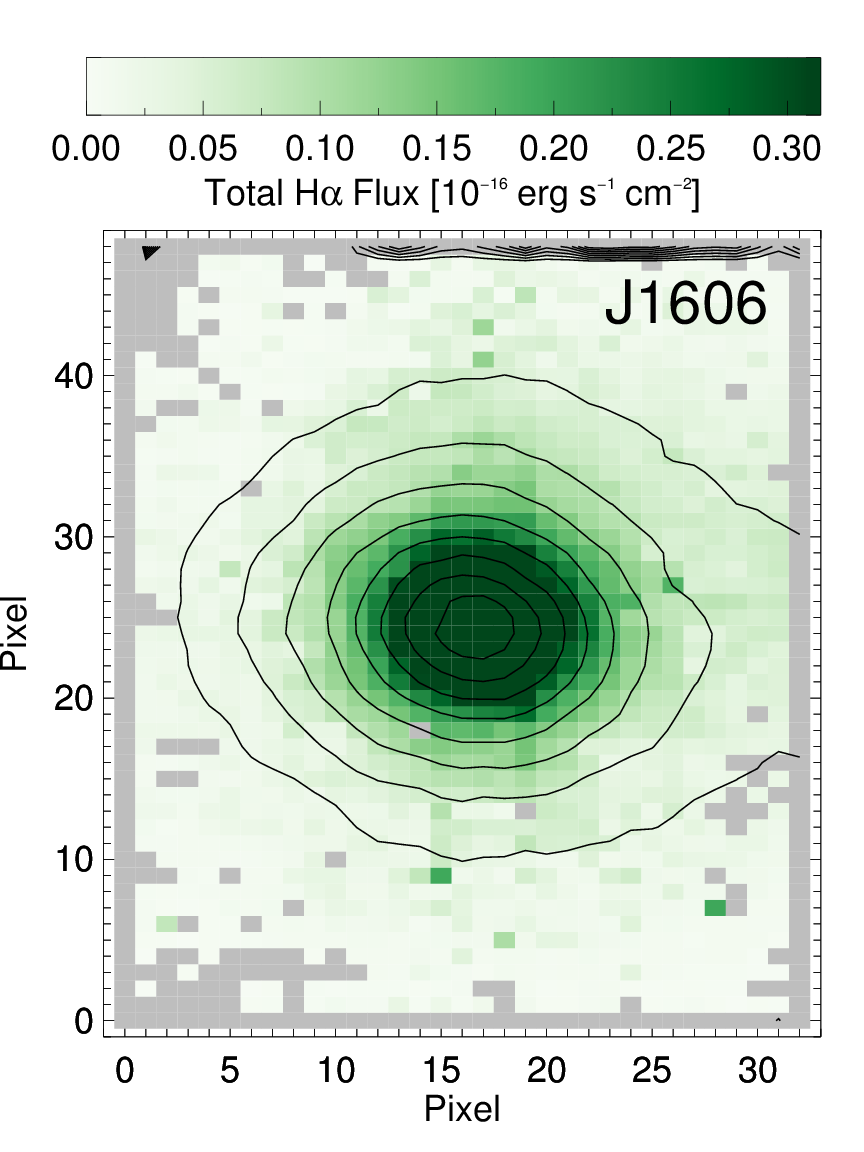}}
\raisebox{-0.5\height}{\includegraphics[width=0.16\textwidth,angle=0,trim={50 60 30 50},clip]{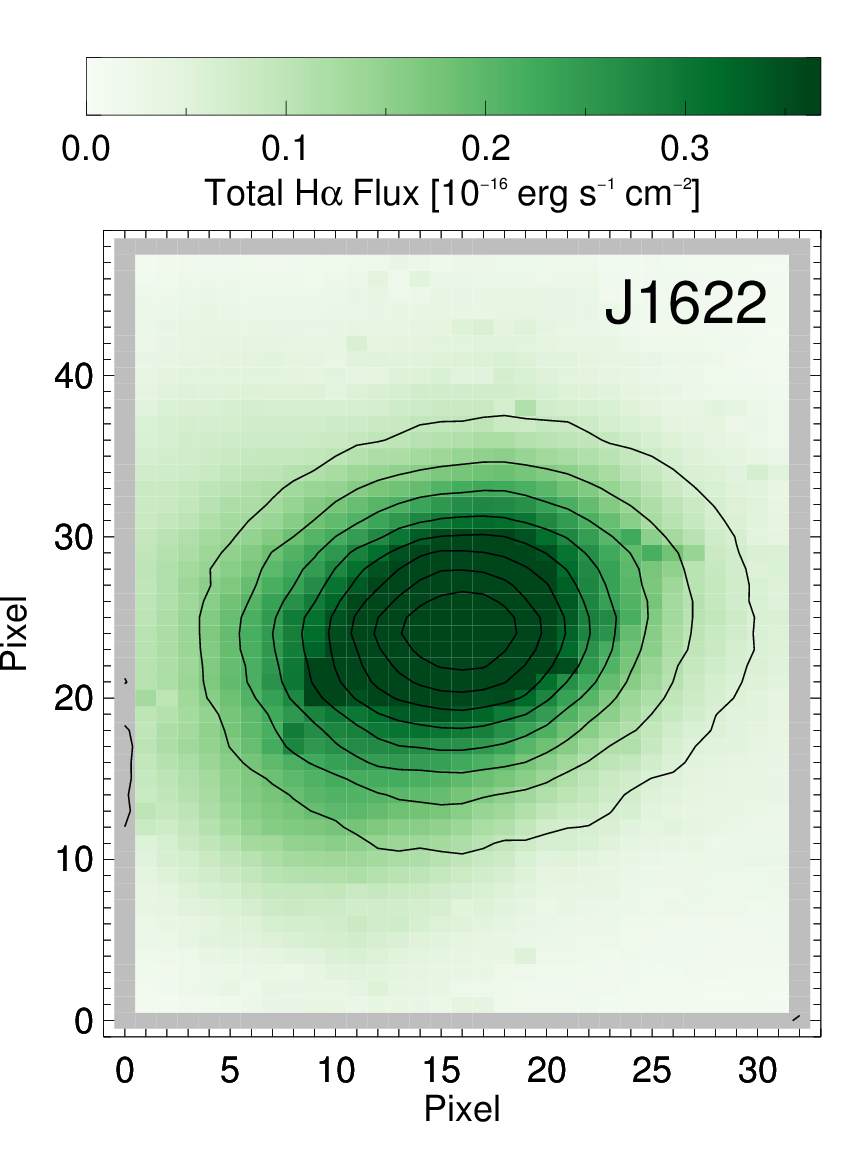}}
\raisebox{-0.5\height}{\includegraphics[width=0.16\textwidth,angle=0,trim={50 60 15 50},clip]{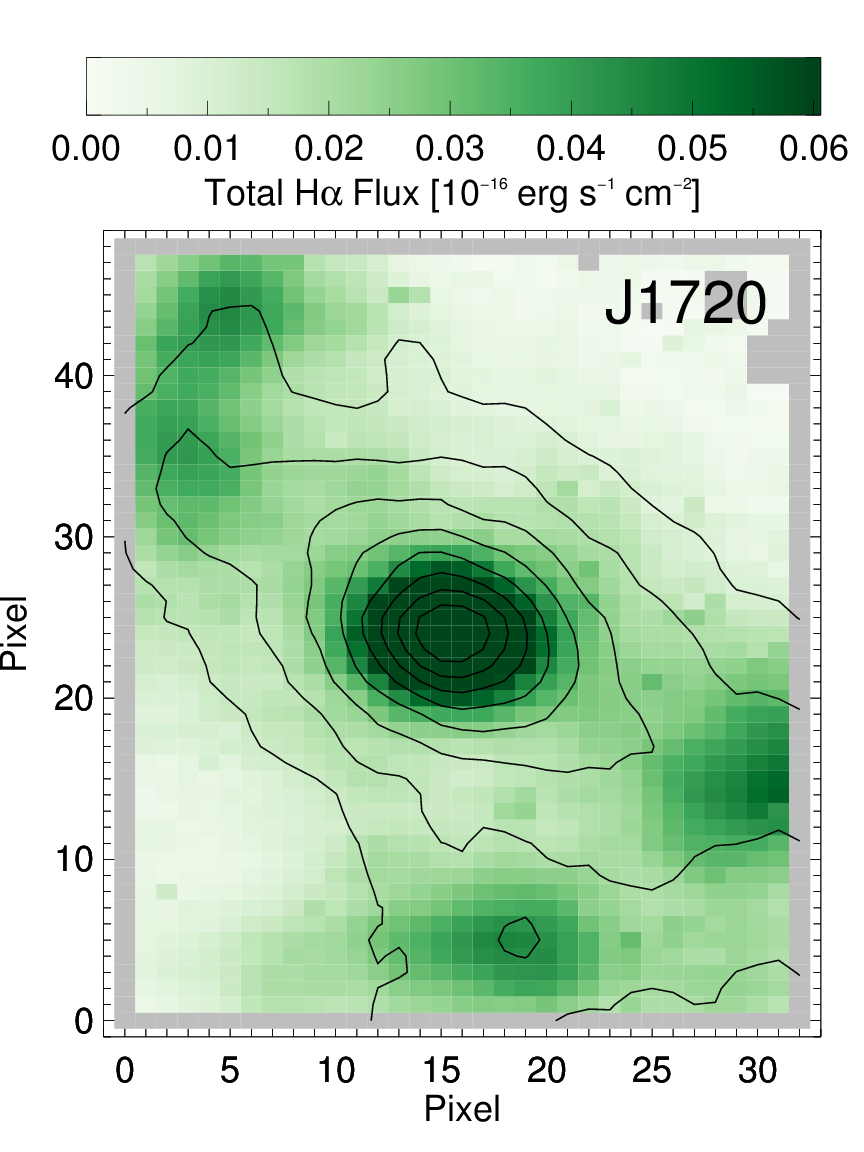}}\\
\raisebox{-0.5\height}{\includegraphics[width=0.16\textwidth,angle=0,trim={50 60 30 50},clip]{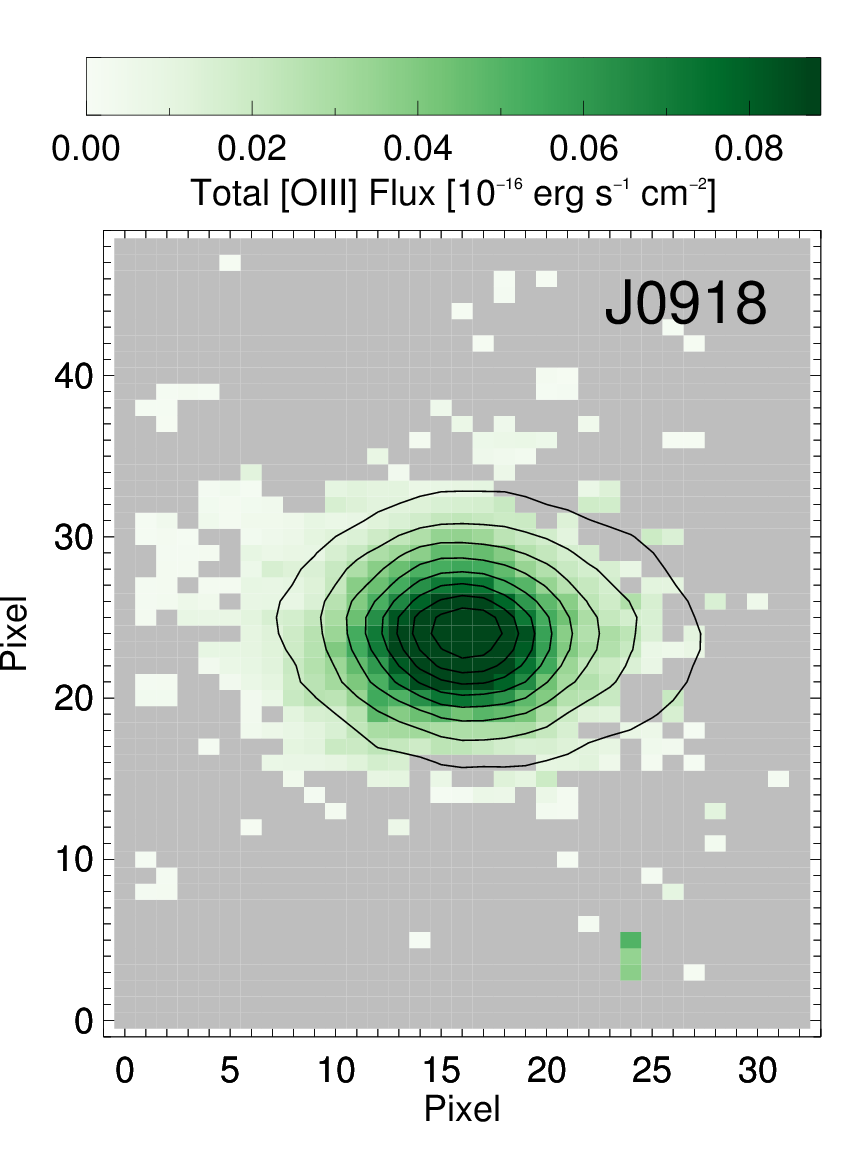}}
\raisebox{-0.5\height}{\includegraphics[width=0.16\textwidth,angle=0,trim={50 60 30 50},clip]{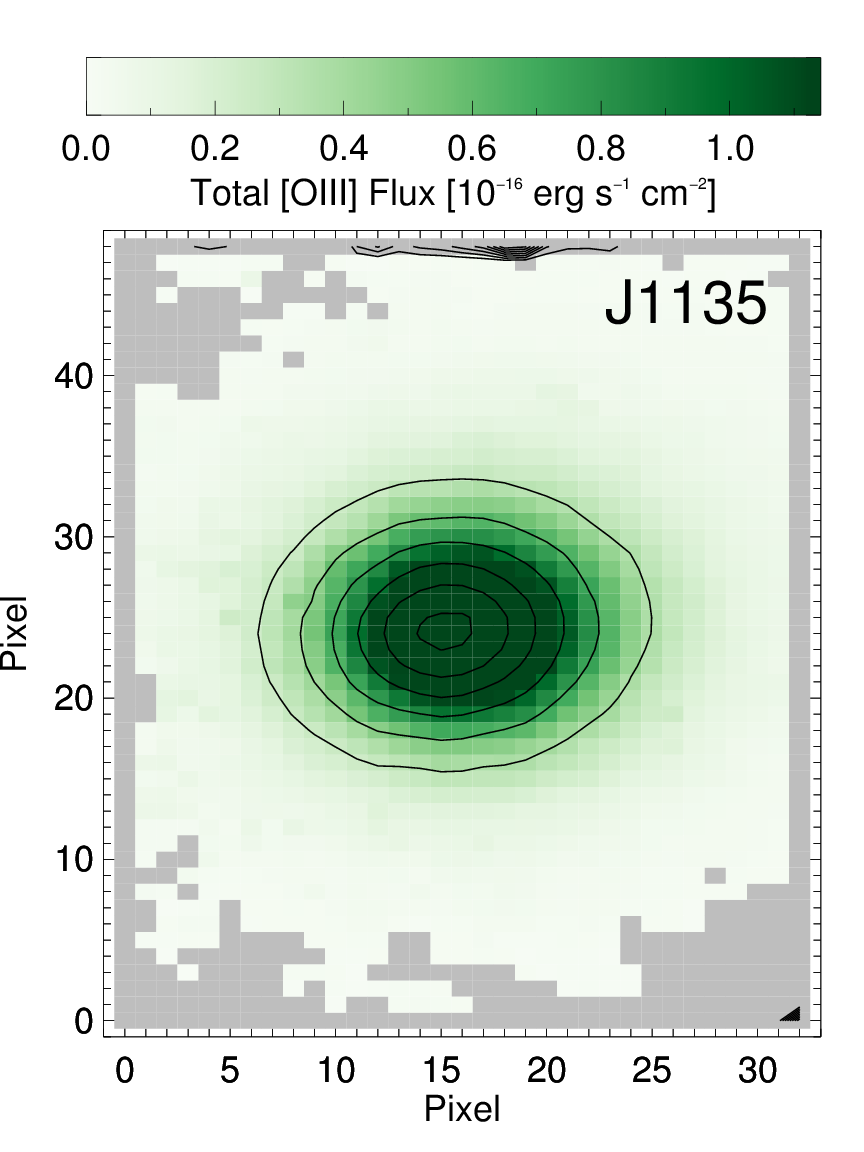}}
\raisebox{-0.5\height}{\includegraphics[width=0.16\textwidth,angle=0,trim={50 60 30 50},clip]{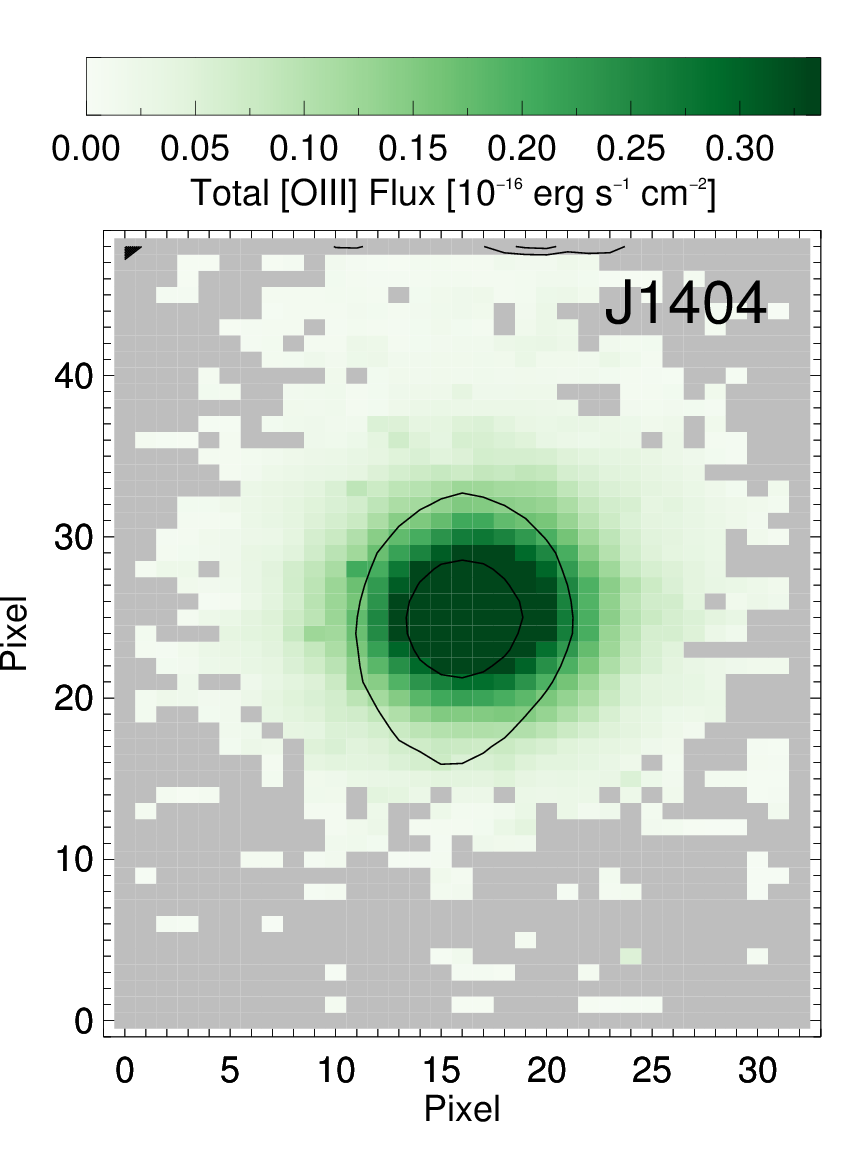}}
\raisebox{-0.5\height}{\includegraphics[width=0.16\textwidth,angle=0,trim={50 60 30 50},clip]{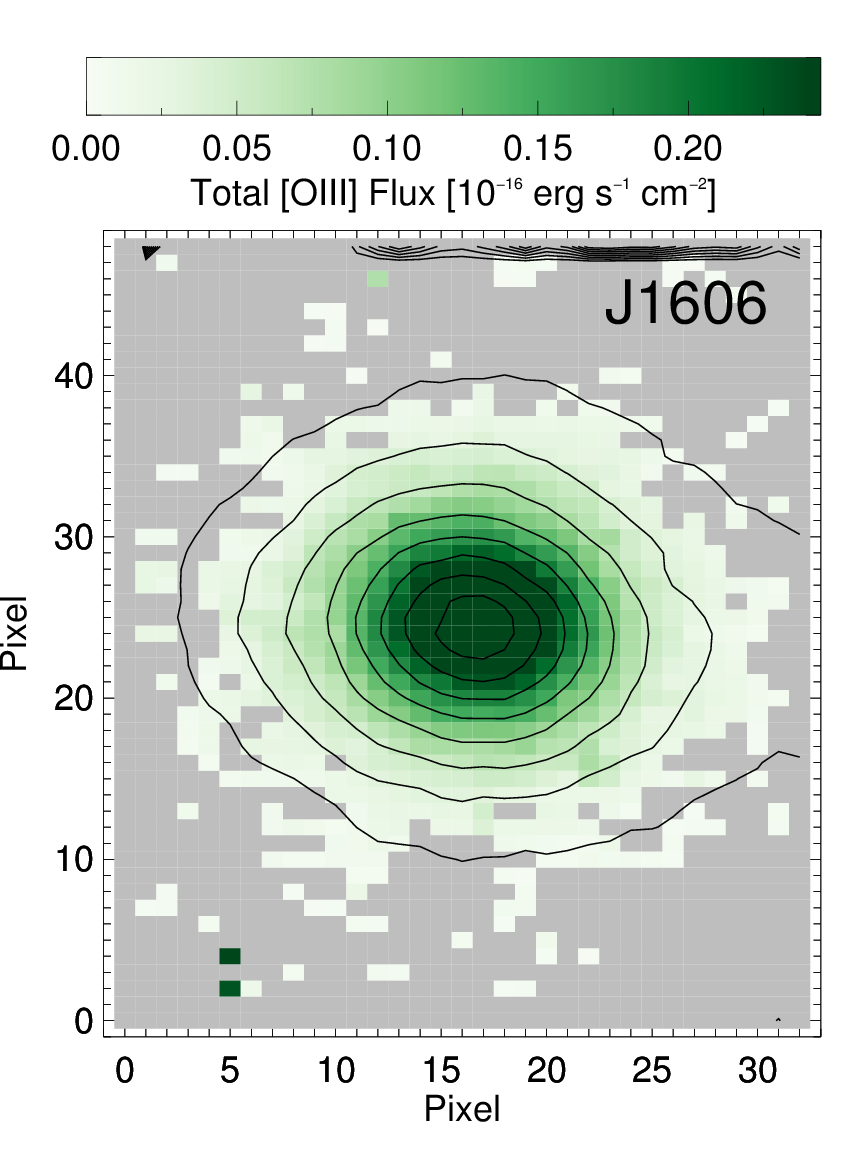}}
\raisebox{-0.5\height}{\includegraphics[width=0.16\textwidth,angle=0,trim={50 60 30 50},clip]{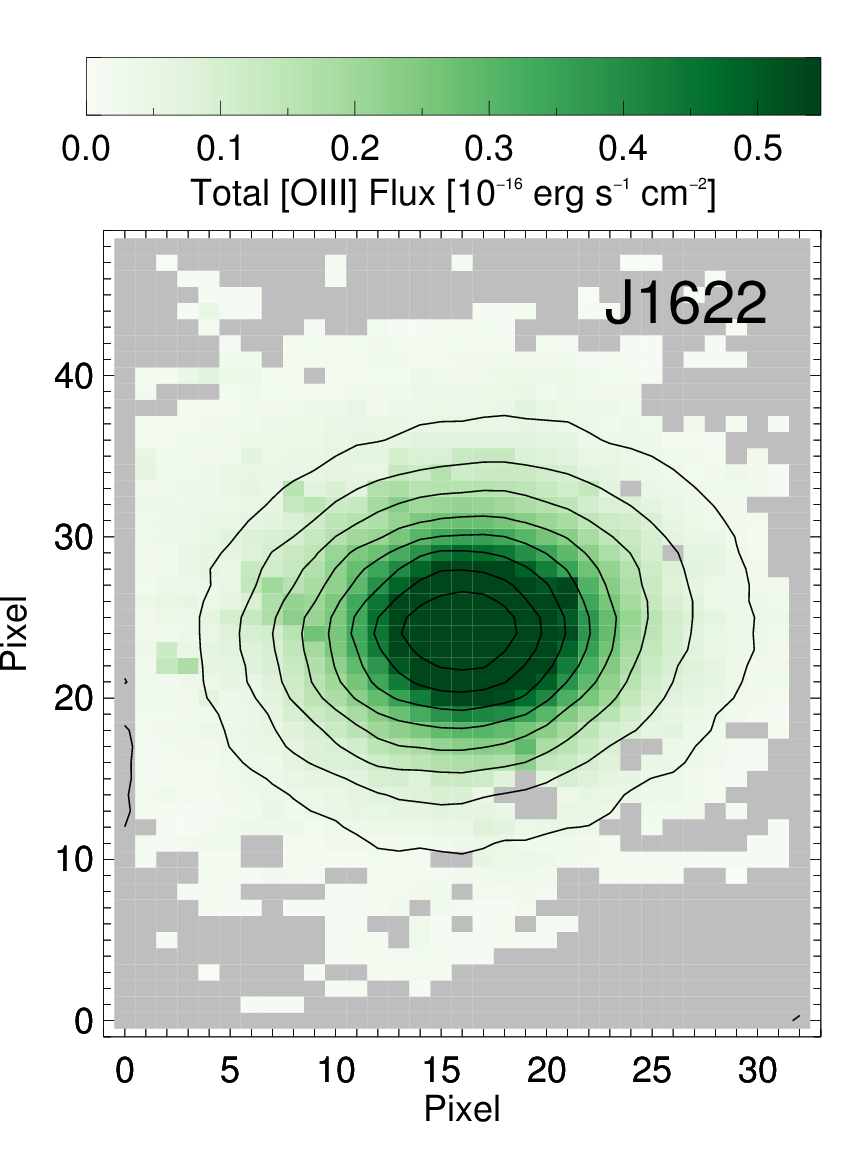}}
\raisebox{-0.5\height}{\includegraphics[width=0.16\textwidth,angle=0,trim={50 60 30 50},clip]{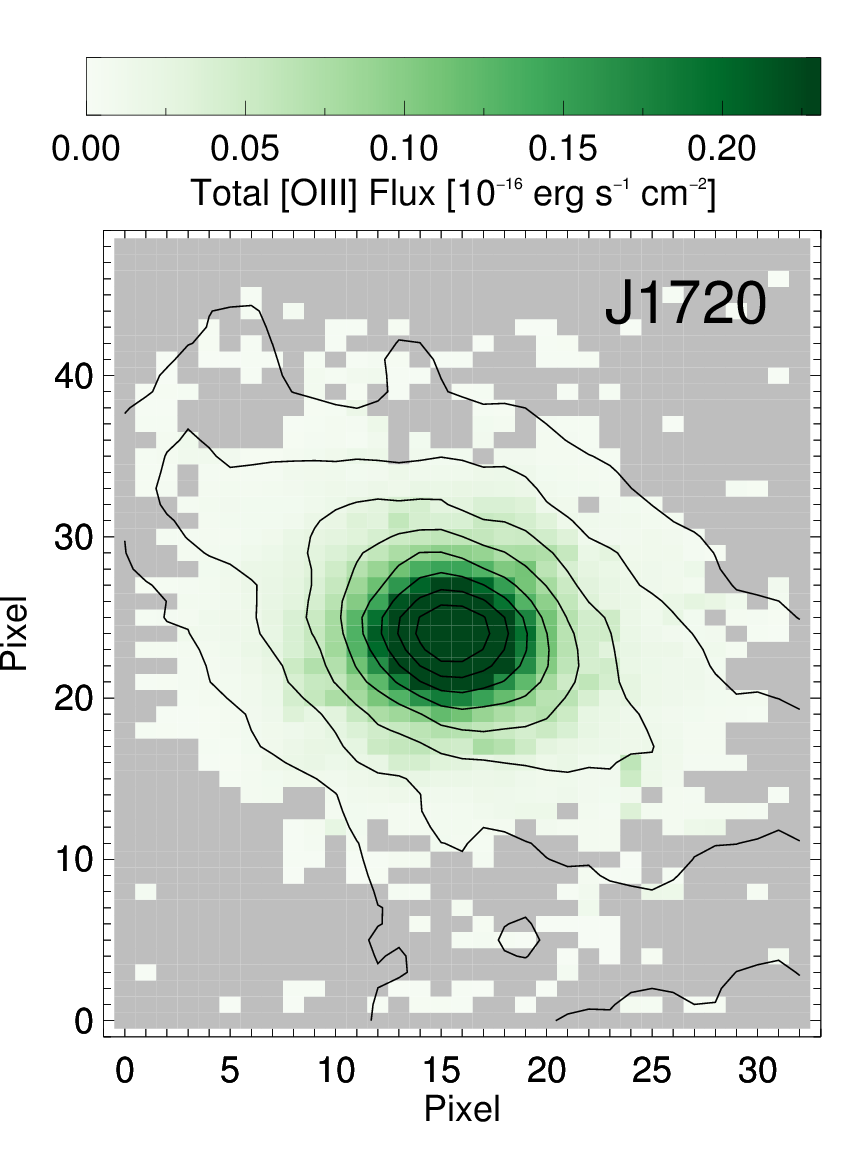}}
\caption{Top: SDSS \textit{gri} composite images and spectra. The FoV of GMOS IFU ($5\farcs0\times3\farcs5$) is shown with blue boxes, while the horizontal blue bars show 5\farcs0 scale. Second row: IFU flux maps integrated over the full wavelength range. 
Third row: H$\alpha$ flux maps from the best fit model profile. Fourth row: \mbox{[O\,\textsc{iii}]} $\lambda$5007\AA\ flux maps from the best fit model profile. For the second to fourth rows, contours show the continuum flux at 10\% intervals from the peak. Grey spaxels indicate weak or non-detection of continuum or emission line with S/N$<$3.}
\label{fig:all_flux}
\end{center}
\end{figure*}

{At the redshift of the targets (z$<0.1$), the GMOS FoV covers $\sim3-9$ kpc diameter with a spatial resolution of $\sim0.5-1$ kpc, given our seeing-limited observations. Based on their SDSS photometry, all six sources show face-on stellar disks with the exception of J1404 that appears strongly inclined (b/a major to minor axis ratio of 0.38).  For at least three sources (J1135, J1606, and J1720) we observe spiral arms in their composite SDSS images (Fig. \ref{fig:all_flux}, top). The IFU continuum flux maps (second row of Fig. \ref{fig:all_flux}) show an elongation in the West-East direction due to PSF effects, as the spectrophotometric star observed during the observing run shows a similar elongation. Integrated H$\alpha$ flux maps (third row of Fig. \ref{fig:all_flux}) show extended emission that follows the continuum emission. In contrast, the} \OIII\ {integrated flux maps appear more concentrated with fairly symmetric spatial flux distribution. For all sources we detect} \OIII\ {emission (SN$>3$) out to several kpc. More detailed comments on the individual sources can be found in the Appendix.}

\begin{figure*}[bpt]
\begin{center}
{\Large Stellar velocity}\\
\raisebox{-0.5\height}{\includegraphics[width=0.16\textwidth,angle=0,trim={50 60 20 50},clip]{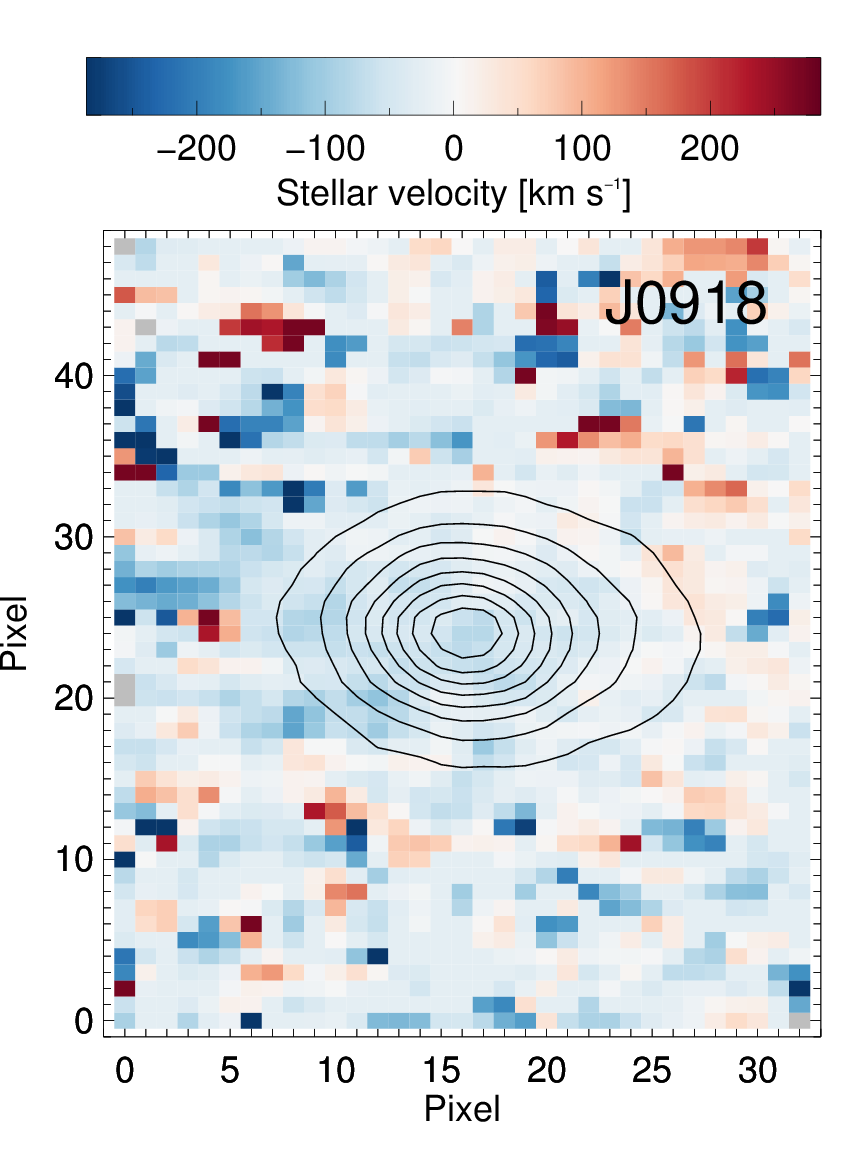}}
\raisebox{-0.5\height}{\includegraphics[width=0.16\textwidth,angle=0,trim={50 60 20 50},clip]{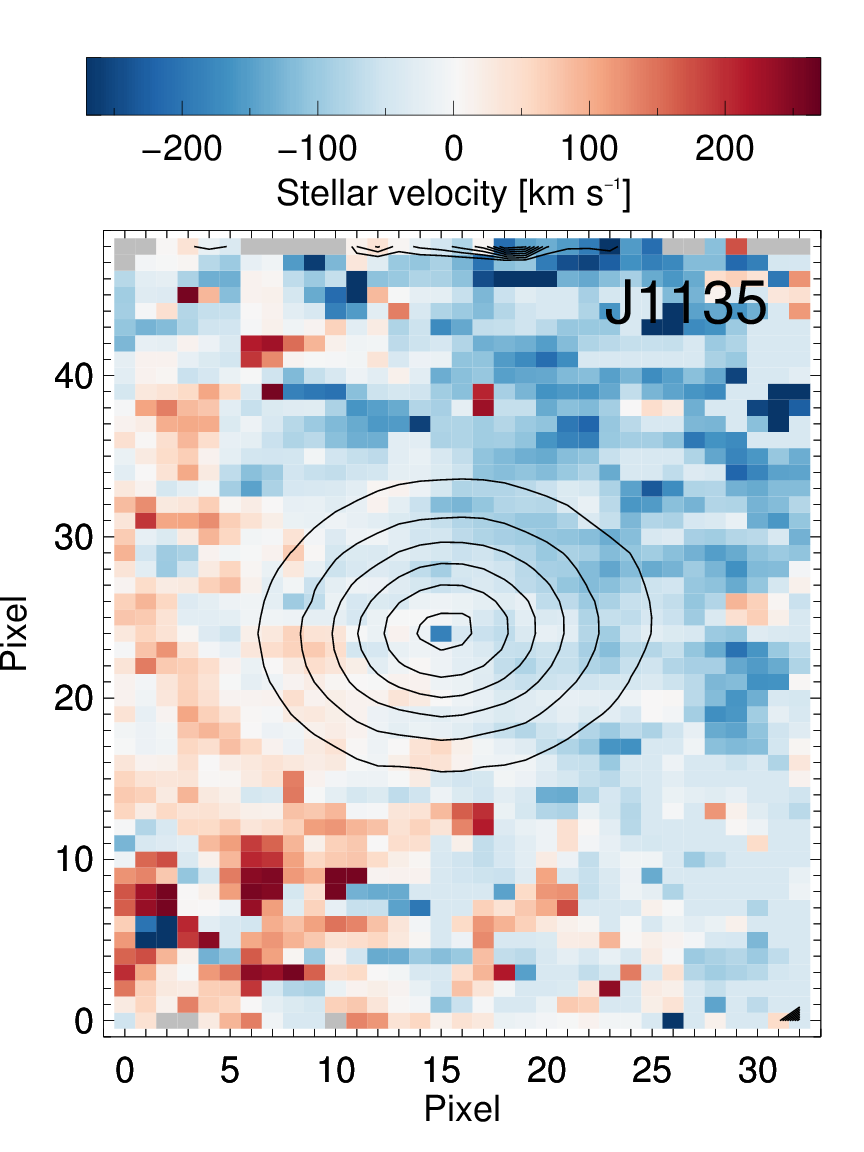}}
\raisebox{-0.5\height}{\includegraphics[width=0.16\textwidth,angle=0,trim={50 60 20 50},clip]{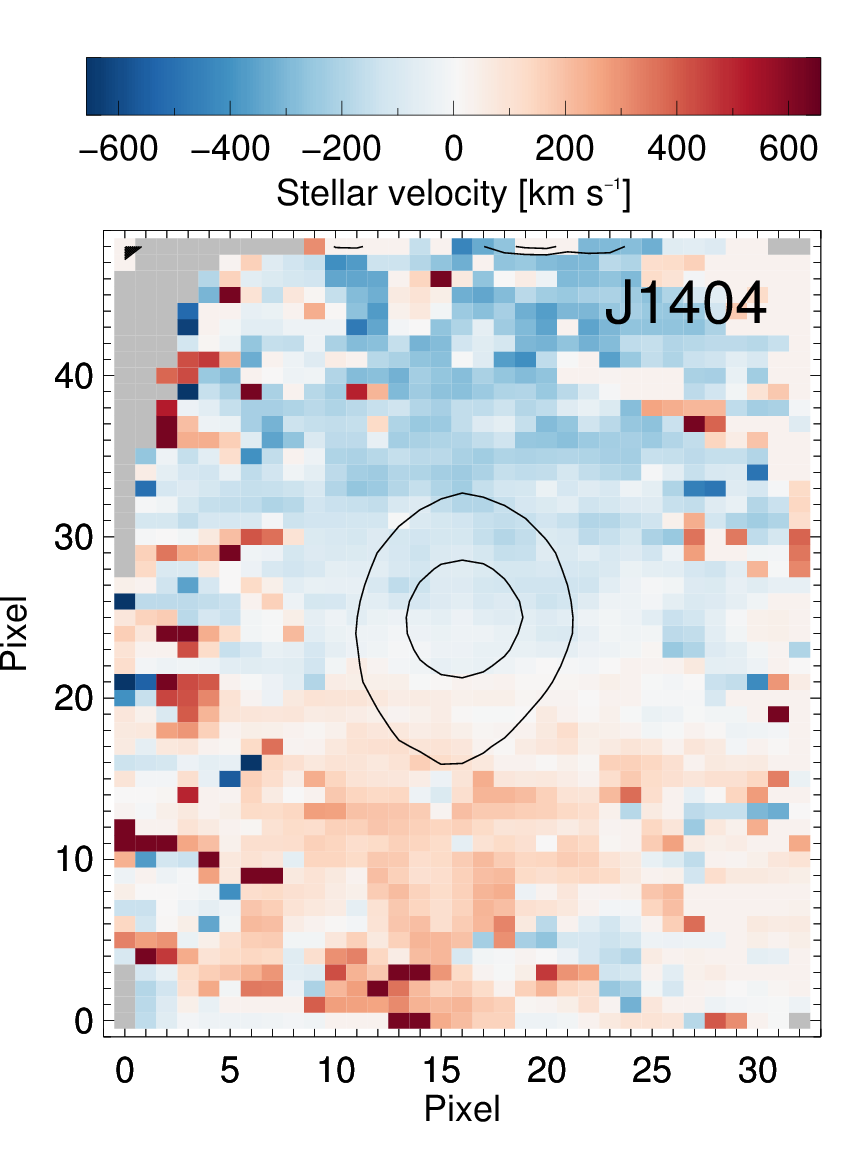}}
\raisebox{-0.5\height}{\includegraphics[width=0.16\textwidth,angle=0,trim={50 60 20 50},clip]{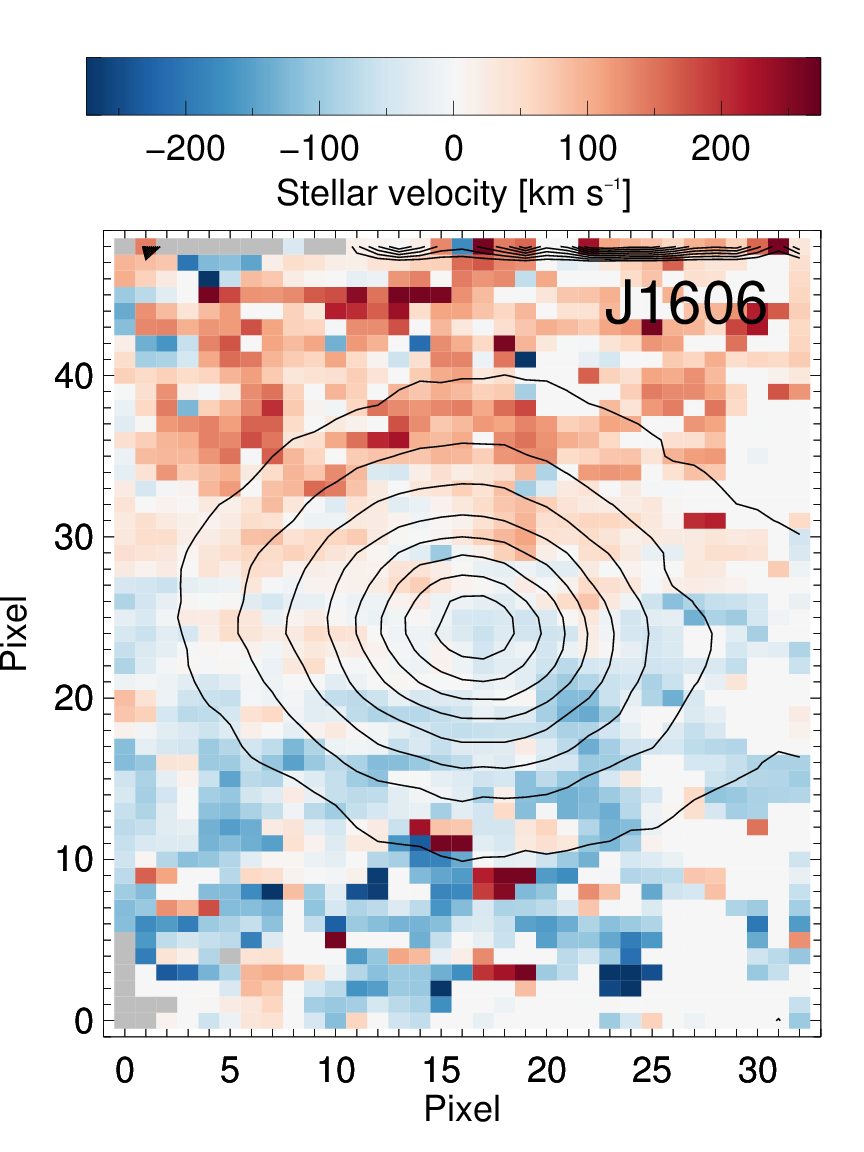}}
\raisebox{-0.5\height}{\includegraphics[width=0.16\textwidth,angle=0,trim={50 60 20 50},clip]{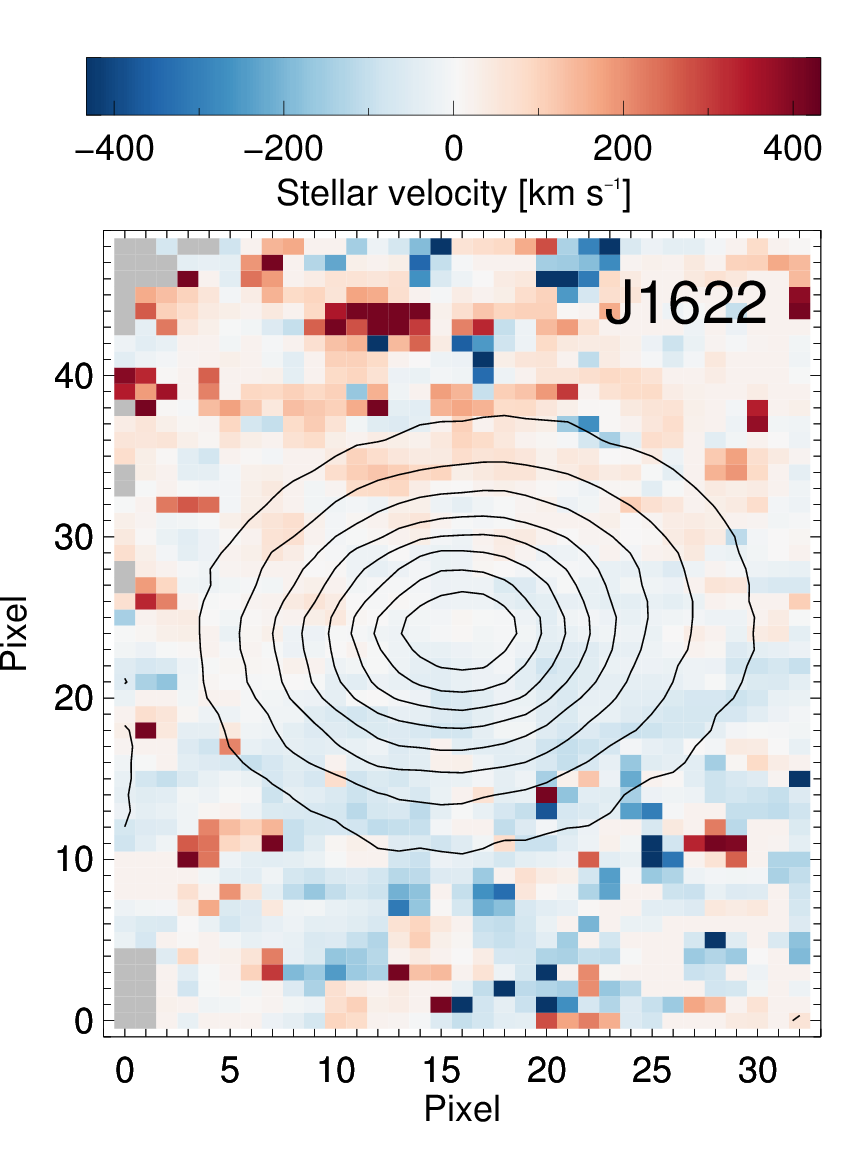}}
\raisebox{-0.5\height}{\includegraphics[width=0.16\textwidth,angle=0,trim={50 60 20 50},clip]{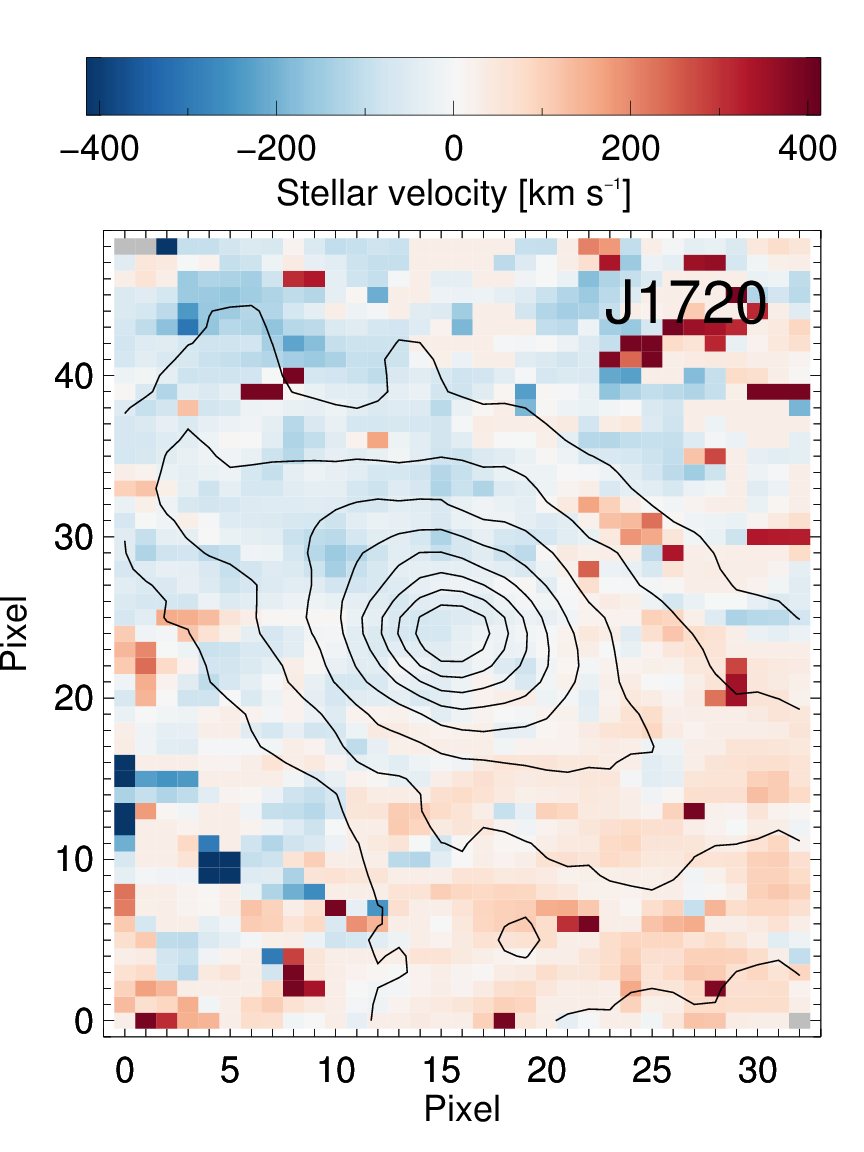}}\\
\vspace{5pt}\hrule\vspace{5pt}
{\Large Velocity}\\
\raisebox{-0.5\height}{\includegraphics[width=0.16\textwidth,angle=0,trim={50 60 40 50},clip]{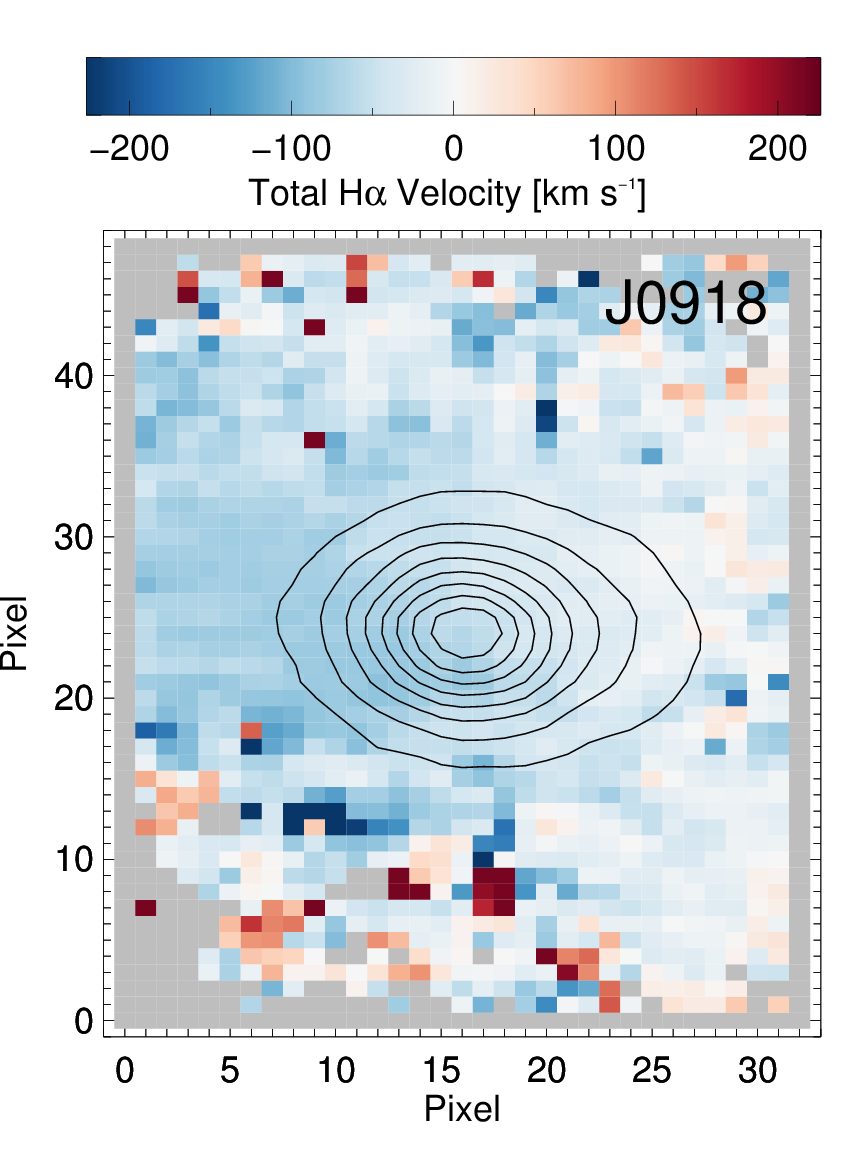}}
\raisebox{-0.5\height}{\includegraphics[width=0.16\textwidth,angle=0,trim={50 60 40 50},clip]{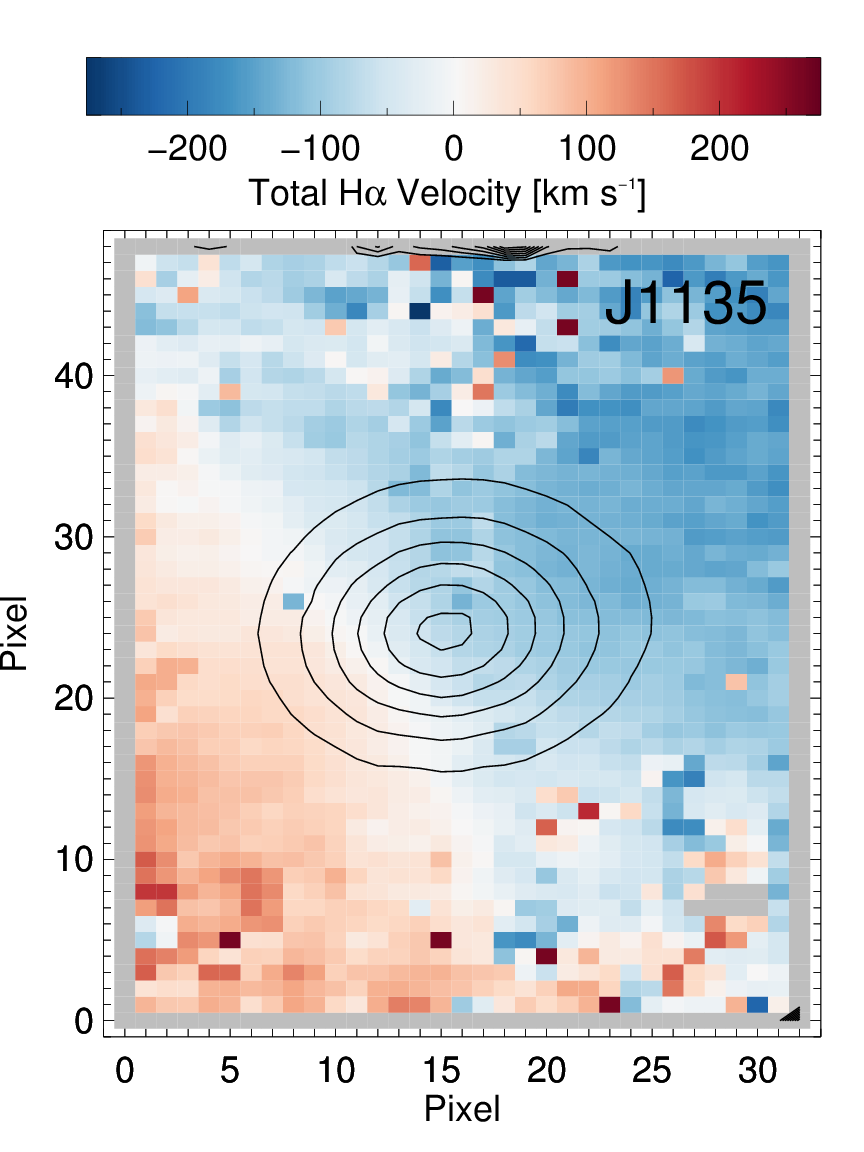}}
\raisebox{-0.5\height}{\includegraphics[width=0.16\textwidth,angle=0,trim={50 60 40 50},clip]{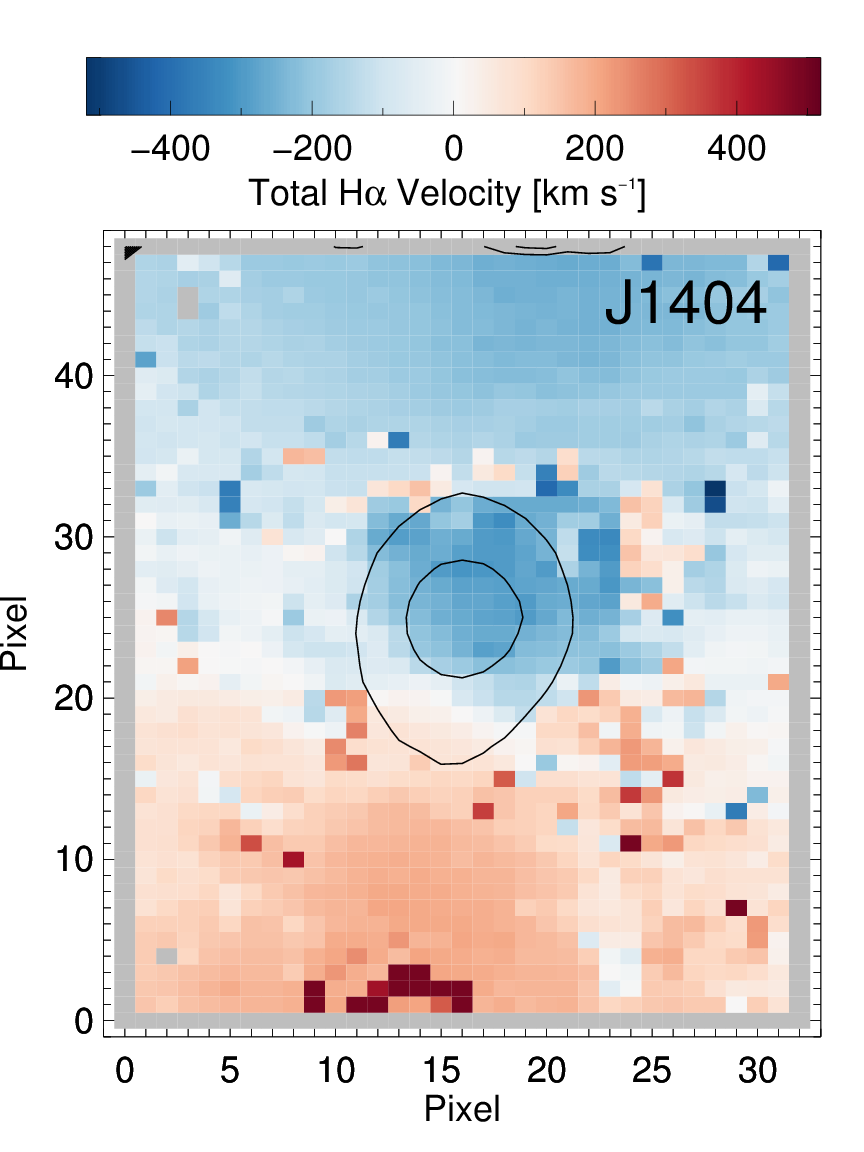}}
\raisebox{-0.5\height}{\includegraphics[width=0.16\textwidth,angle=0,trim={50 60 40 50},clip]{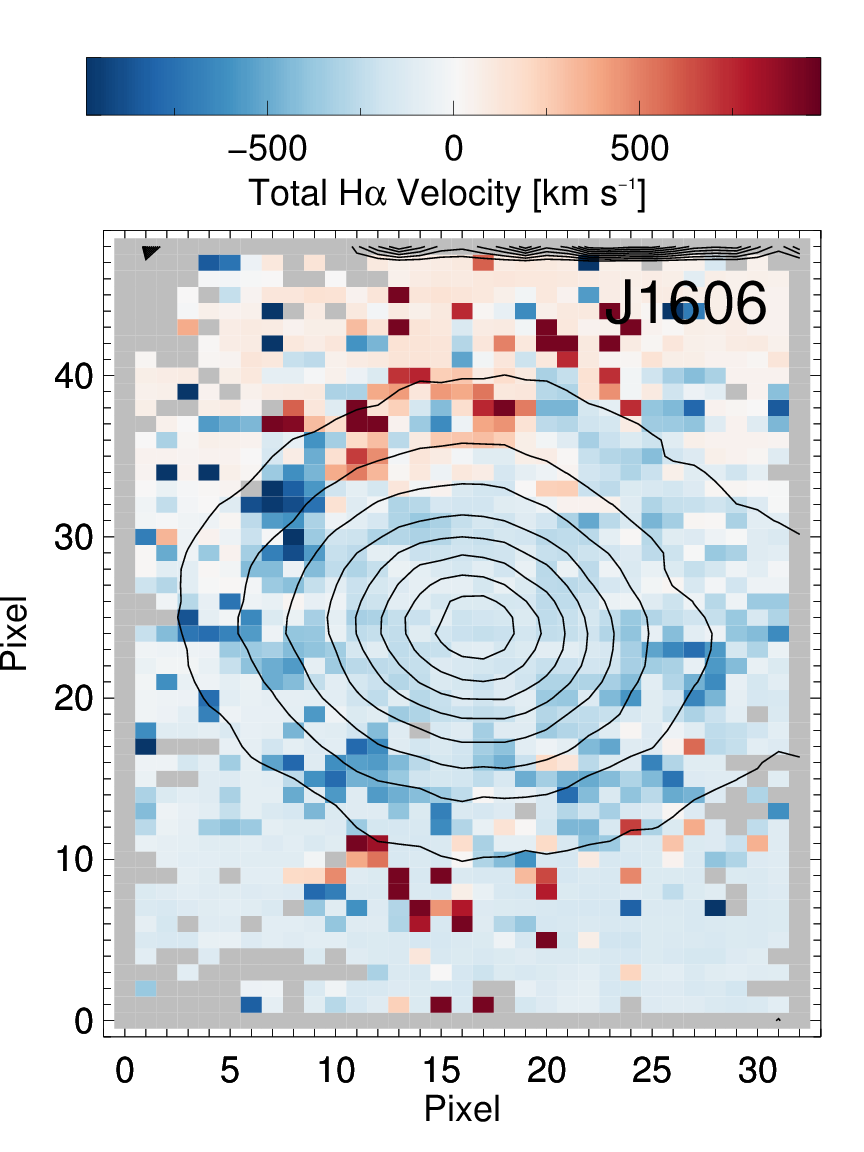}}
\raisebox{-0.5\height}{\includegraphics[width=0.16\textwidth,angle=0,trim={50 60 40 50},clip]{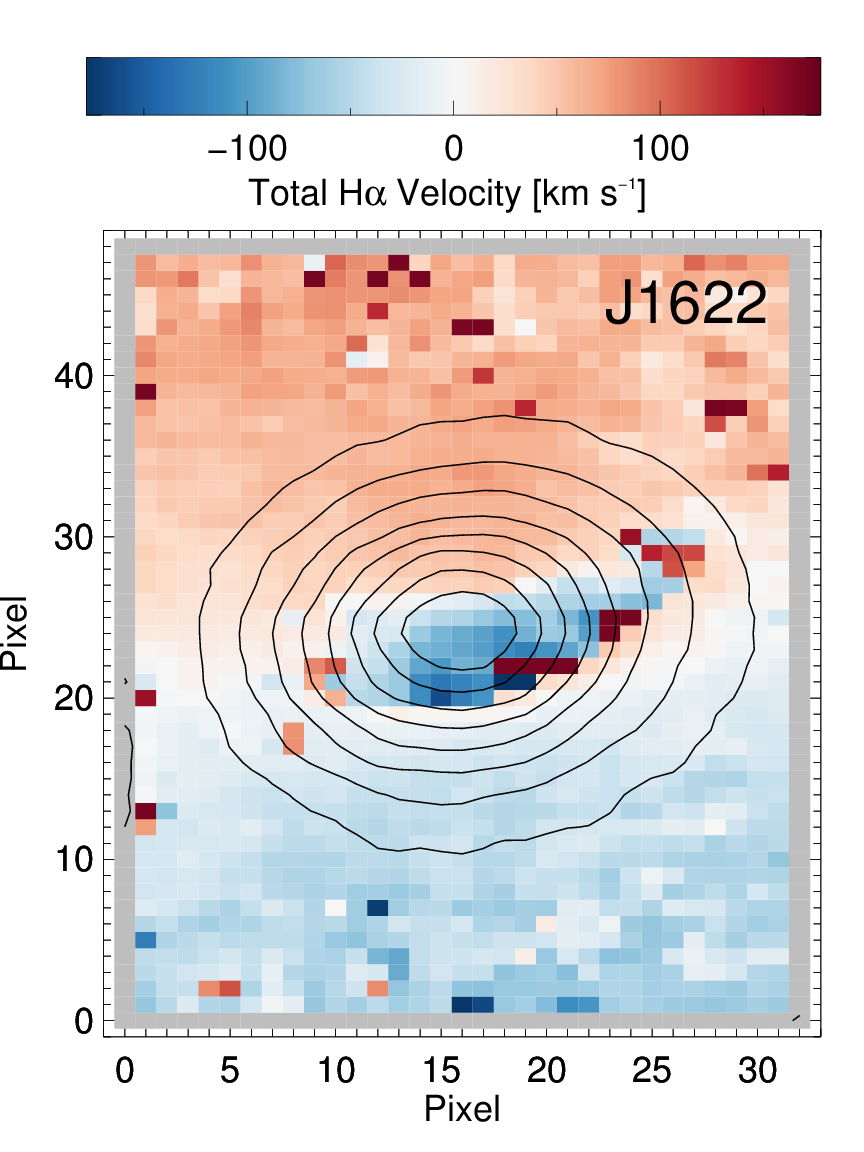}}
\raisebox{-0.5\height}{\includegraphics[width=0.16\textwidth,angle=0,trim={50 60 40 50},clip]{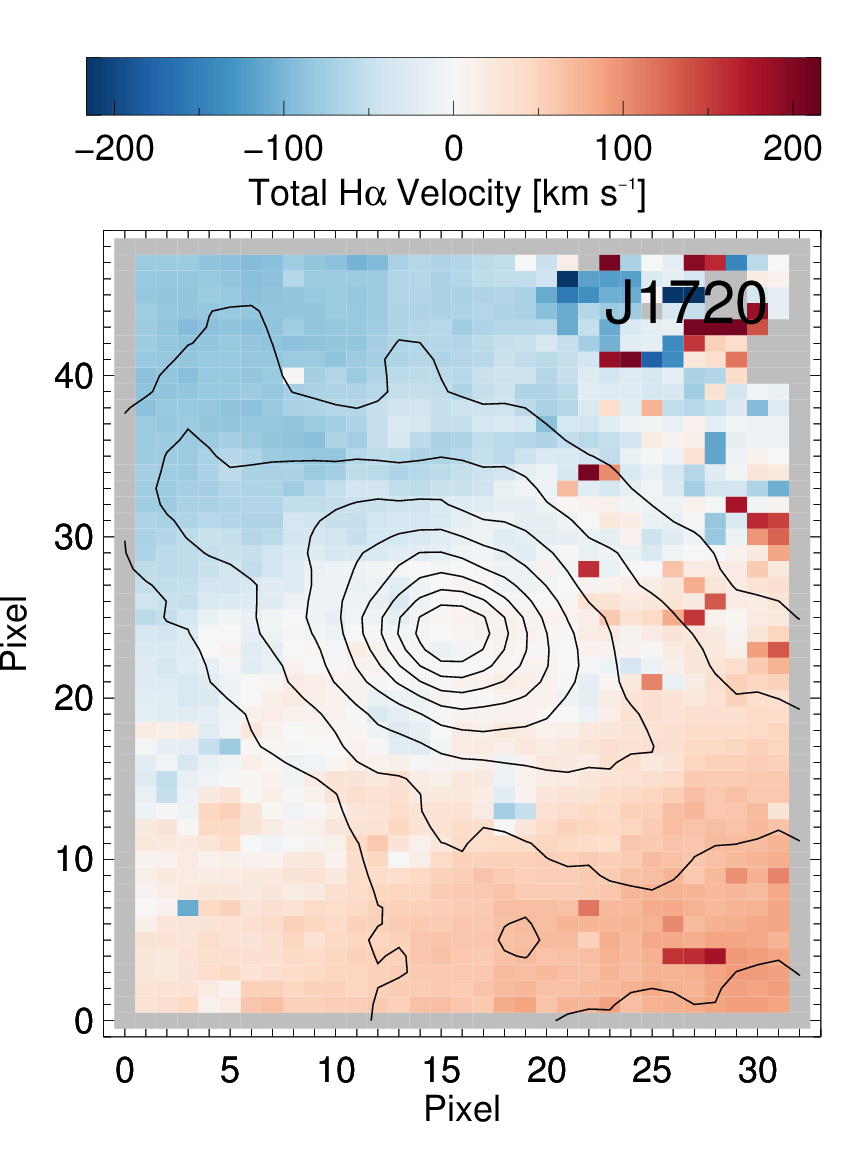}}\\
\raisebox{-0.5\height}{\includegraphics[width=0.16\textwidth,angle=0,trim={50 60 40 50},clip]{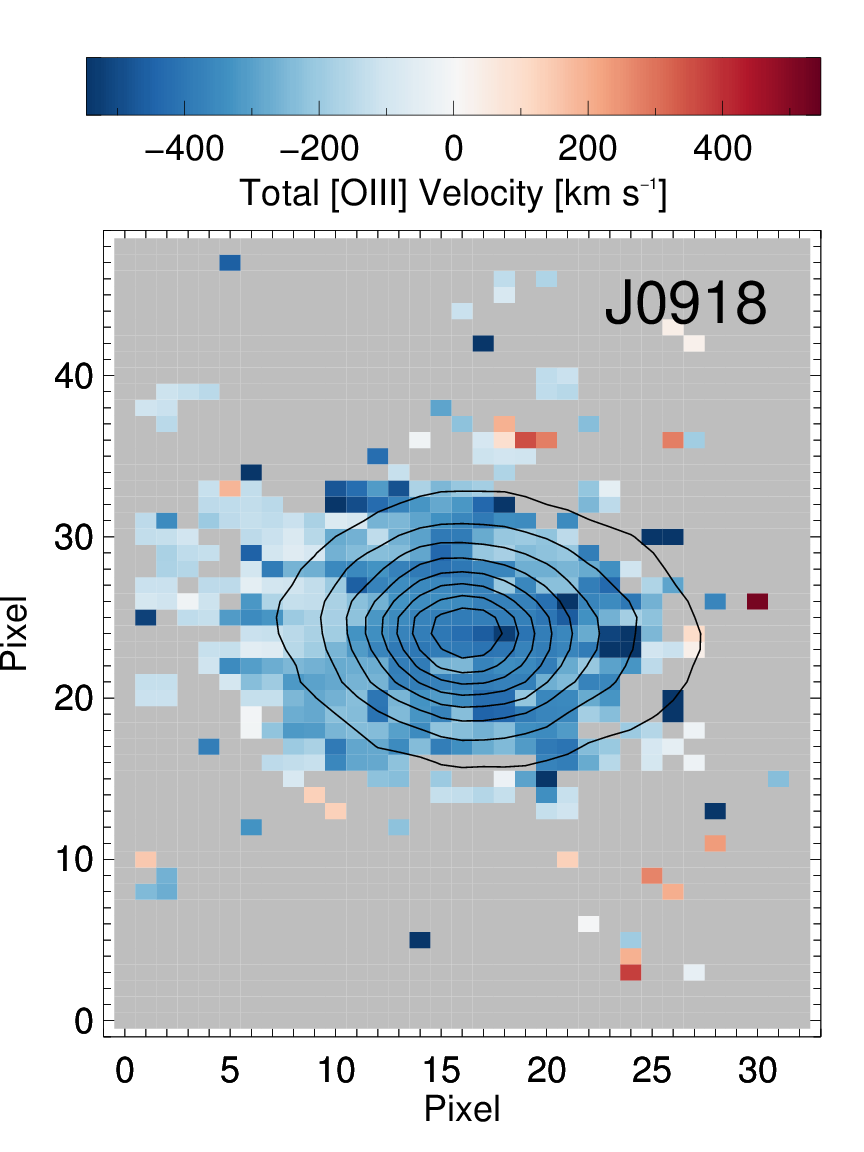}}
\raisebox{-0.5\height}{\includegraphics[width=0.16\textwidth,angle=0,trim={50 60 40 50},clip]{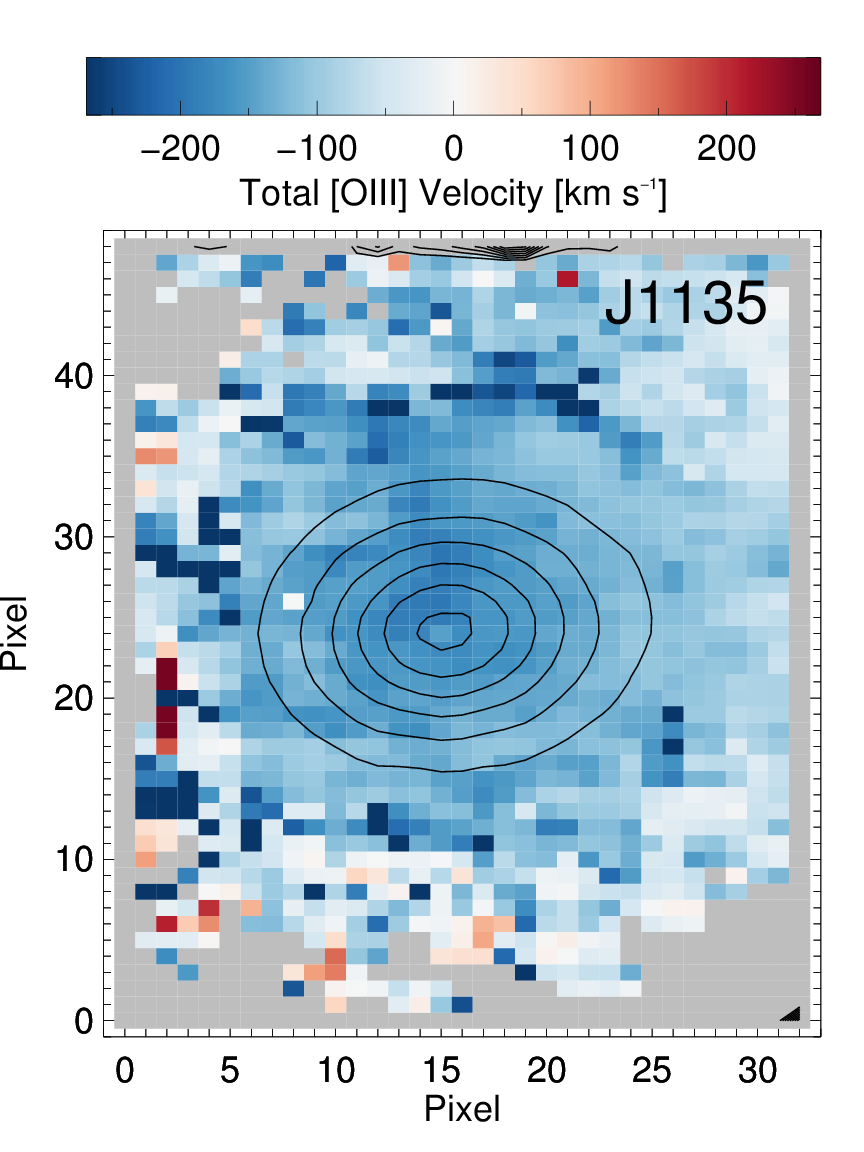}}
\raisebox{-0.5\height}{\includegraphics[width=0.16\textwidth,angle=0,trim={50 60 40 50},clip]{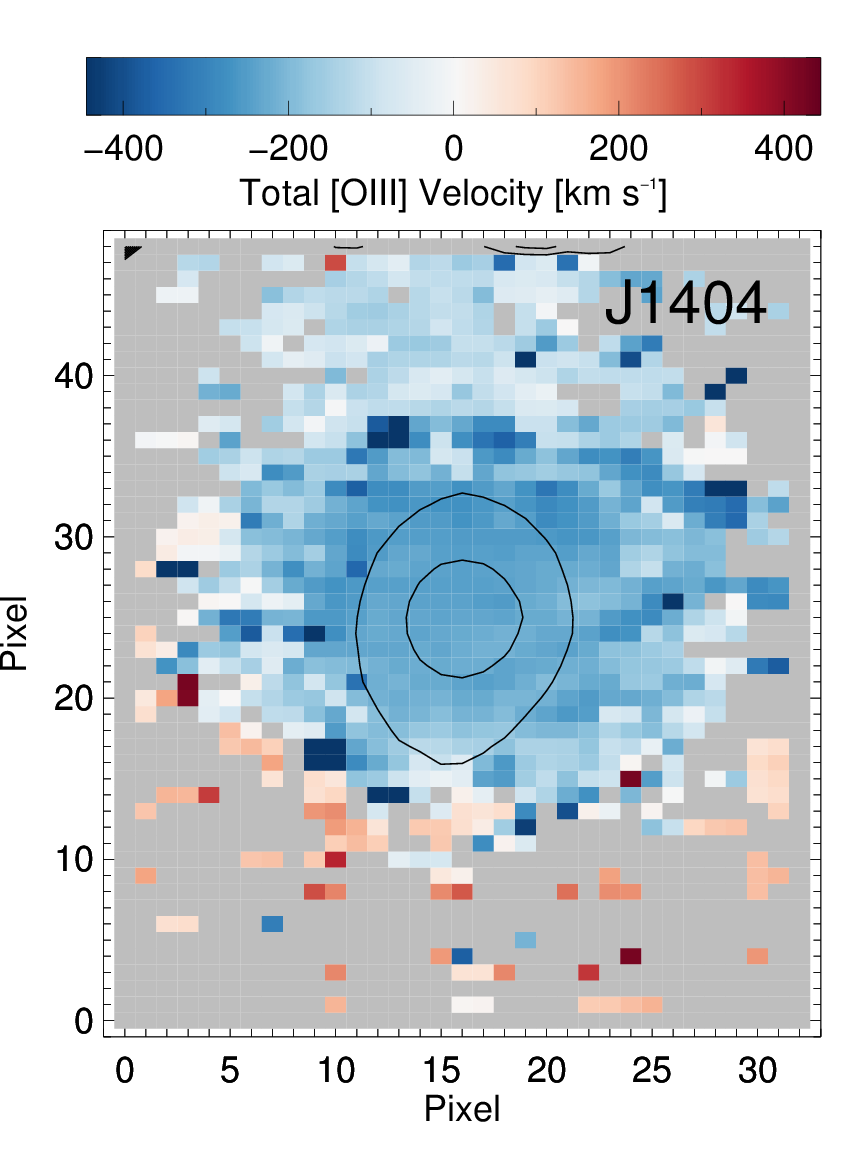}}
\raisebox{-0.5\height}{\includegraphics[width=0.16\textwidth,angle=0,trim={50 60 40 50},clip]{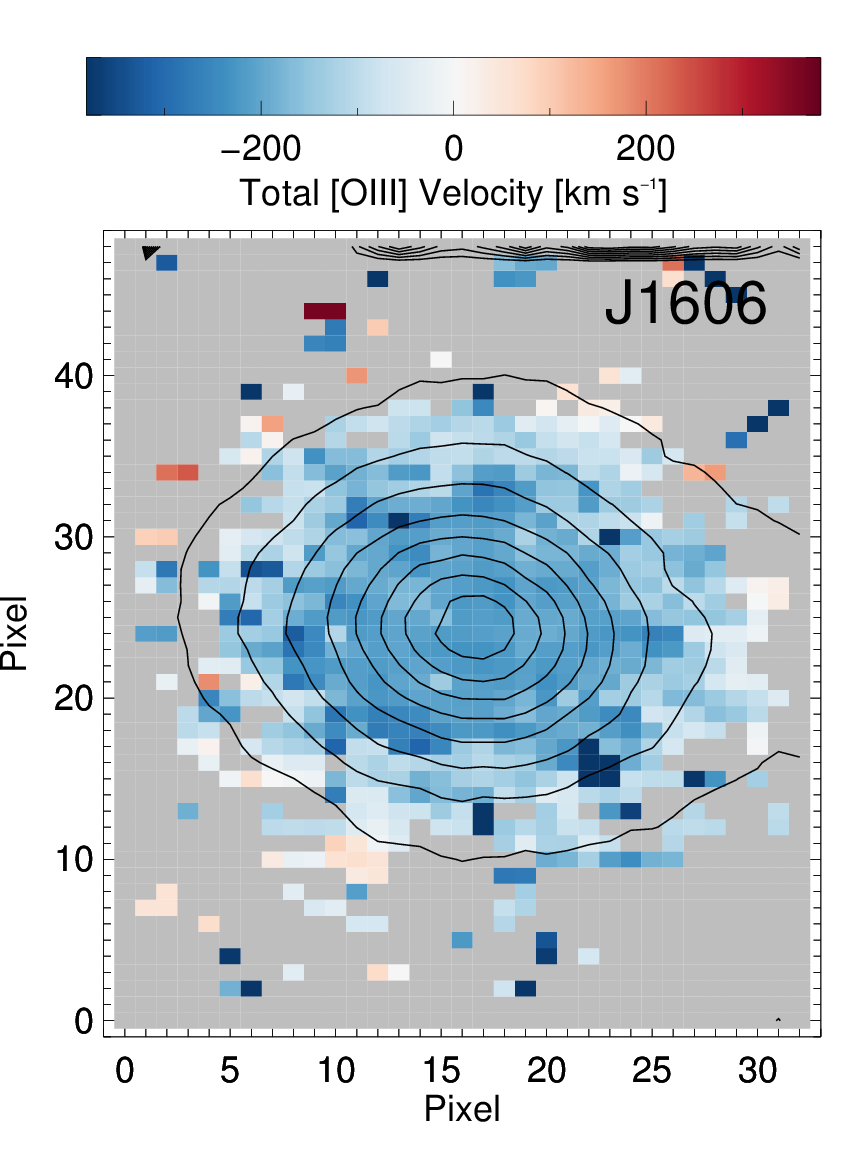}}
\raisebox{-0.5\height}{\includegraphics[width=0.16\textwidth,angle=0,trim={50 60 40 50},clip]{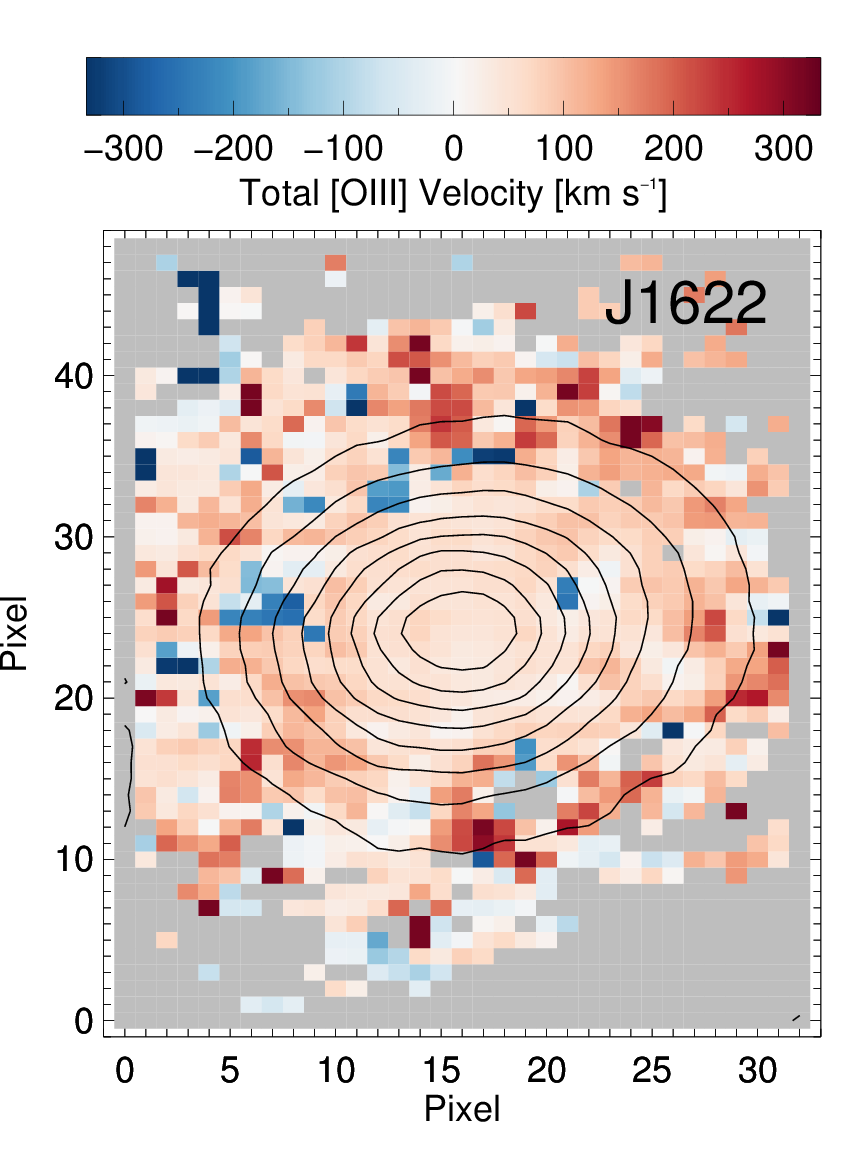}}
\raisebox{-0.5\height}{\includegraphics[width=0.16\textwidth,angle=0,trim={50 60 40 50},clip]{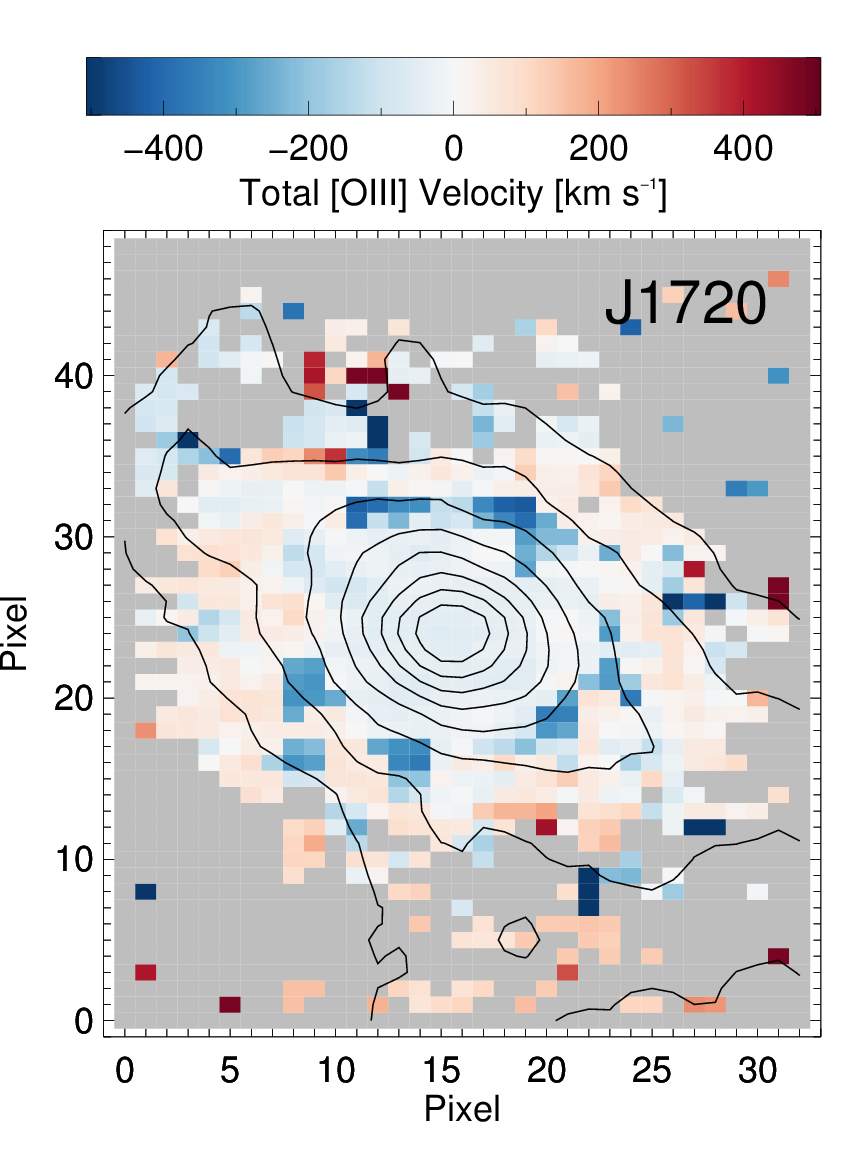}}\\
\vspace{5pt}\hrule\vspace{5pt}
{\Large Velocity dispersion}\\
\raisebox{-0.5\height}{\includegraphics[width=0.16\textwidth,angle=0,trim={50 60 20 50},clip]{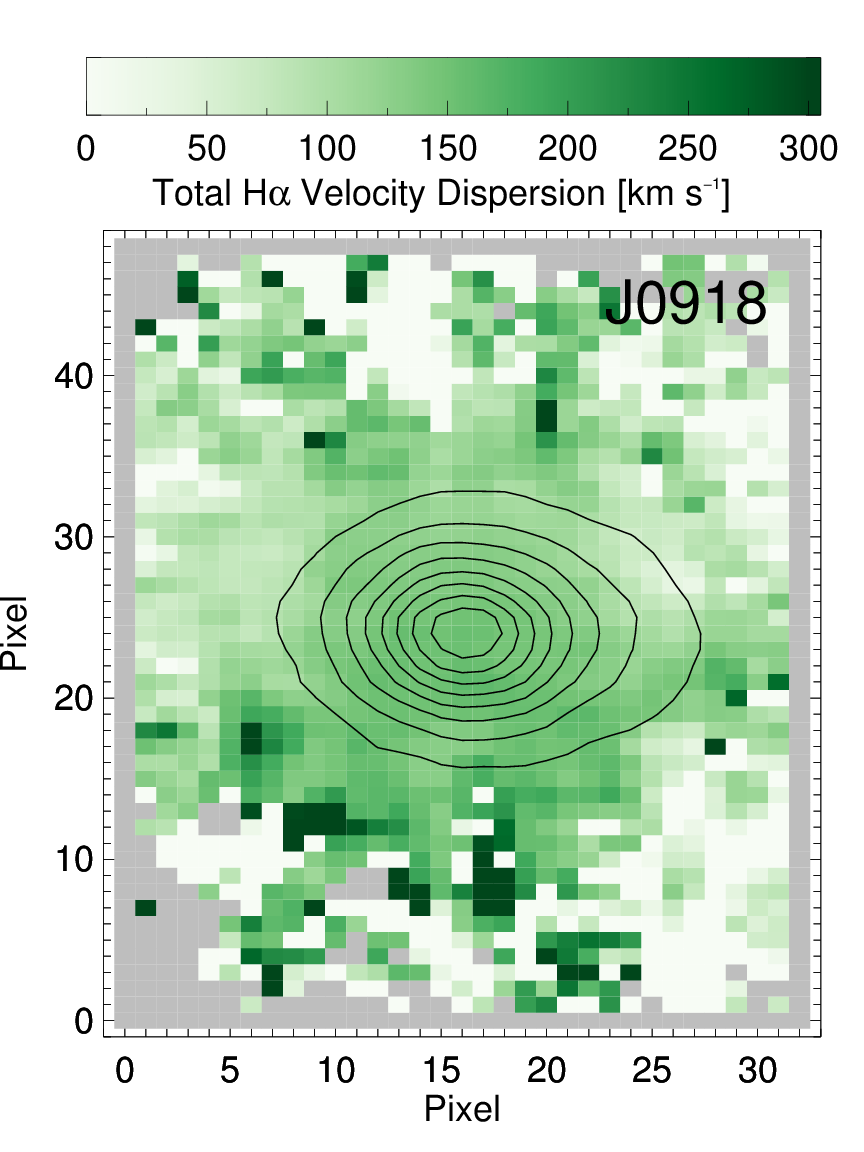}}
\raisebox{-0.5\height}{\includegraphics[width=0.16\textwidth,angle=0,trim={50 60 20 50},clip]{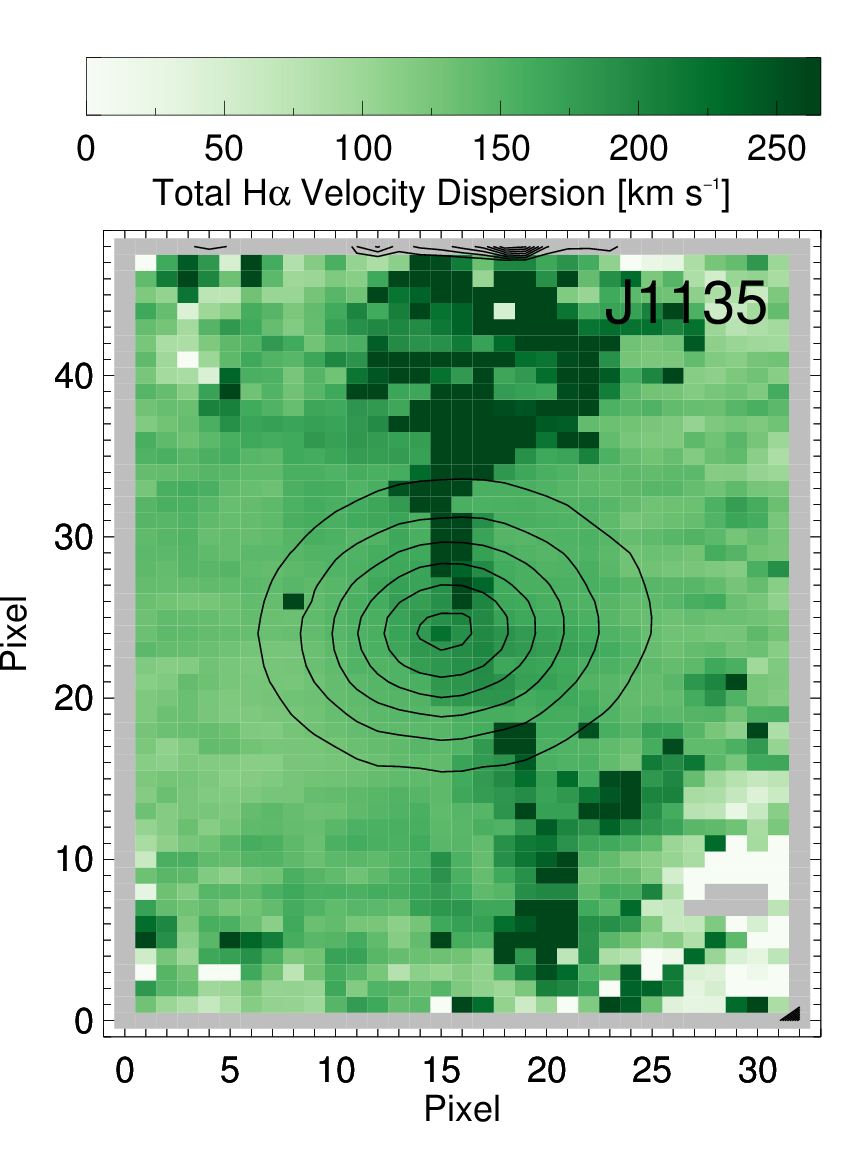}}
\raisebox{-0.5\height}{\includegraphics[width=0.16\textwidth,angle=0,trim={50 60 20 50},clip]{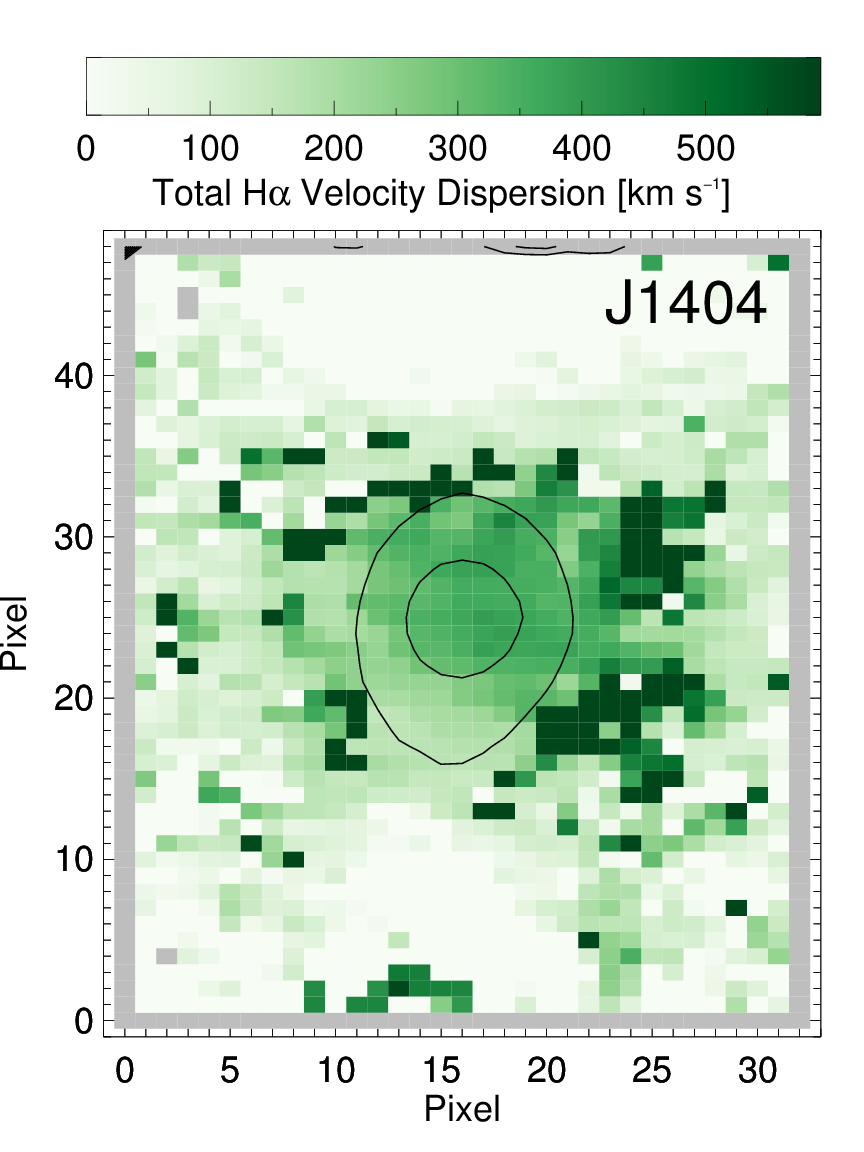}}
\raisebox{-0.5\height}{\includegraphics[width=0.16\textwidth,angle=0,trim={50 60 20 50},clip]{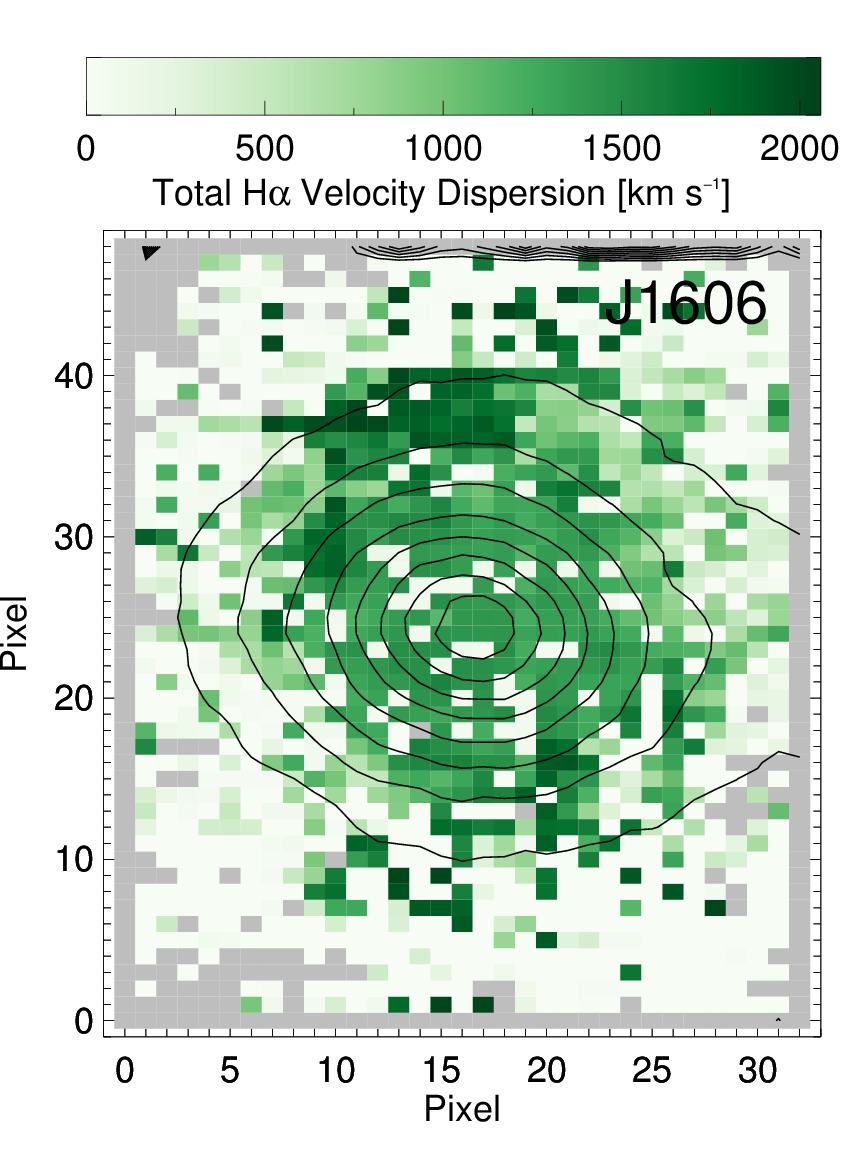}}
\raisebox{-0.5\height}{\includegraphics[width=0.16\textwidth,angle=0,trim={50 60 20 50},clip]{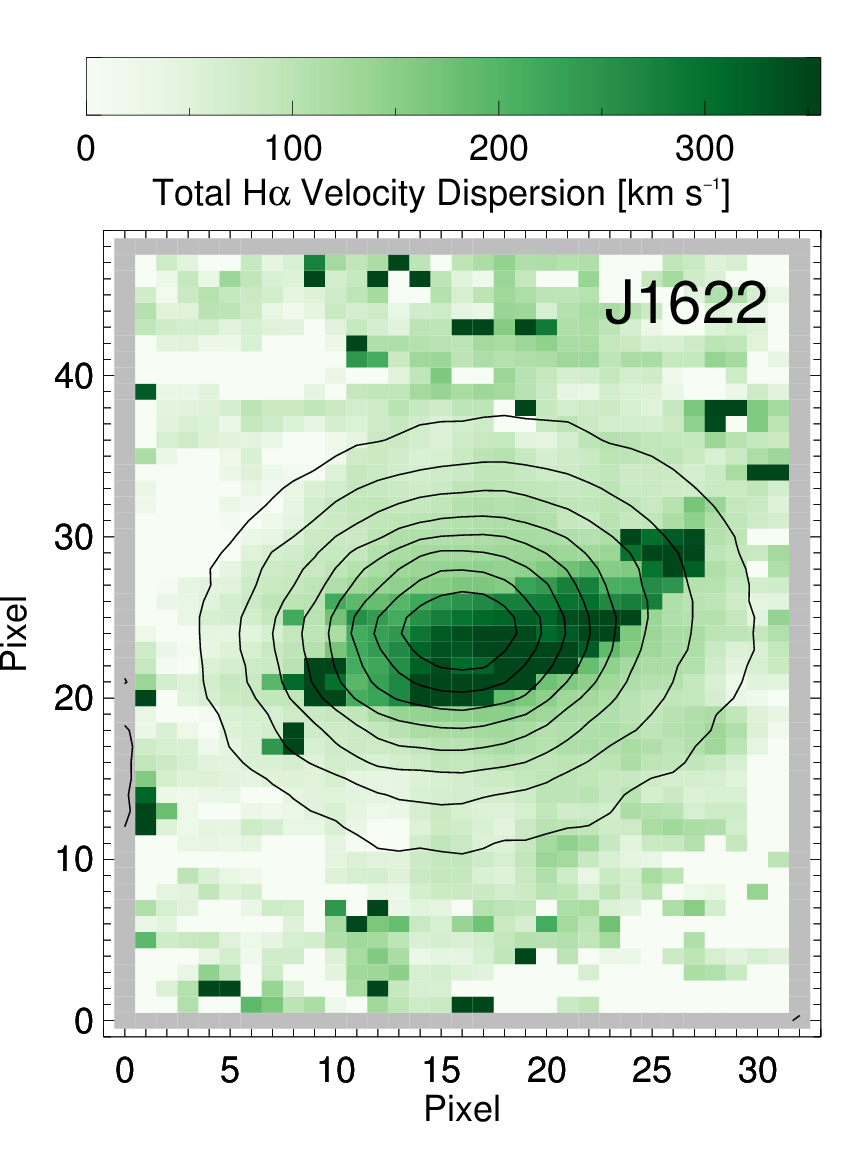}}
\raisebox{-0.5\height}{\includegraphics[width=0.16\textwidth,angle=0,trim={50 60 20 50},clip]{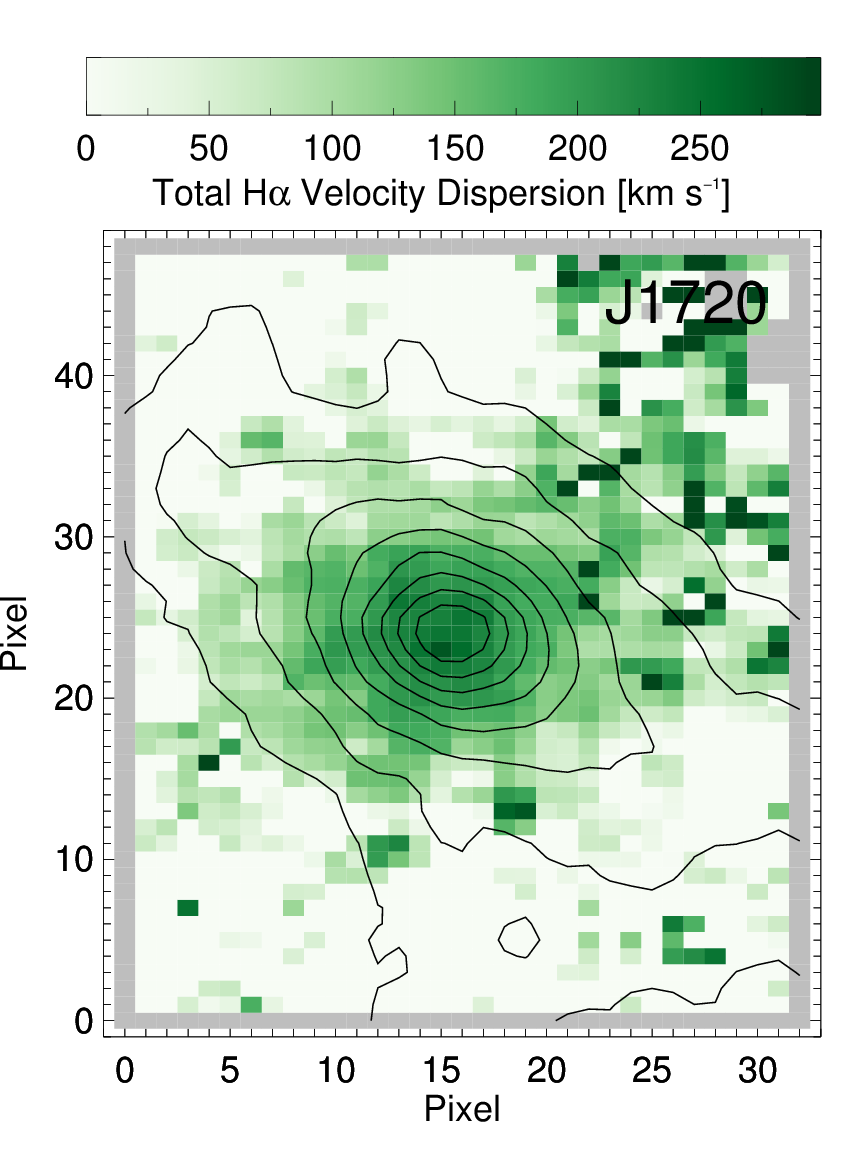}}
\raisebox{-0.5\height}{\includegraphics[width=0.16\textwidth,angle=0,trim={50 60 15 50},clip]{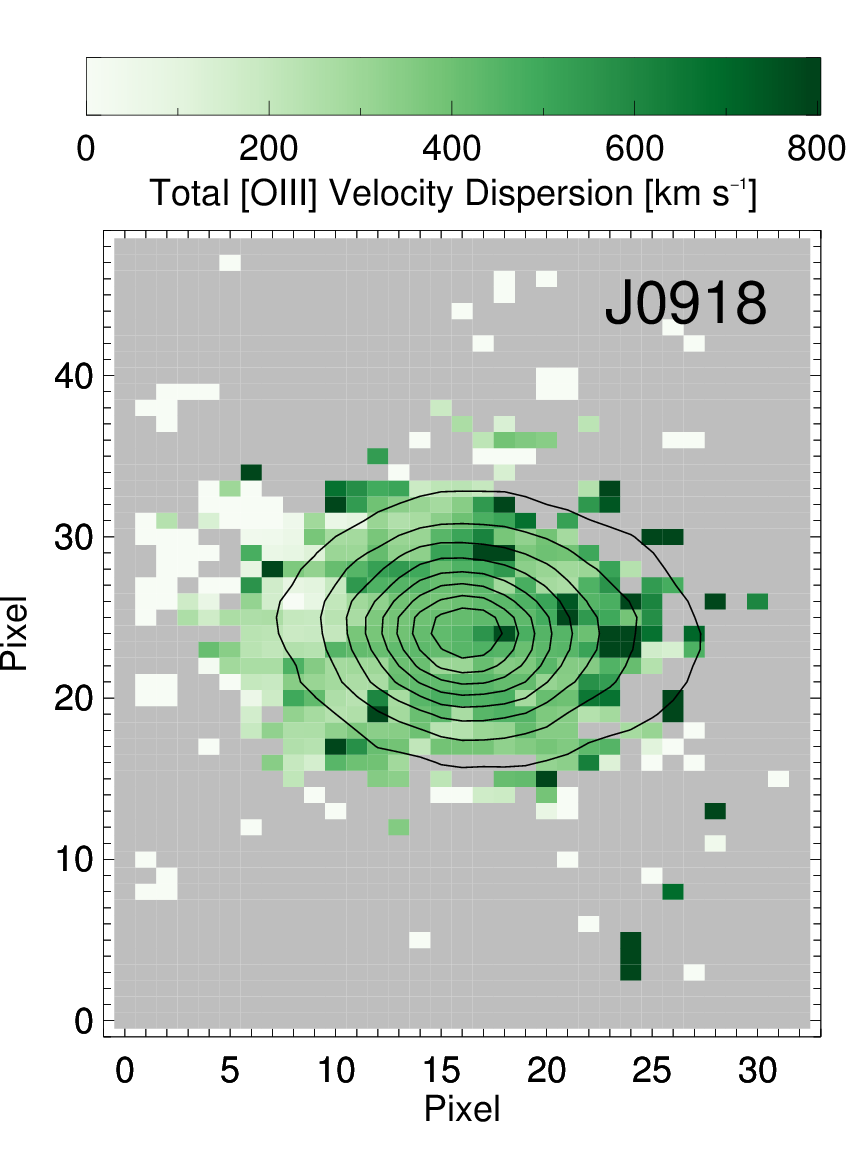}}
\raisebox{-0.5\height}{\includegraphics[width=0.16\textwidth,angle=0,trim={50 60 20 50},clip]{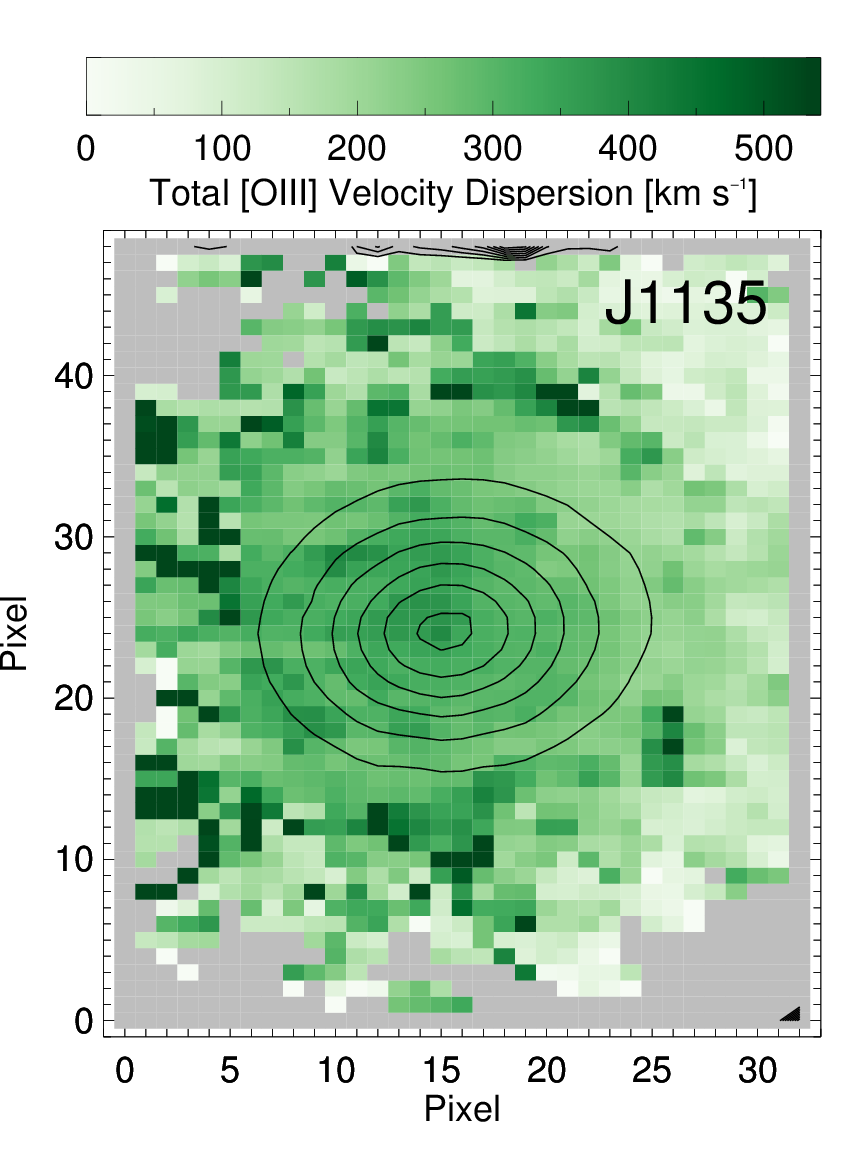}}
\raisebox{-0.5\height}{\includegraphics[width=0.16\textwidth,angle=0,trim={50 60 20 50},clip]{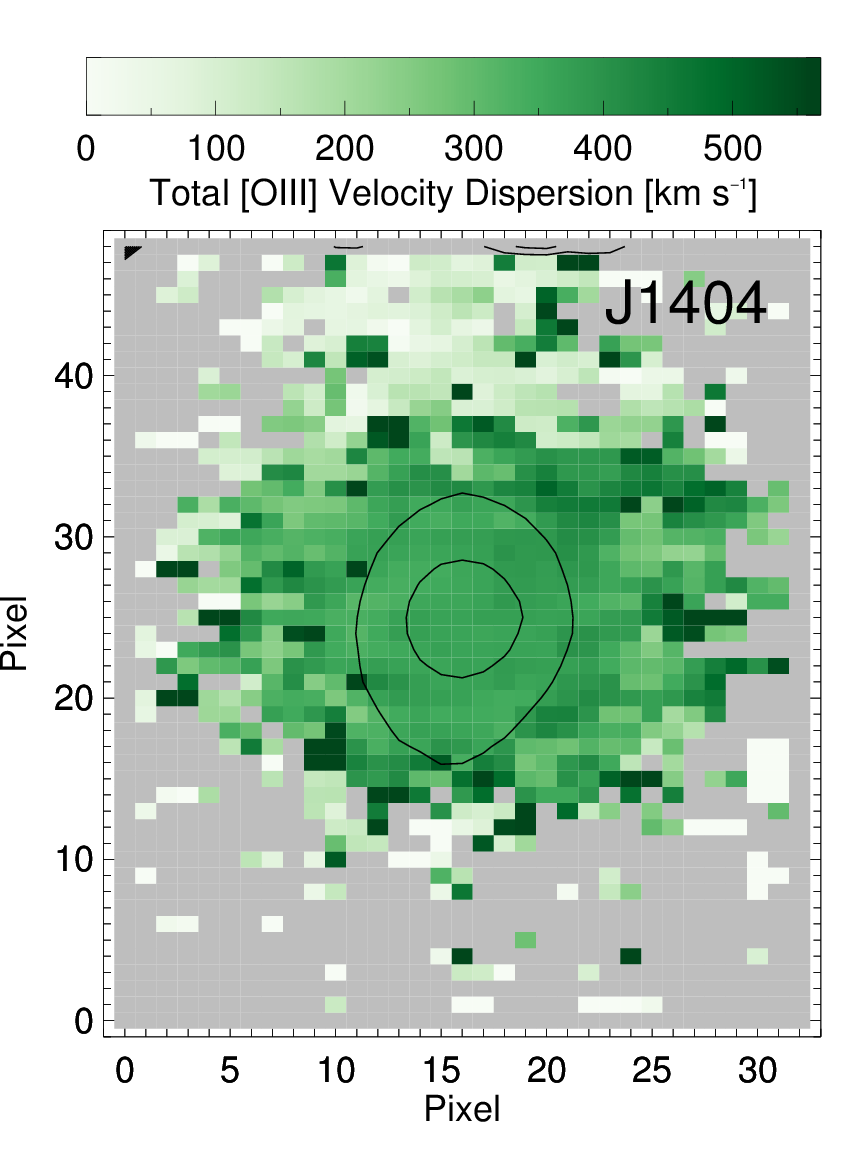}}
\raisebox{-0.5\height}{\includegraphics[width=0.16\textwidth,angle=0,trim={50 60 20 50},clip]{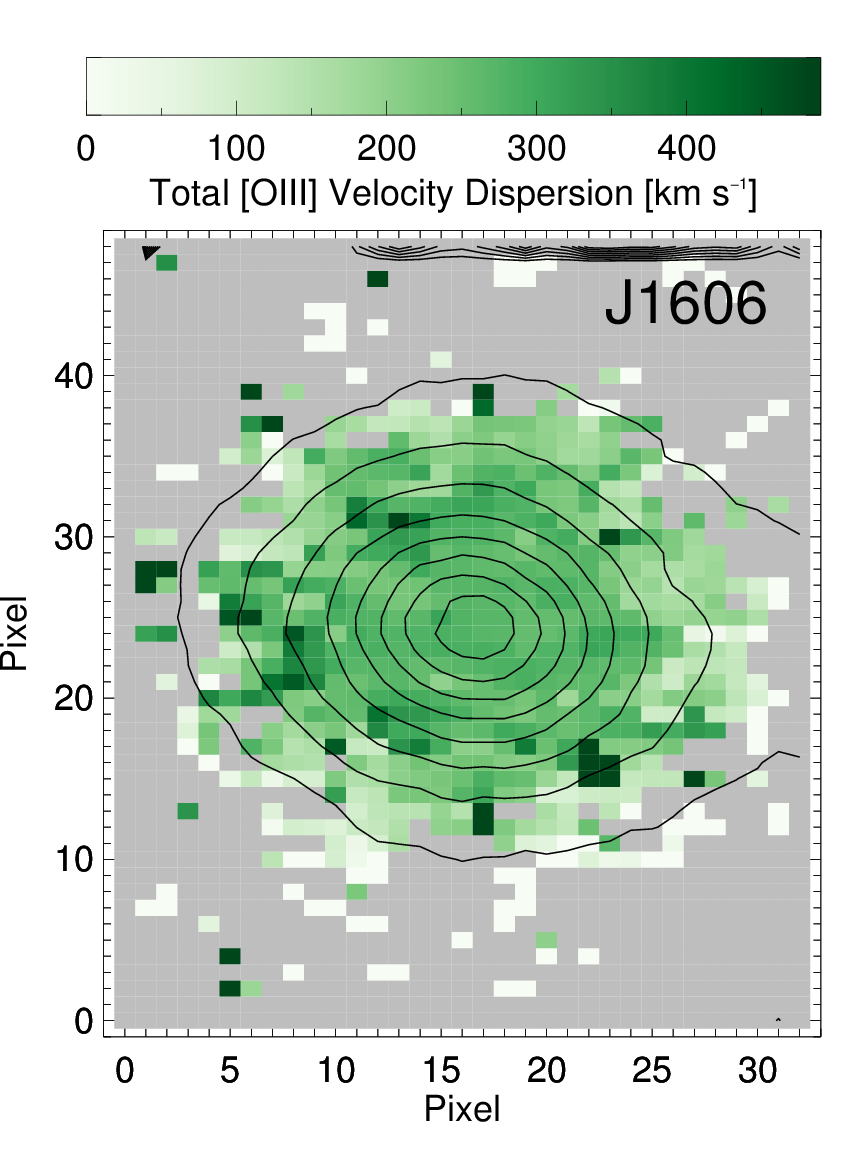}}
\raisebox{-0.5\height}{\includegraphics[width=0.16\textwidth,angle=0,trim={50 60 20 50},clip]{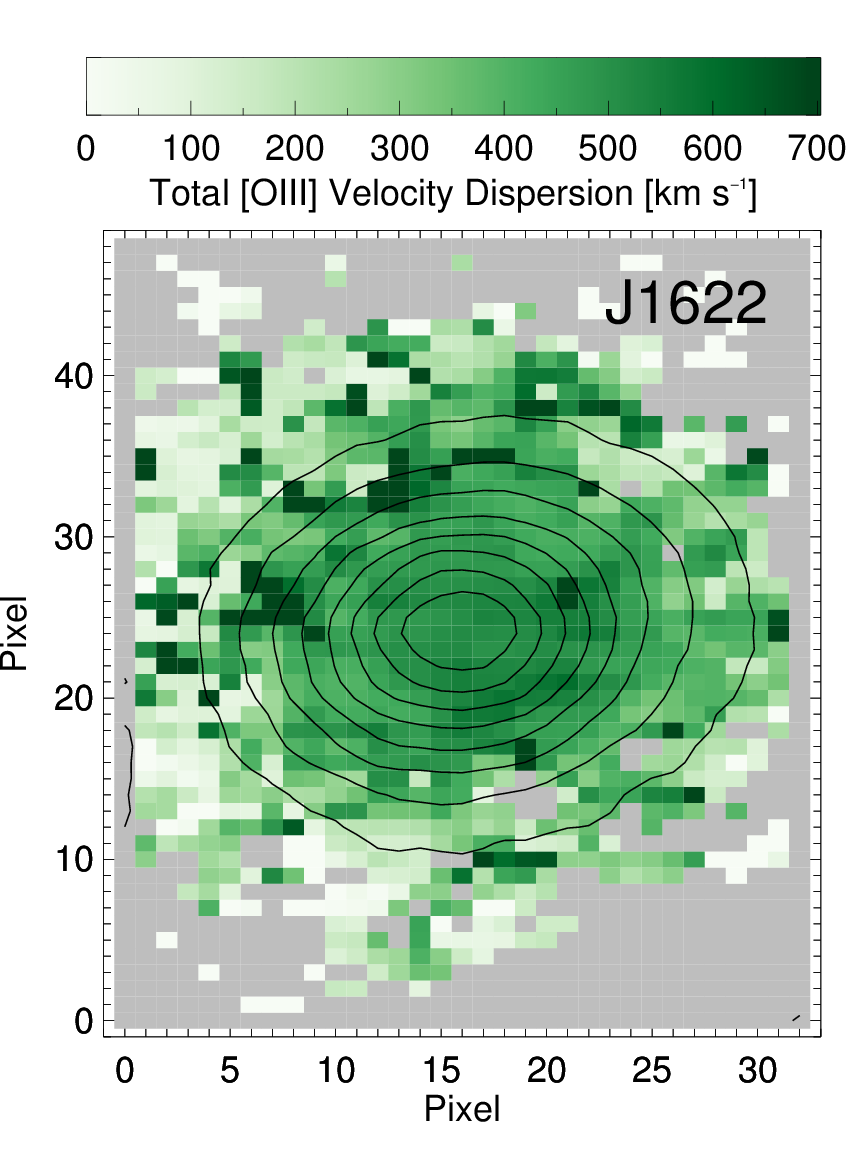}}
\raisebox{-0.5\height}{\includegraphics[width=0.16\textwidth,angle=0,trim={50 60 20 50},clip]{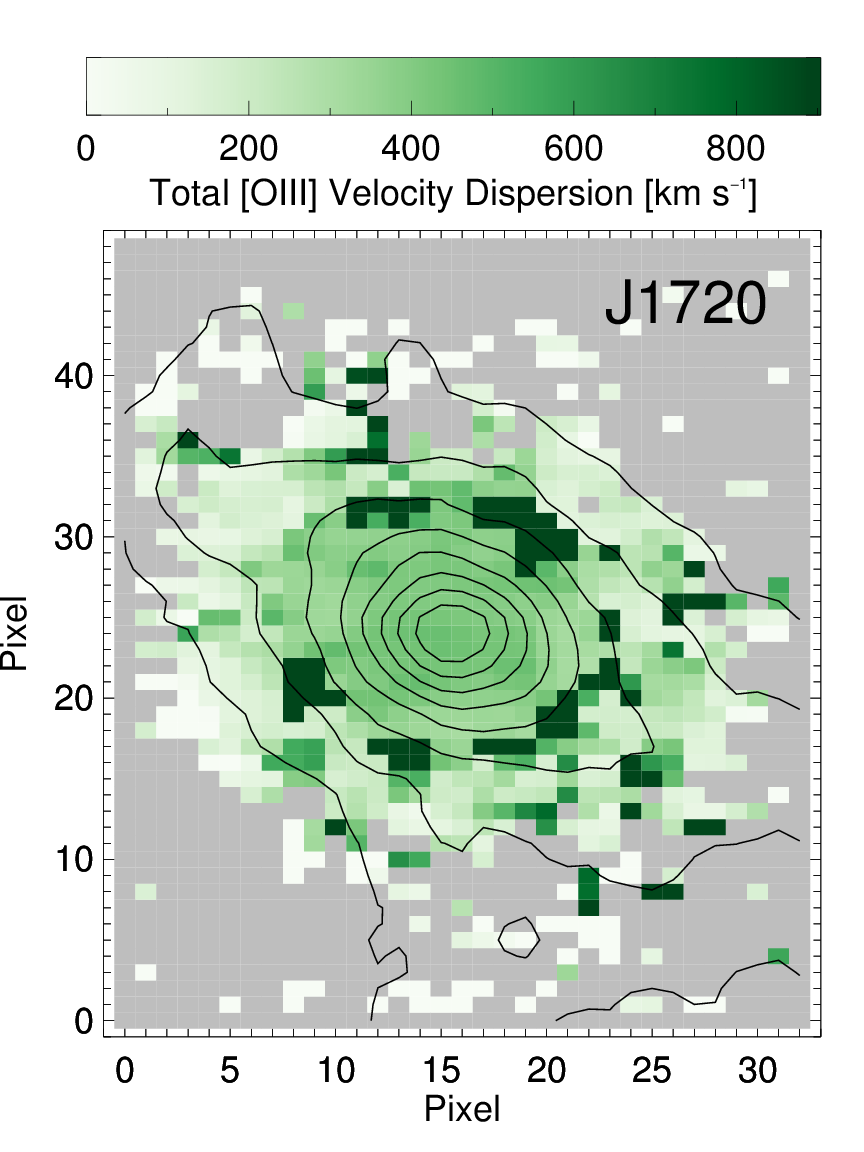}}\\
\caption{Velocity maps of stars {based on the best-fit stellar spectrum from the pPXF fitting} (top), H$\alpha$ (2nd row), and  \mbox{[O\,\textsc{iii}]} (3rd row) with respect to the
systemic velocity measured from stellar absorption lines within the central 3\arcsec\,
and velocity dispersion maps of H$\alpha$ (4th row) and  \mbox{[O\,\textsc{iii}]} (5th row).
Contours are the same as in Fig. \ref{fig:all_flux}. Grey spaxels indicate non-detection of the emission lines (i.e., S/N$<$3).}\label{fig:all_kins_total}
\end{center}
\end{figure*}

\subsection{Kinematics of the ionized gas}
\label{sec:kins}

In this section we investigate the kinematics of the ionized gas using the spatial distribution of velocity and velocity dispersion of H$\alpha$ and \mbox{[O\,\textsc{iii}]}.

\subsubsection{Ionized gas velocity}

All sources show rotation patterns in the stellar velocity maps with maximum velocities ranging between 200 and 250 km s$^{-1}$ (Fig. \ref{fig:all_kins_total}), indicating the presence of a central stellar disk. J0918 may be an exception as its stellar velocity map is very noisy , which is potentially due to its inclination (this is a face-on galaxy)\footnote{{We remind that stellar velocities are derived based on the fitting of the full spectrum of each spaxel (including the G-band and the Mgb absorption triplet) using a combination of stellar population models.}}.

In contrast, the ionized gas maps show much more diverse kinematic patterns both in their velocity and velocity dispersion distributions. The velocity maps appear to be a superposition of two different kinematic components. While the outer parts of the H$\alpha$ velocity maps (second row of Fig. \ref{fig:all_kins_total}) appear to trace the stellar kinematics, the inner parts show blueshifted emission (e.g., J1404). For \mbox{[O\,\textsc{iii}]}, we do not recover any traces of the stellar rotation but instead find blueshifted emission. The only exception is J1622, which shows redshifted \mbox{[O\,\textsc{iii}]} emission. The velocity dispersion maps show similar behavior, with features of very high dispersion (compared to $\sigma_{*}$) at the center of the galaxies but values similar to $\sigma_{*}$ or consistent with zero (unresolved emission line) in the outer parts of the FoV. 

\begin{figure*}[tbp]
\begin{center}
{\Large Narrow component}\\
\raisebox{-0.5\height}{\includegraphics[width=0.16\textwidth,angle=0,trim={50 60 40 50},clip]{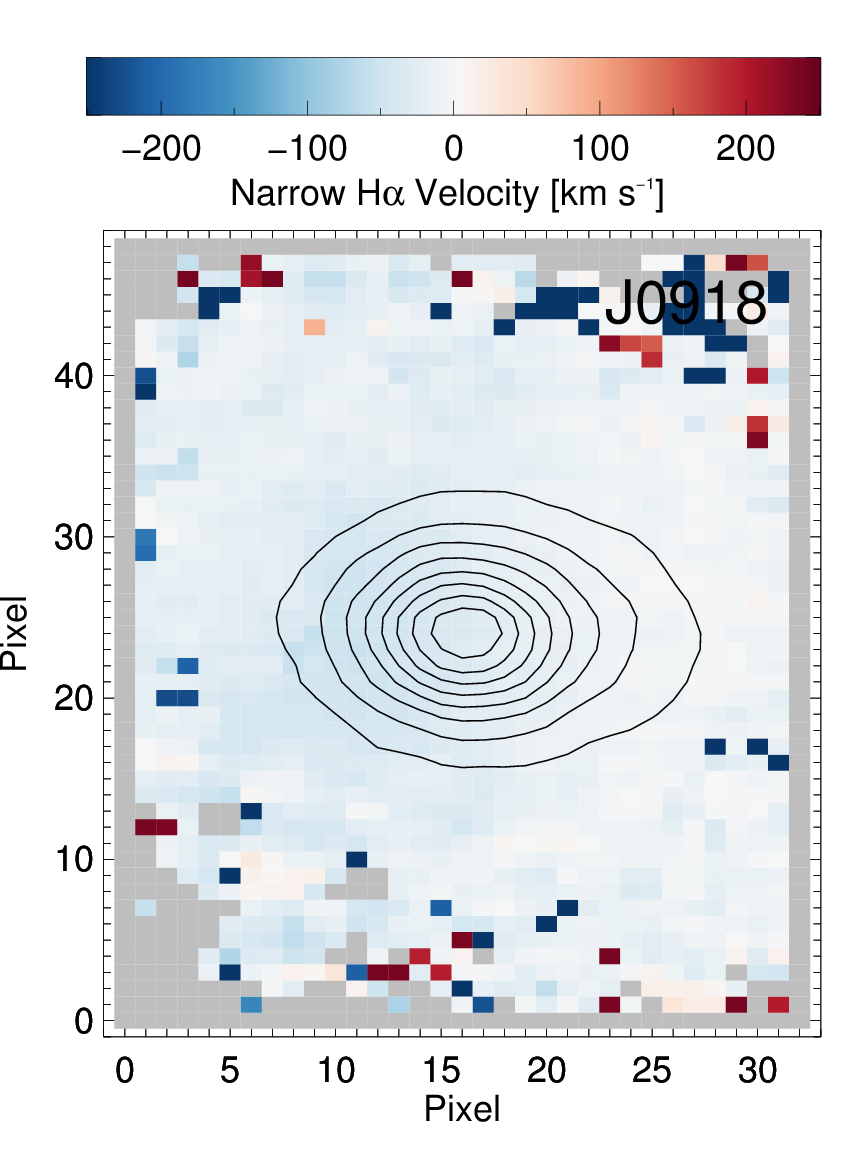}}
\raisebox{-0.5\height}{\includegraphics[width=0.16\textwidth,angle=0,trim={50 60 40 50},clip]{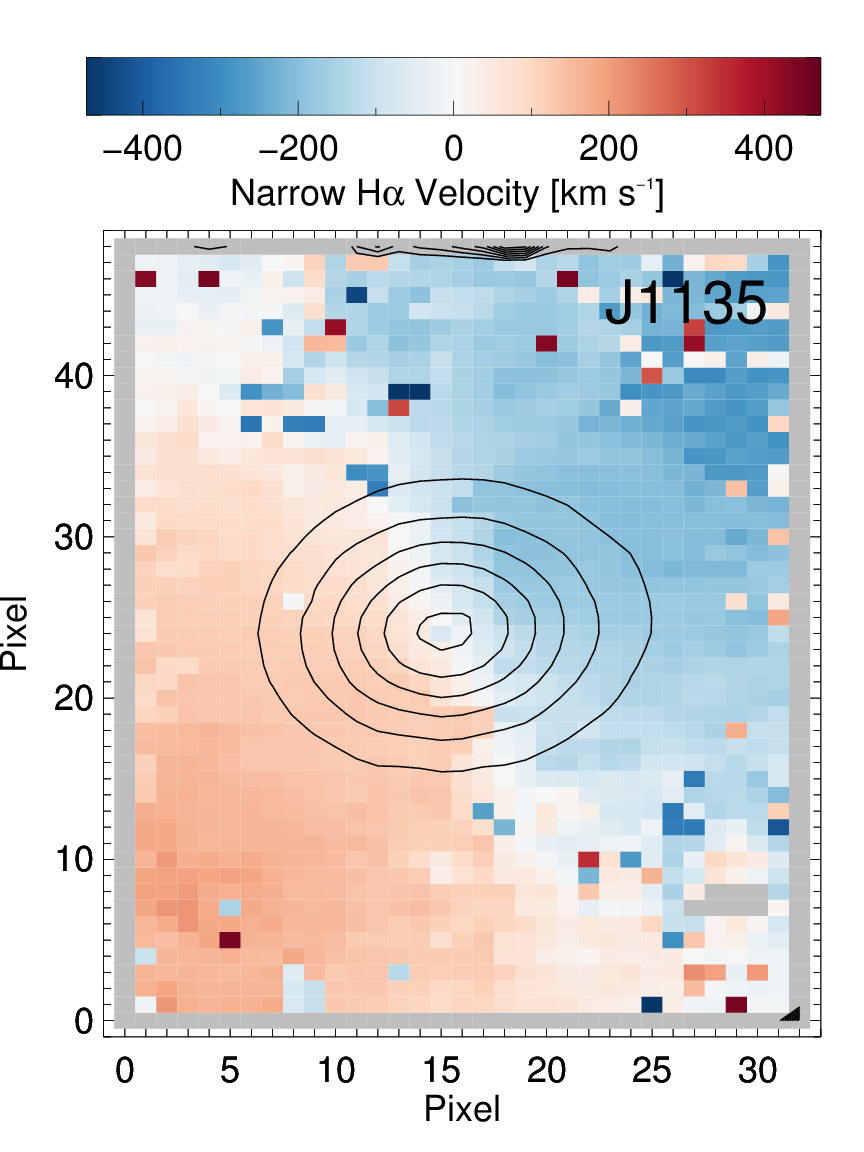}}
\raisebox{-0.5\height}{\includegraphics[width=0.16\textwidth,angle=0,trim={50 60 40 50},clip]{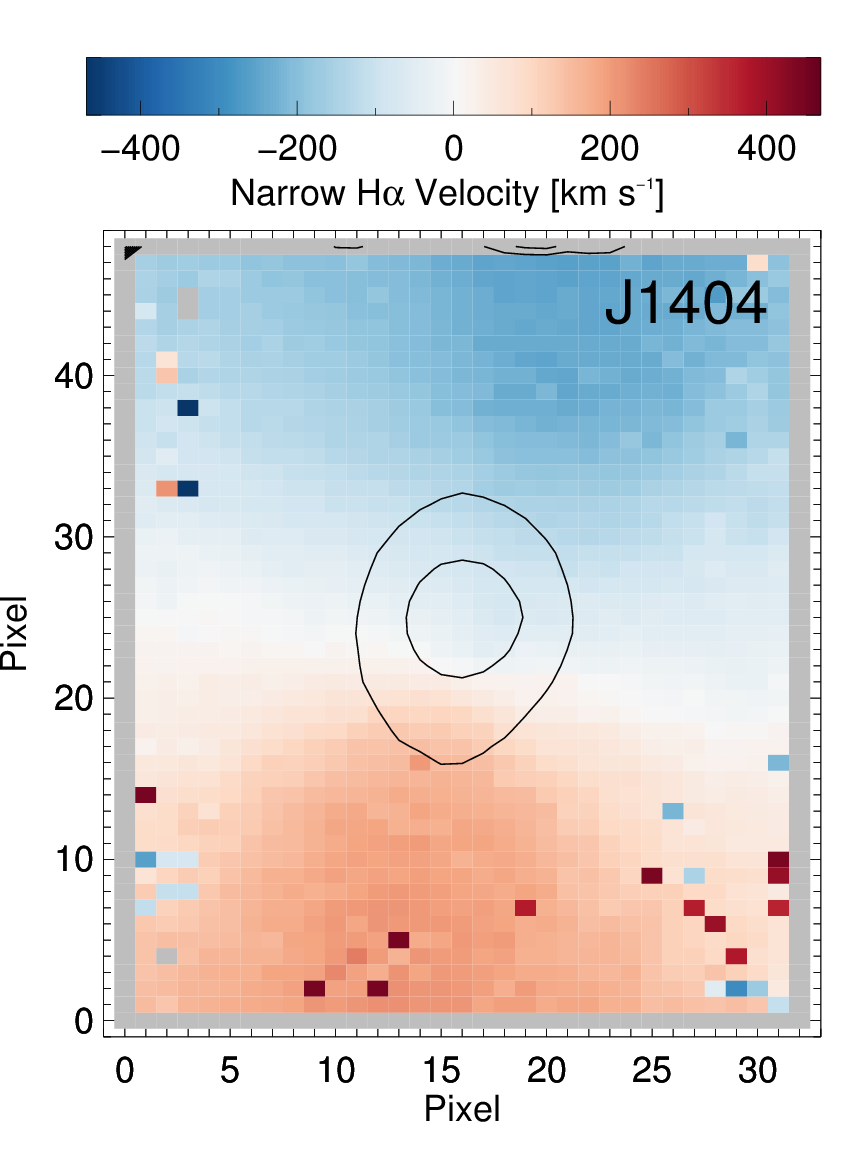}}
\raisebox{-0.5\height}{\includegraphics[width=0.16\textwidth,angle=0,trim={50 60 40 50},clip]{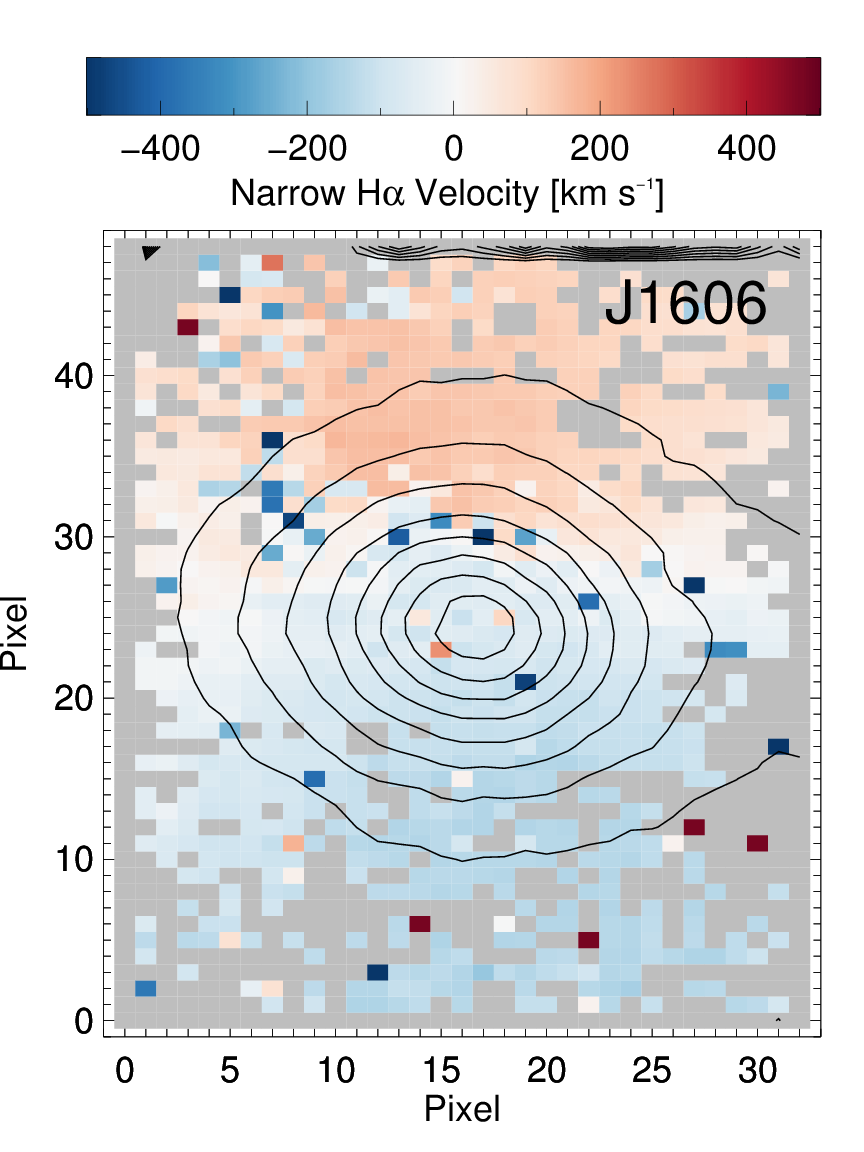}}
\raisebox{-0.5\height}{\includegraphics[width=0.16\textwidth,angle=0,trim={50 60 40 50},clip]{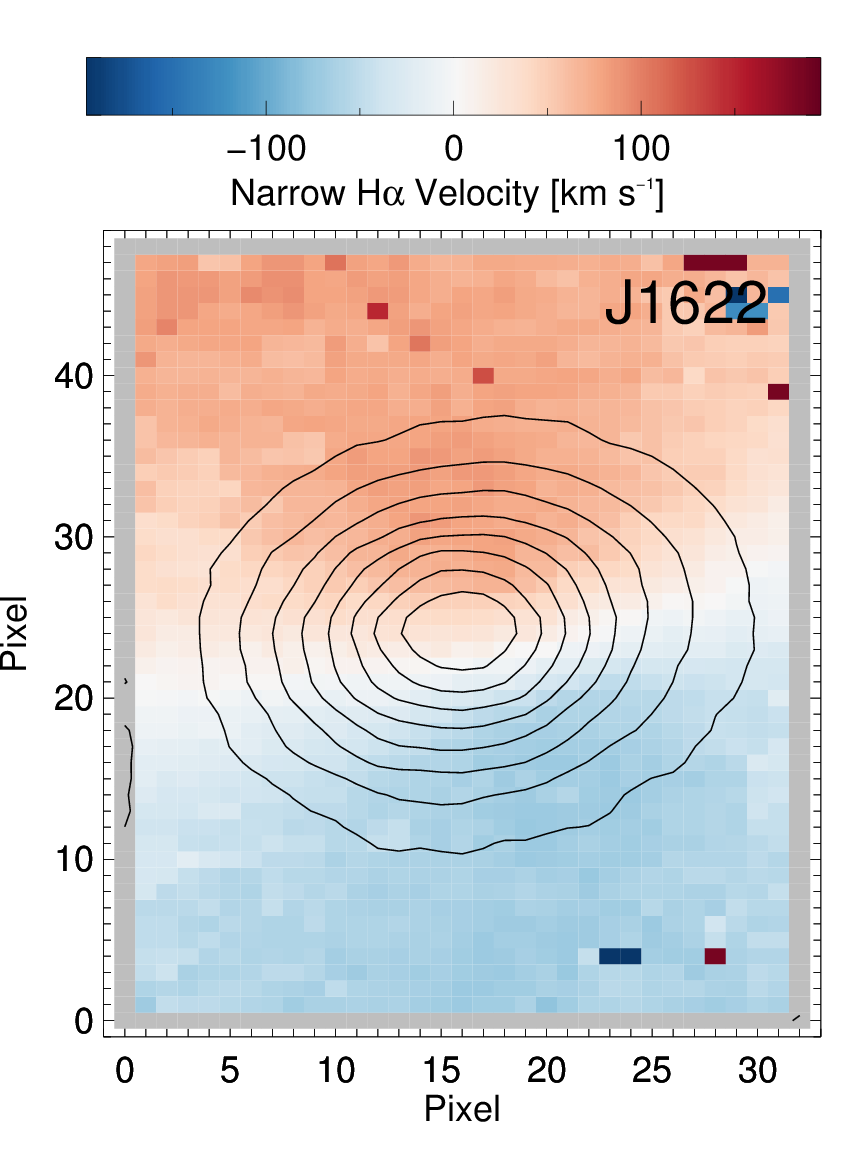}}
\raisebox{-0.5\height}{\includegraphics[width=0.16\textwidth,angle=0,trim={50 60 40 50},clip]{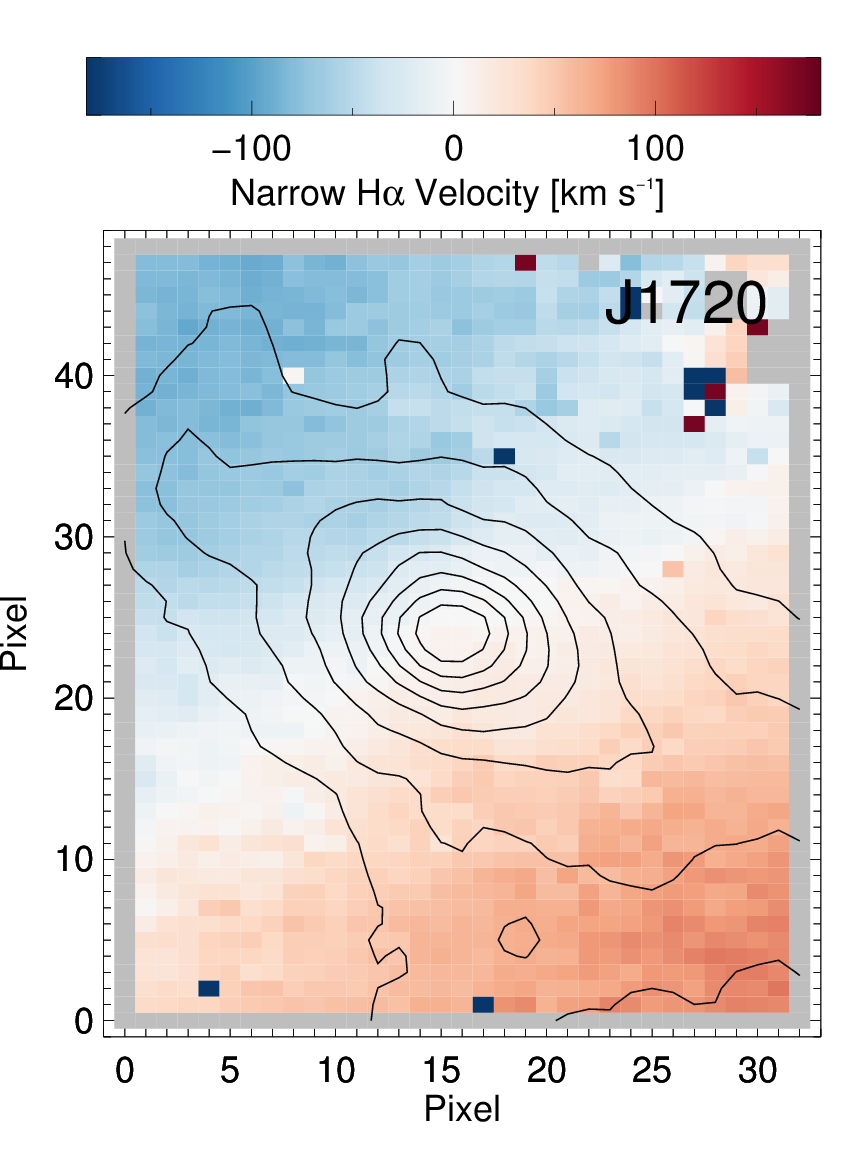}}\\
\raisebox{-0.5\height}{\includegraphics[width=0.16\textwidth,angle=0,trim={50 60 40 50},clip]{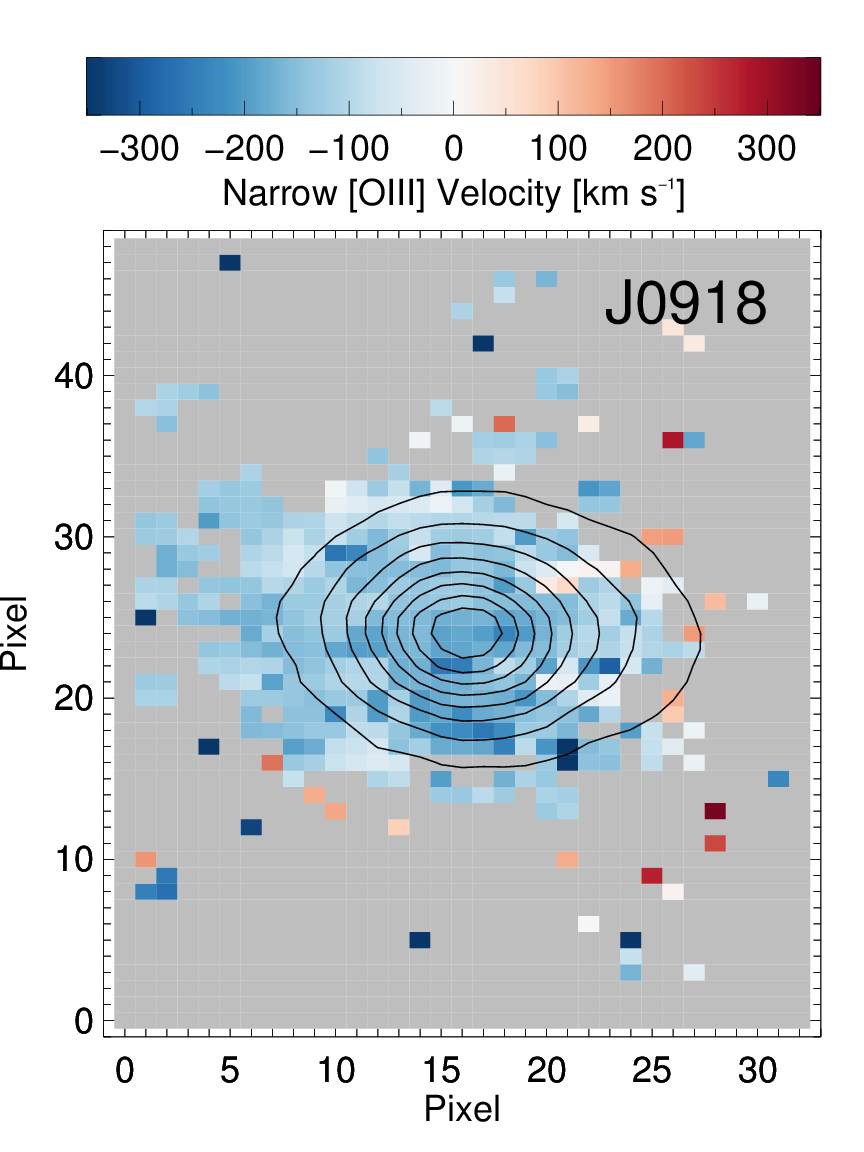}}
\raisebox{-0.5\height}{\includegraphics[width=0.16\textwidth,angle=0,trim={50 60 40 50},clip]{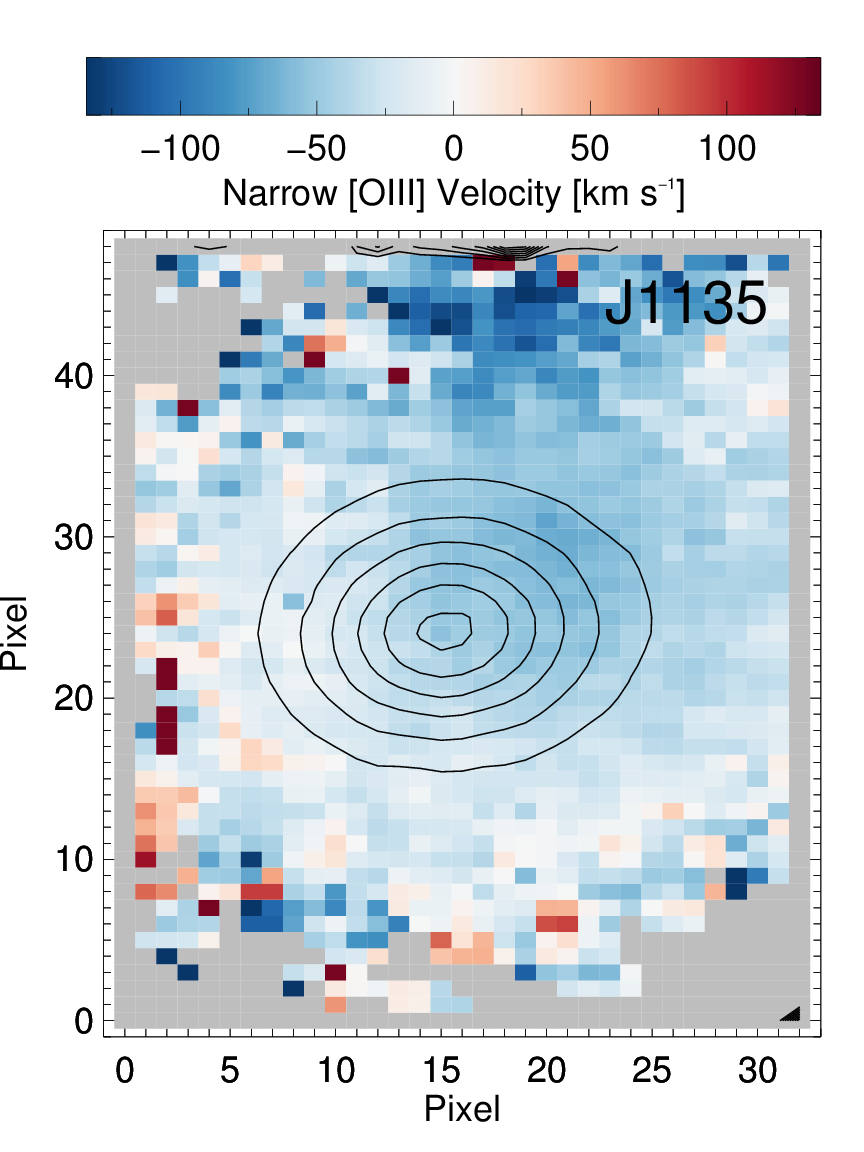}}
\raisebox{-0.5\height}{\includegraphics[width=0.16\textwidth,angle=0,trim={50 60 40 50},clip]{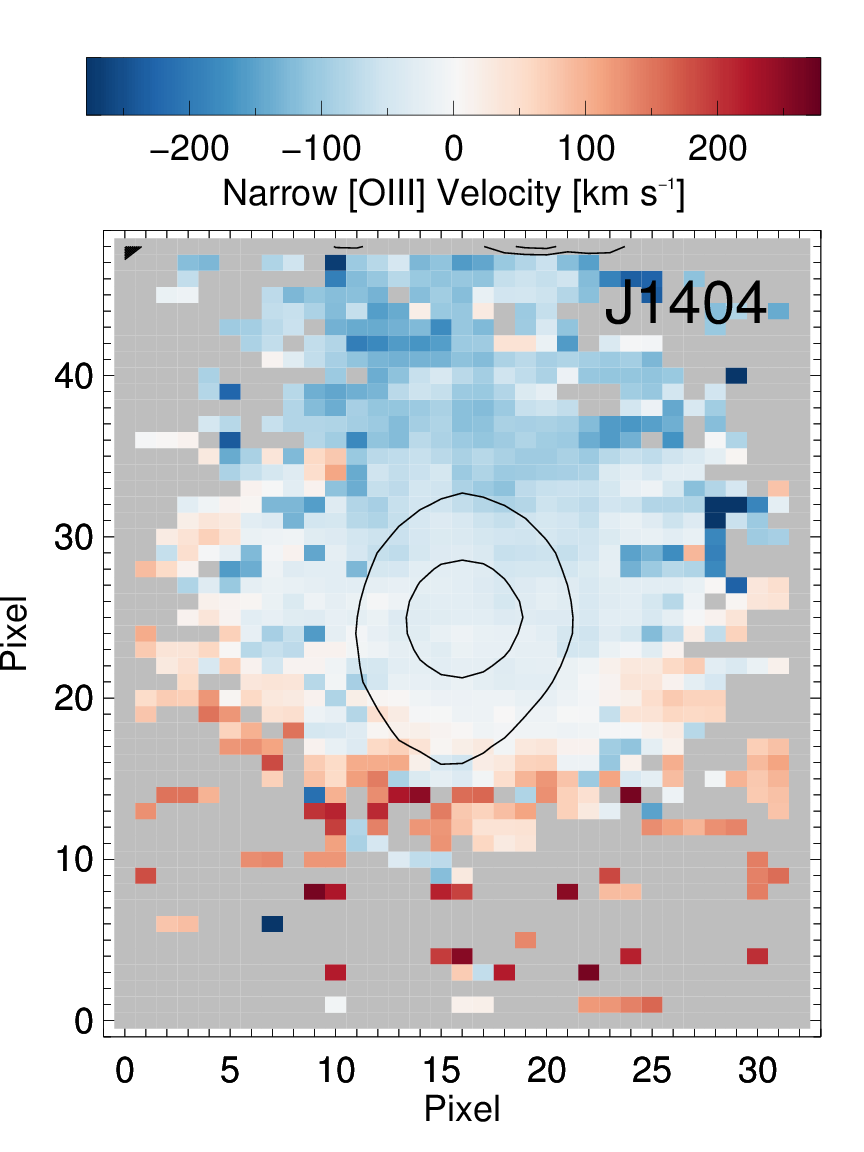}}
\raisebox{-0.5\height}{\includegraphics[width=0.16\textwidth,angle=0,trim={50 60 40 50},clip]{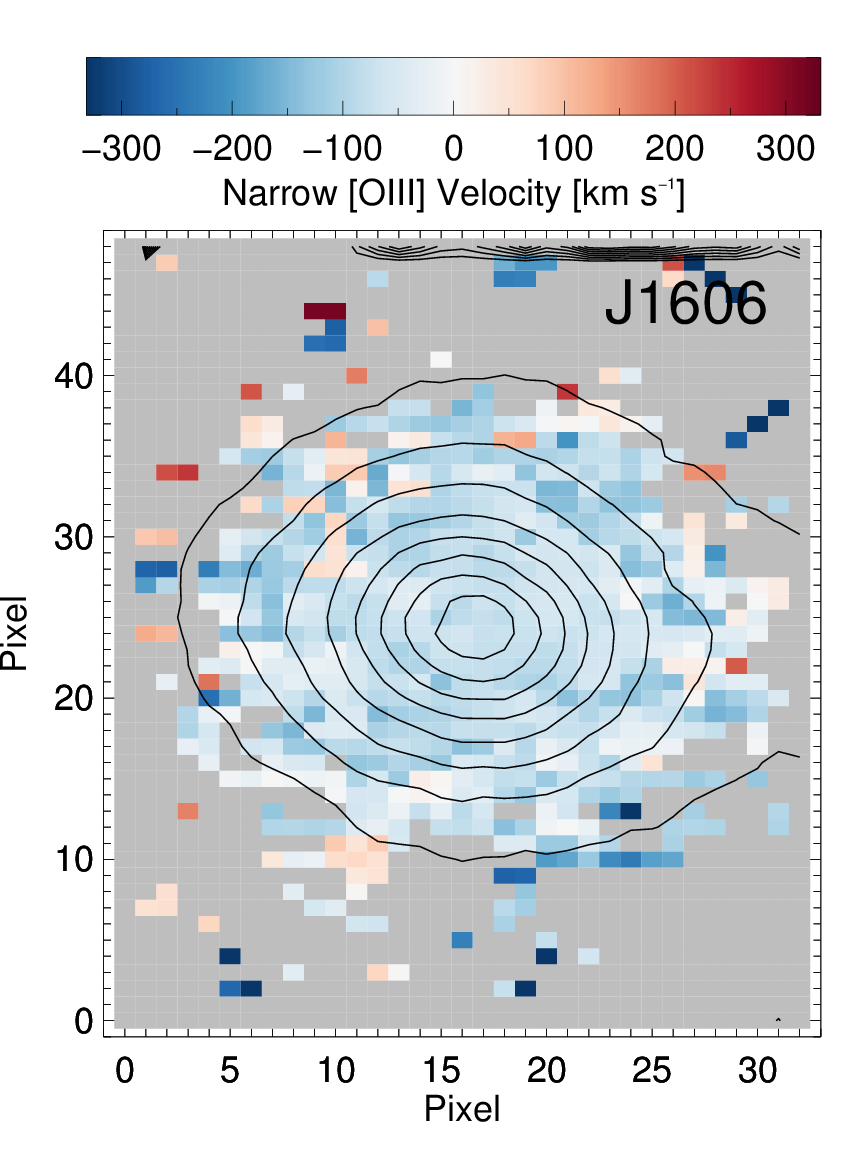}}
\raisebox{-0.5\height}{\includegraphics[width=0.16\textwidth,angle=0,trim={50 60 30 50},clip]{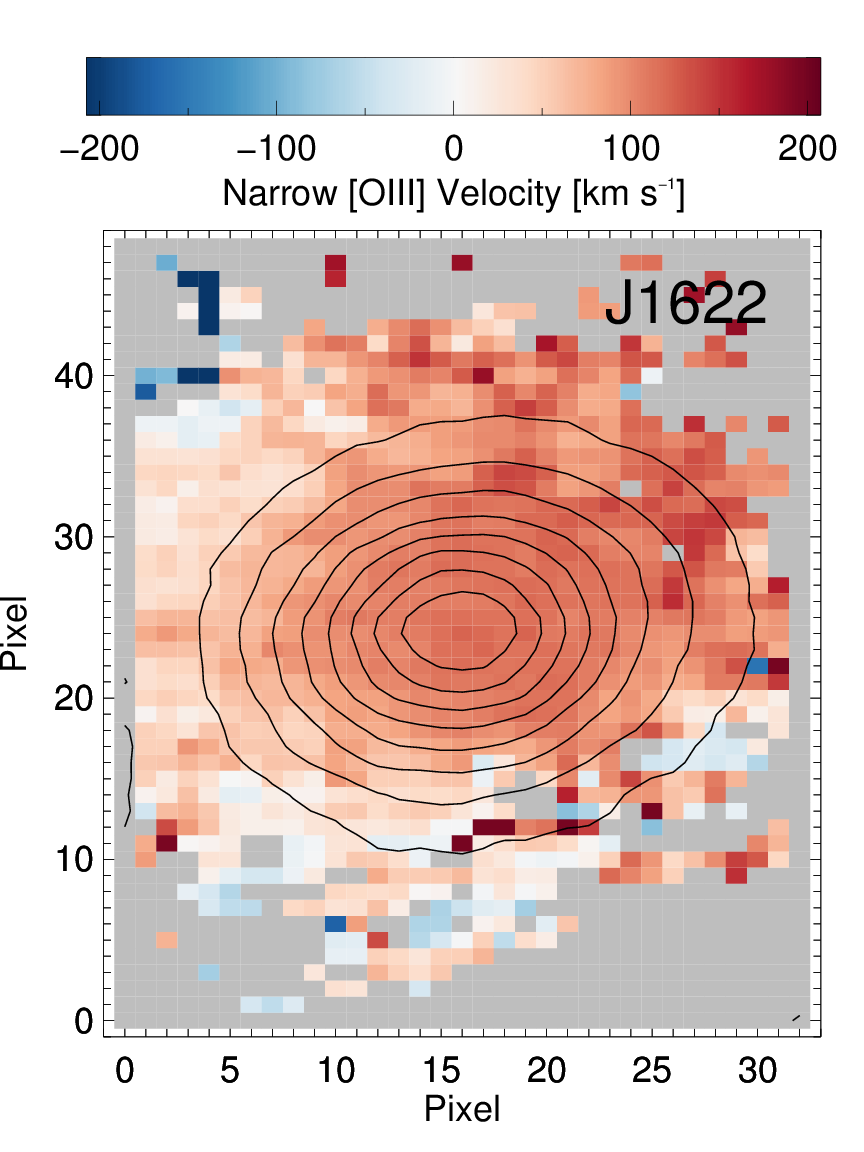}}
\raisebox{-0.5\height}{\includegraphics[width=0.16\textwidth,angle=0,trim={50 60 40 50},clip]{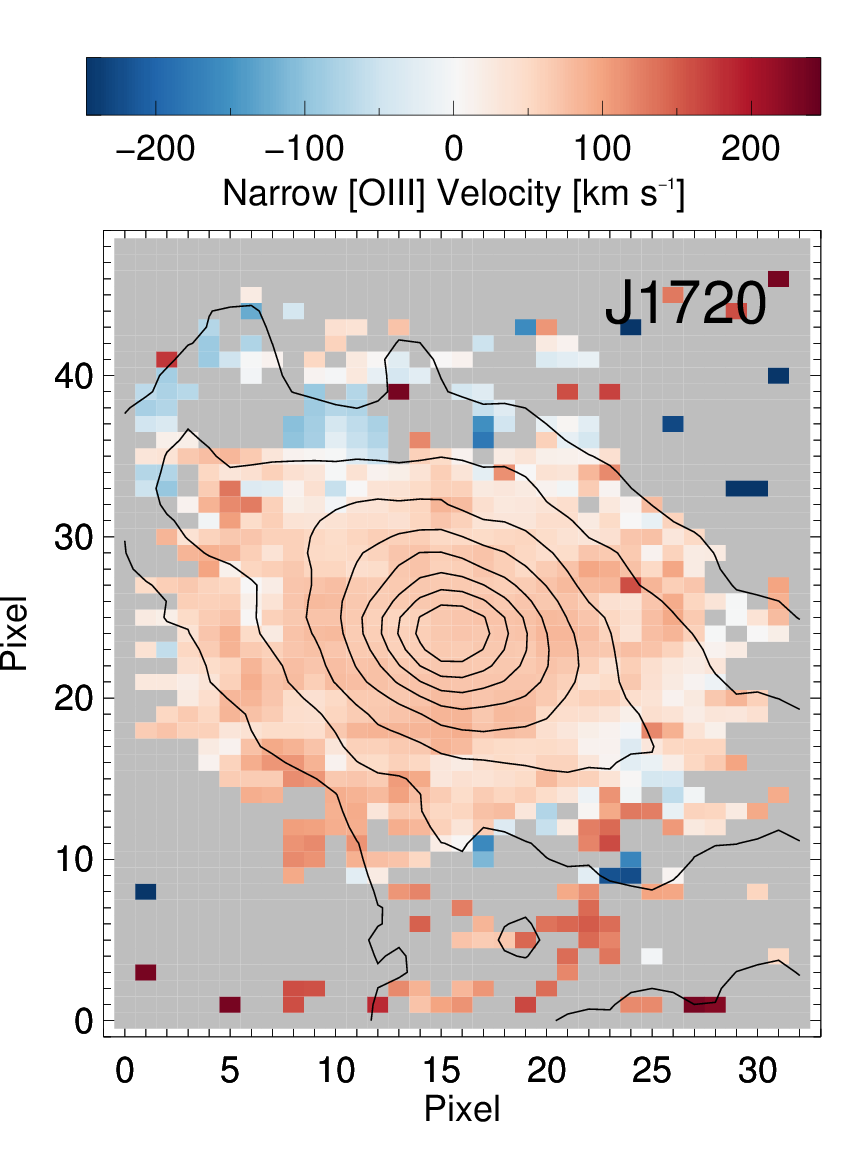}}\\
\vspace{5pt}\hrule\vspace{5pt}
{\Large Broad component}\\
\raisebox{-0.5\height}{\includegraphics[width=0.16\textwidth,angle=0,trim={50 60 40 50},clip]{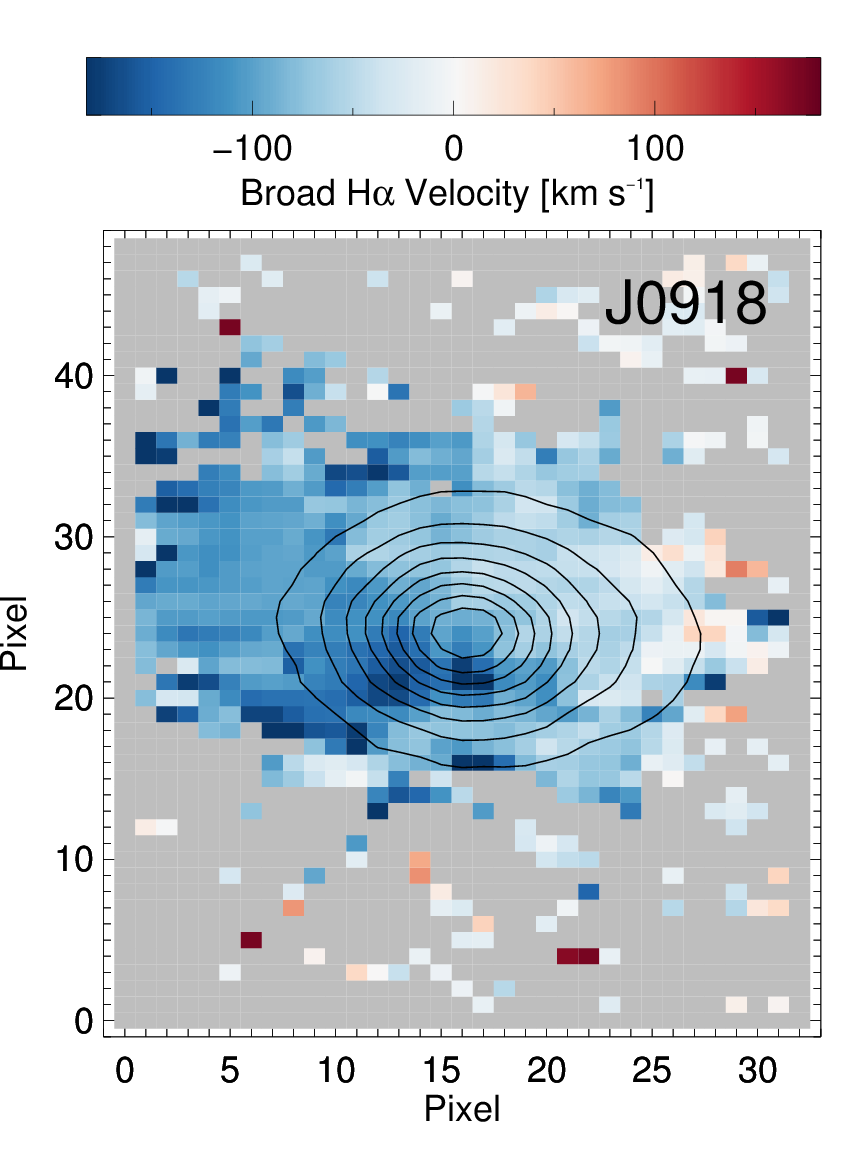}}
\raisebox{-0.5\height}{\includegraphics[width=0.16\textwidth,angle=0,trim={50 60 40 50},clip]{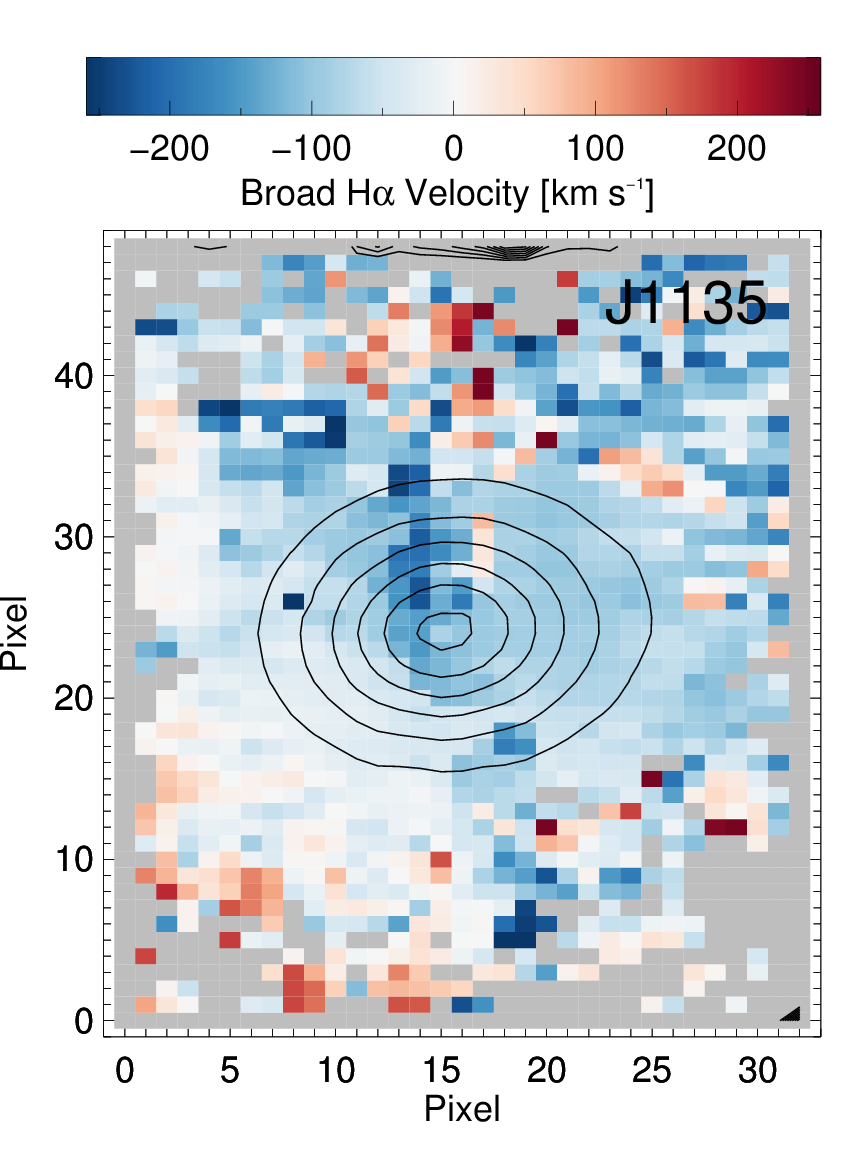}}
\raisebox{-0.5\height}{\includegraphics[width=0.16\textwidth,angle=0,trim={50 60 40 50},clip]{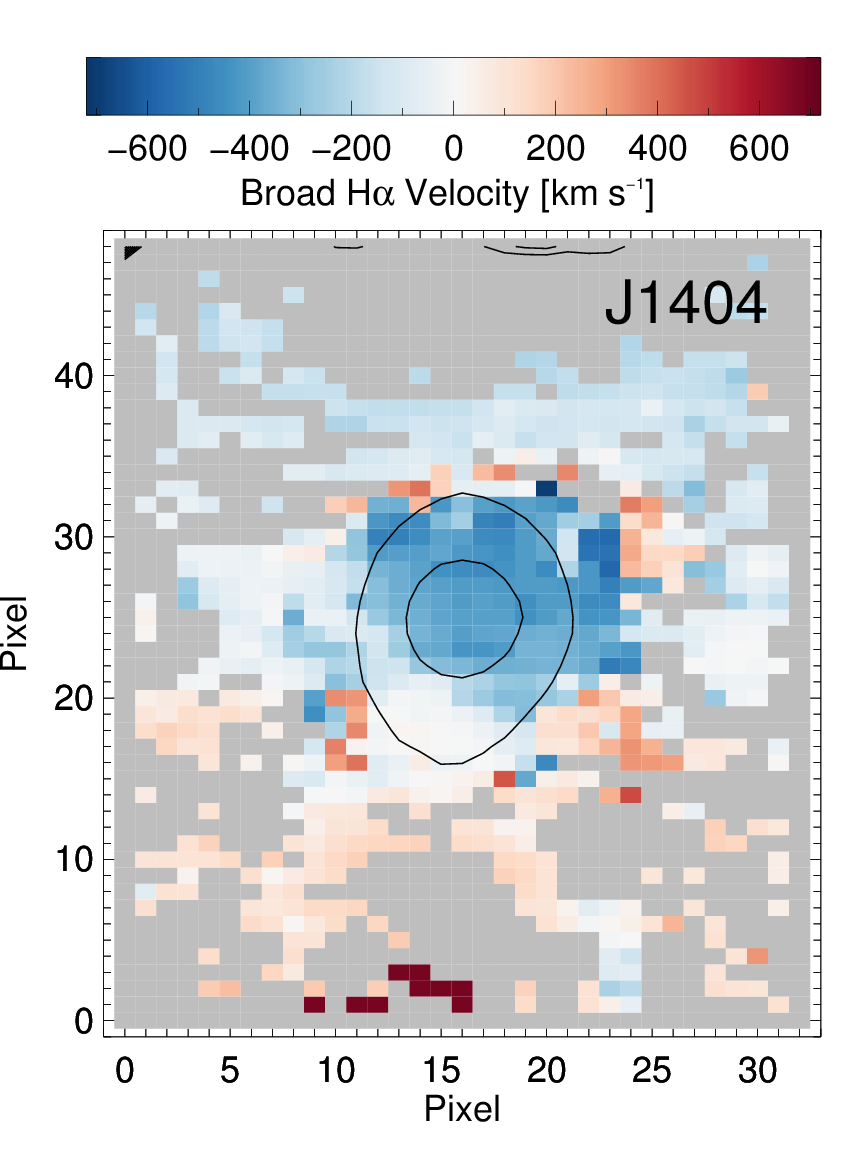}}
\raisebox{-0.5\height}{\includegraphics[width=0.16\textwidth,angle=0,trim={50 60 40 50},clip]{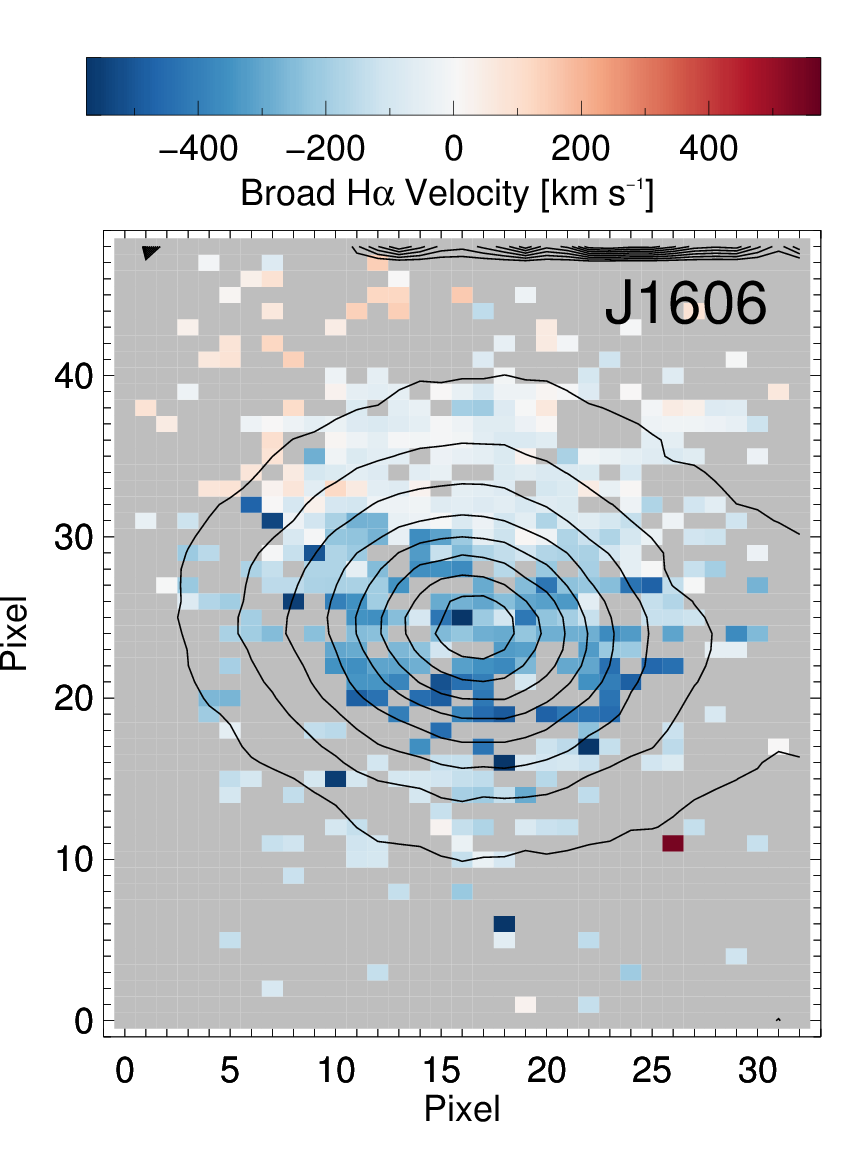}}
\raisebox{-0.5\height}{\includegraphics[width=0.16\textwidth,angle=0,trim={50 60 40 50},clip]{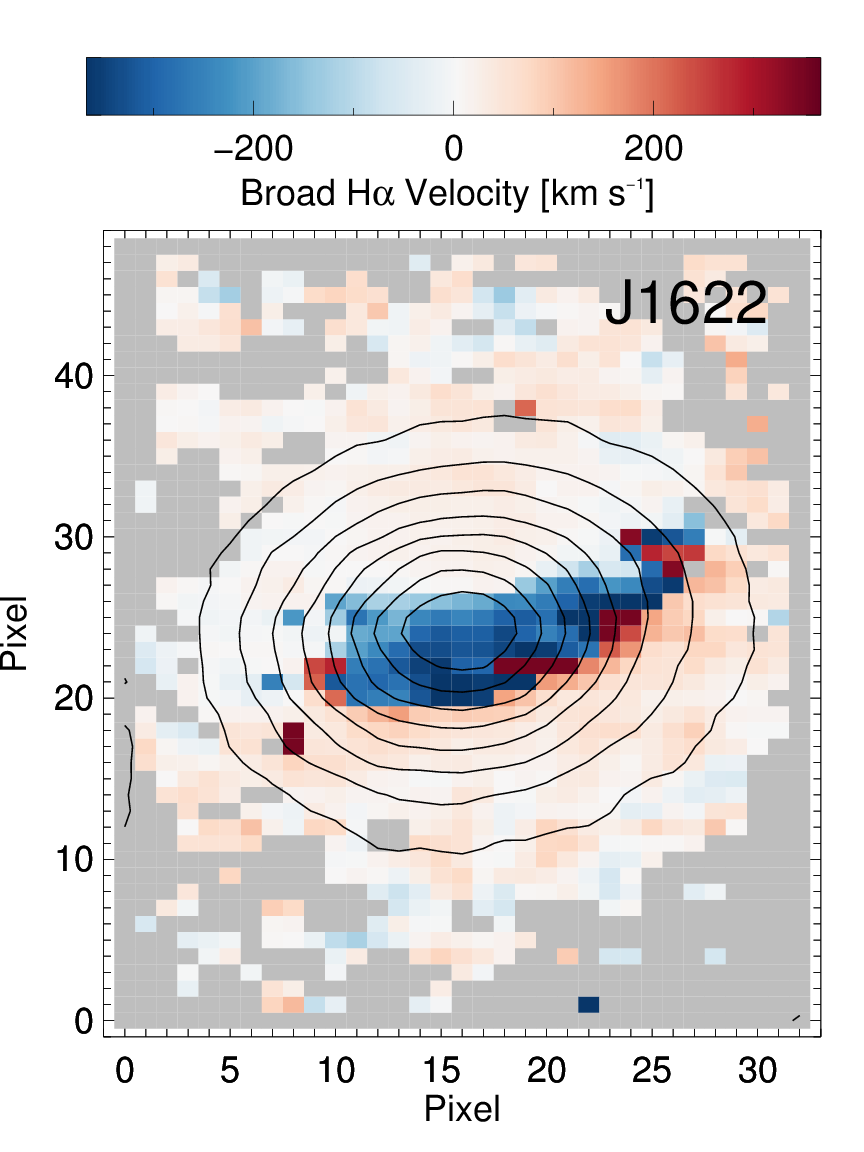}}
\raisebox{-0.5\height}{\includegraphics[width=0.16\textwidth,angle=0,trim={50 60 40 50},clip]{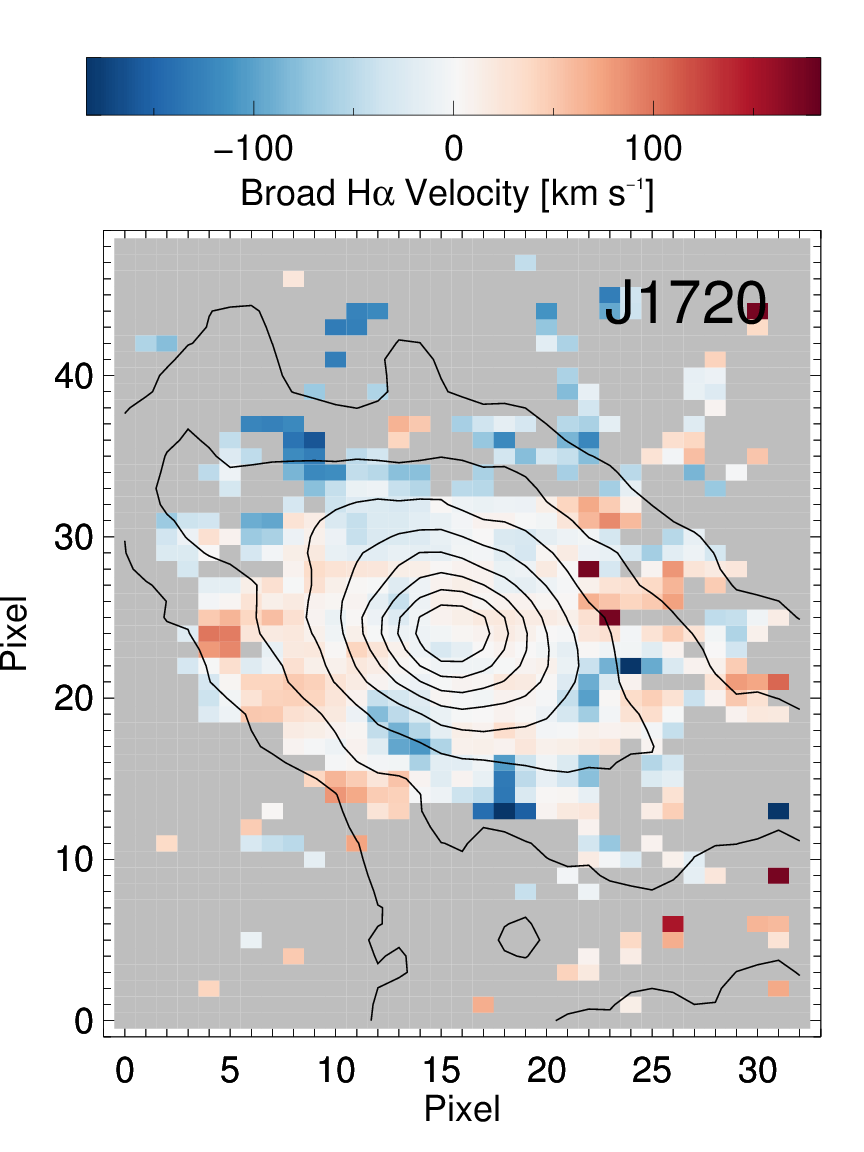}}\\
\raisebox{-0.5\height}{\includegraphics[width=0.16\textwidth,angle=0,trim={50 60 40 50},clip]{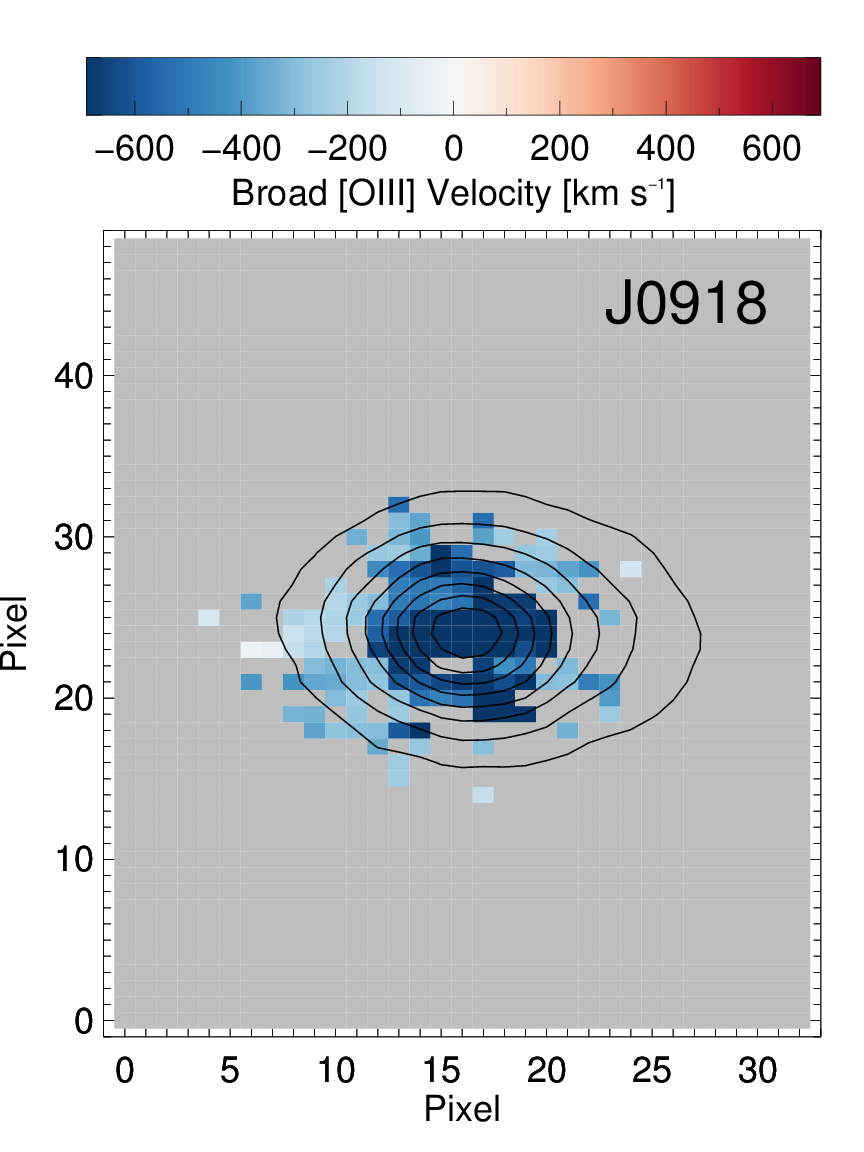}}
\raisebox{-0.5\height}{\includegraphics[width=0.16\textwidth,angle=0,trim={50 60 40 50},clip]{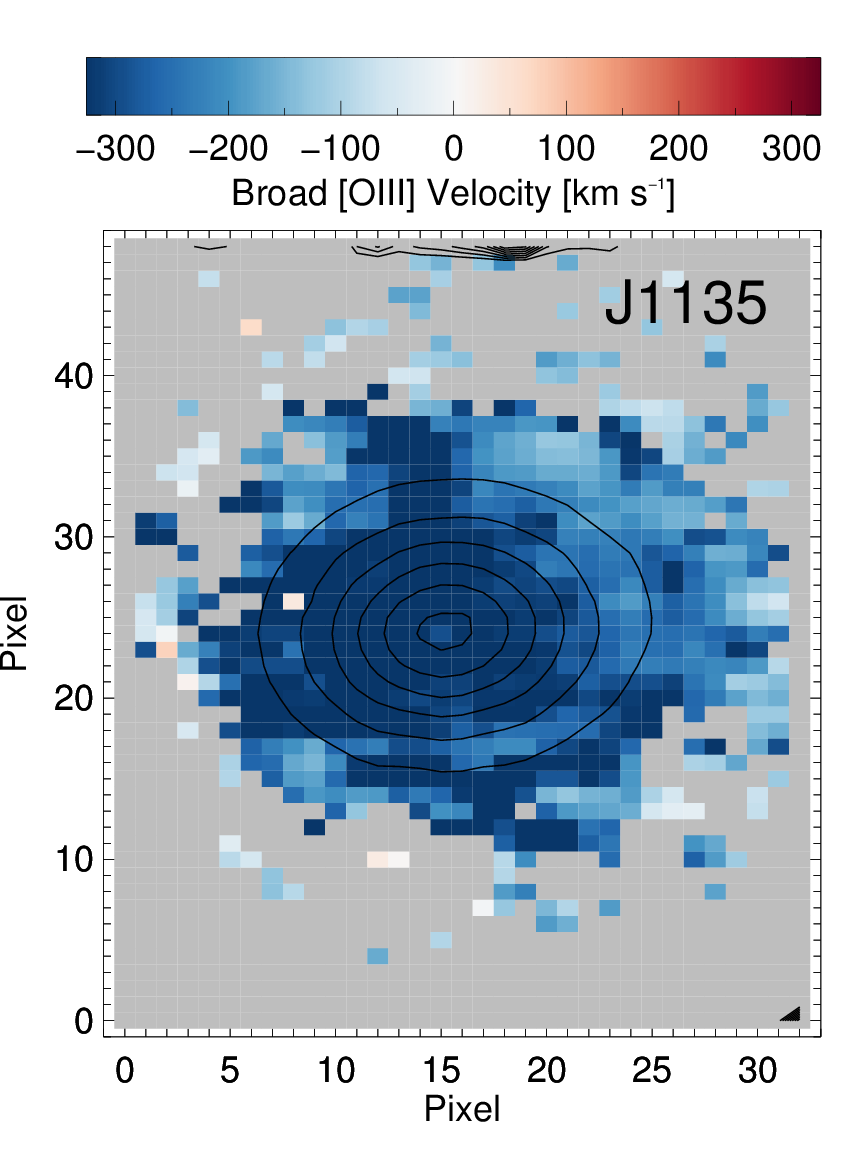}}
\raisebox{-0.5\height}{\includegraphics[width=0.16\textwidth,angle=0,trim={50 60 40 50},clip]{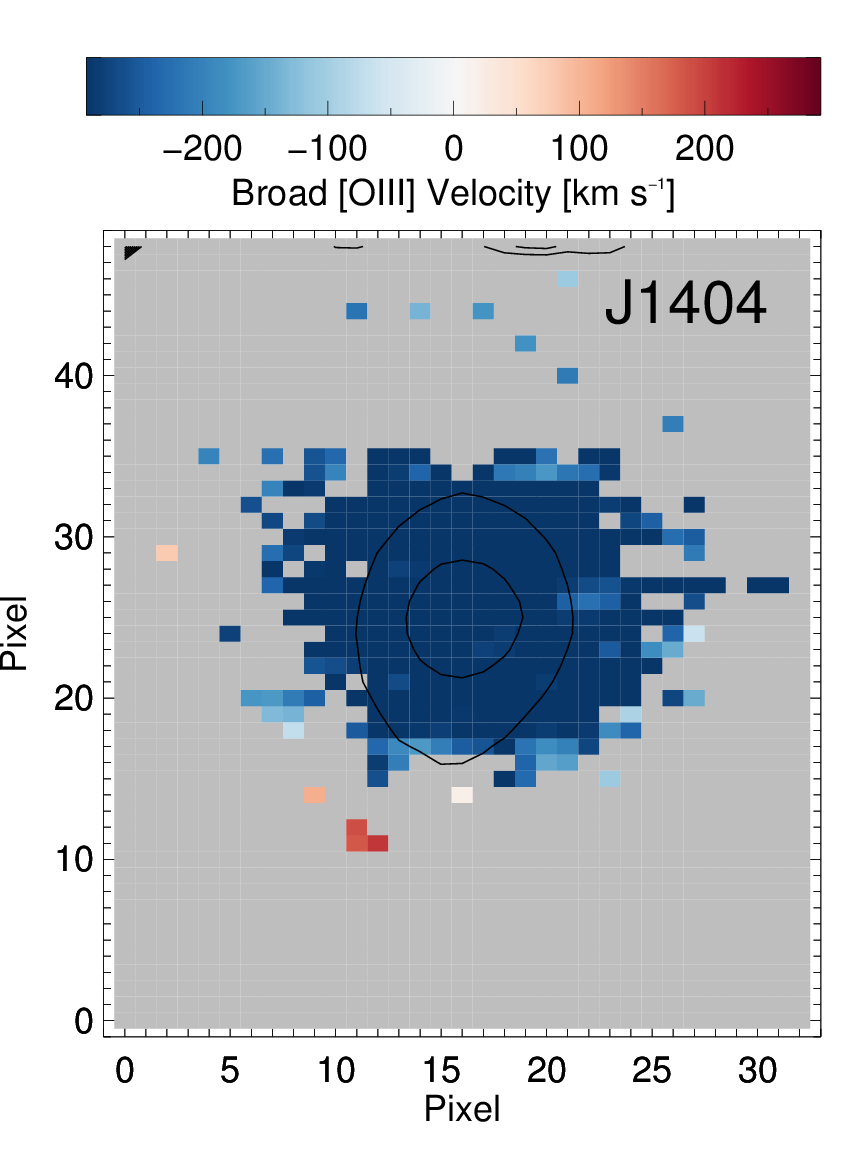}}
\raisebox{-0.5\height}{\includegraphics[width=0.16\textwidth,angle=0,trim={50 60 40 50},clip]{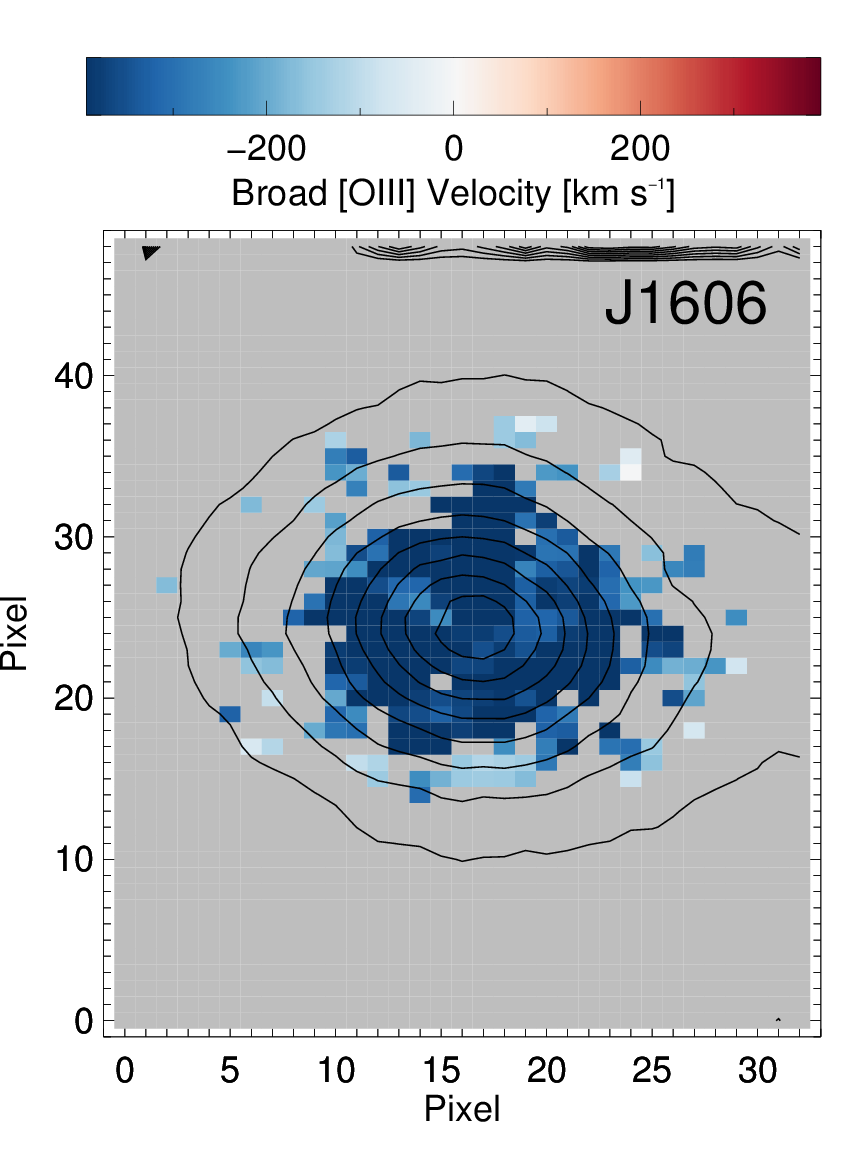}}
\raisebox{-0.5\height}{\includegraphics[width=0.16\textwidth,angle=0,trim={50 60 40 50},clip]{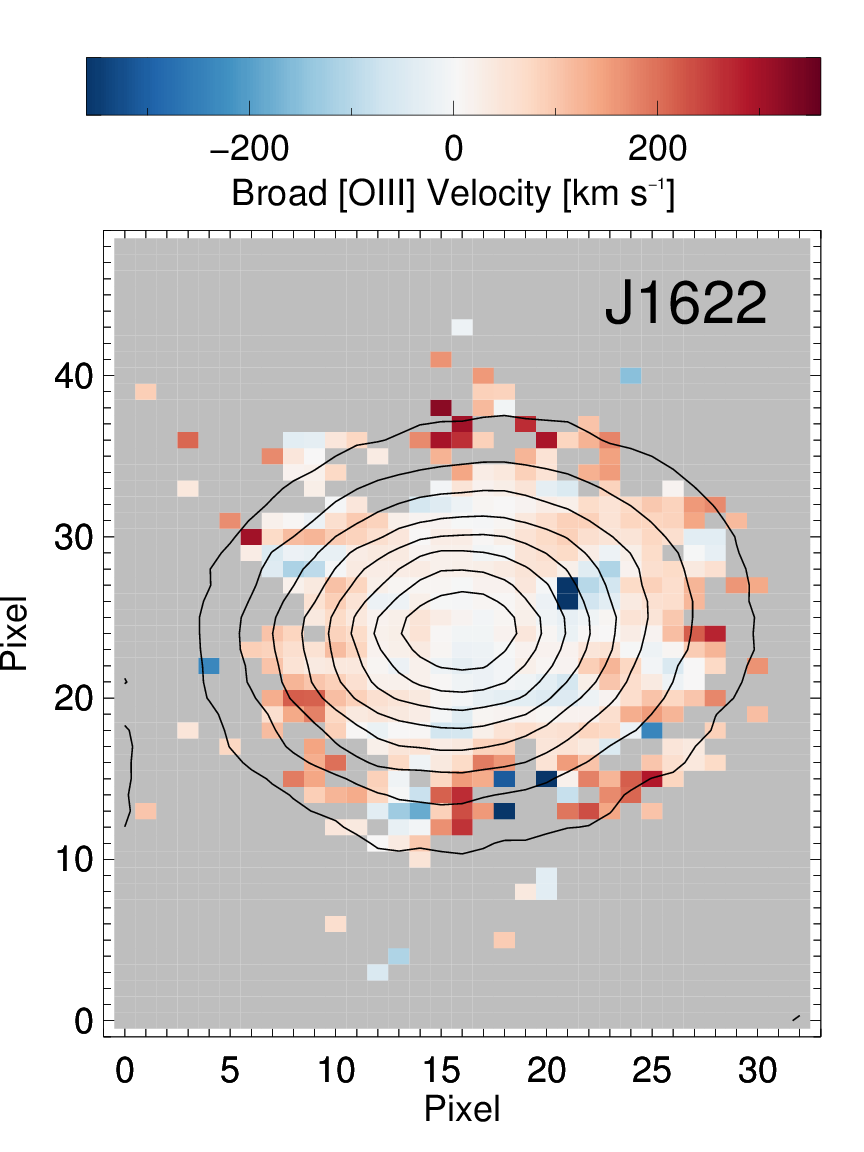}}
\raisebox{-0.5\height}{\includegraphics[width=0.16\textwidth,angle=0,trim={50 60 40 50},clip]{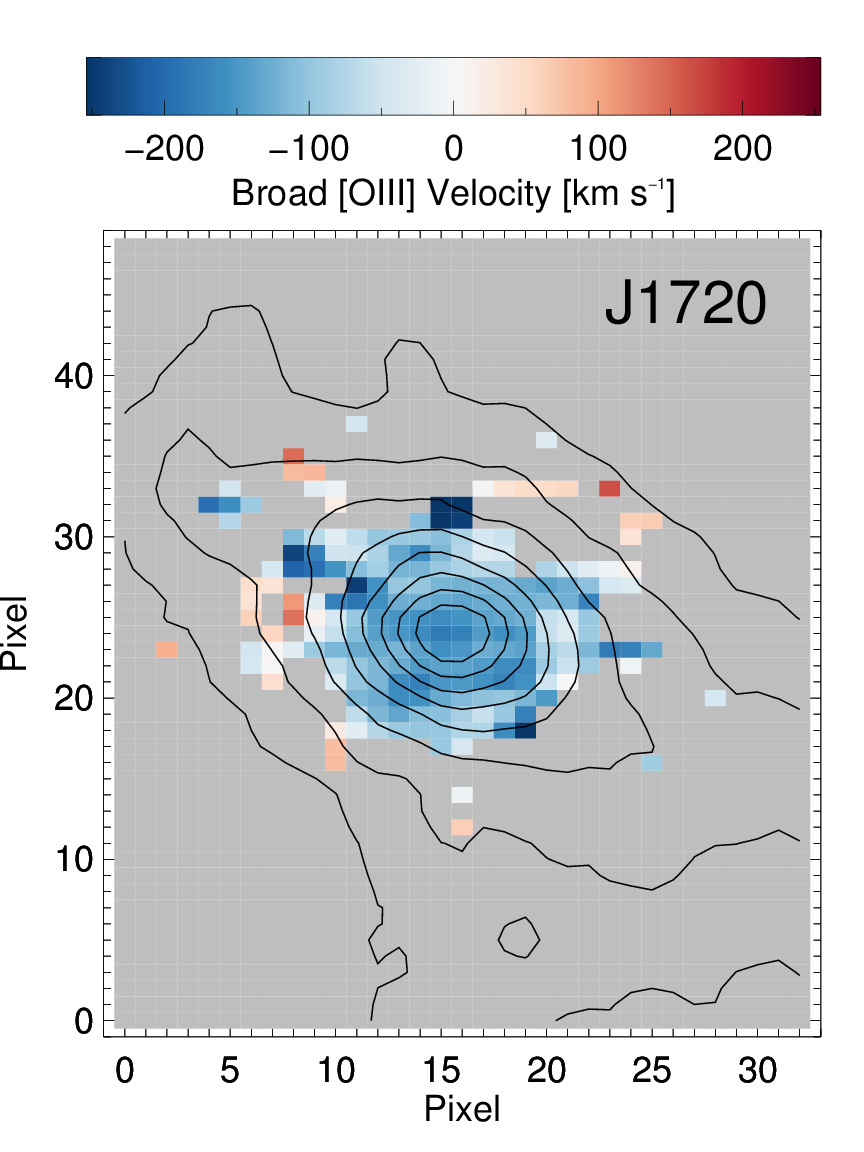}}
\caption{Velocity maps of the narrow components of H$\alpha$ and \mbox{[O\,\textsc{iii}]} (top) and the broad components of H$\alpha$ and \mbox{[O\,\textsc{iii}]} (bottom) from spectral decomposition in each spaxel.  Contours are the same as in Fig. \ref{fig:all_flux}. Grey spaxels indicate non-detection of the emission lines (i.e., S/N$<$3).}
\label{fig:all_kins_dv}
\end{center}
\end{figure*}

As discussed in Section \ref{sec:method}, we find that emission lines require two Gaussian components to be adequately described, due to strong, and sometimes very asymmetric wing components (see Fig. \ref{fig:examplefit}). \citet{Woo2015b}, through a rigorous statistical study of a sample of $\sim39,000$ Type 2 AGNs from the SDSS, concluded that this secondary Gaussian component is a strong indication of the presence of outflows since it is considerably broader than stellar absorption lines and kinematically offset from the systemic velocity of the galaxy. At the same time, its kinematic properties broadly correlate with the AGN luminosity and Eddington ratio. Conversely, the narrow component is found to mostly reflect the gravitational potential of the galaxy. 
The velocity maps in Fig. \ref{fig:all_kins_total}, particularly for H$\alpha$, reveal that there is a spatial dependence, with the gravitational (rotation) kinematic component dominating in the outer parts of the FoV. The complex kinematic structure of the emission line profiles and their spatial trend as well as the statistical results of \citet{Woo2015b} motivate us to look at kinematics maps of each Gaussian component separately, instead of the total emission profile. 

\begin{figure*}[tbp]
\begin{center}
{\Large Narrow component}\\
\raisebox{-0.5\height}{\includegraphics[width=0.16\textwidth,angle=0,trim={50 60 20 50},clip]{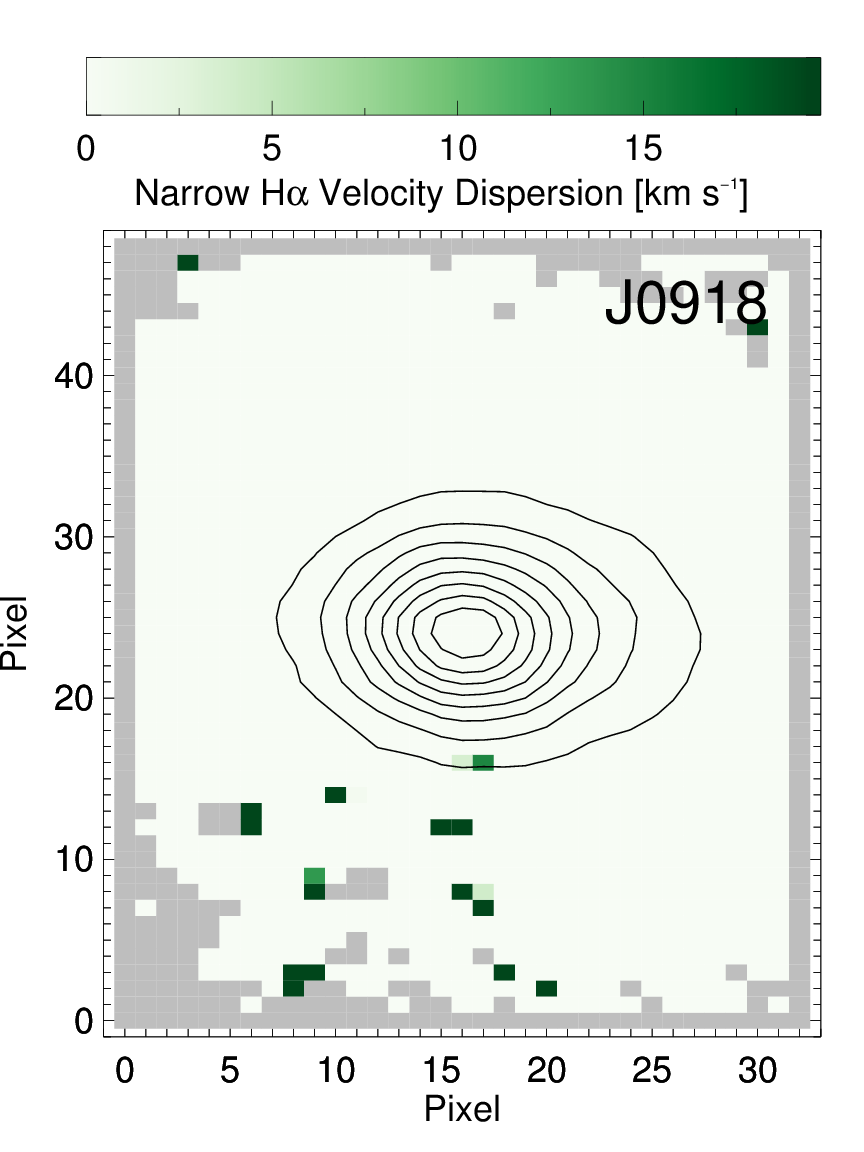}}
\raisebox{-0.5\height}{\includegraphics[width=0.16\textwidth,angle=0,trim={50 60 20 50},clip]{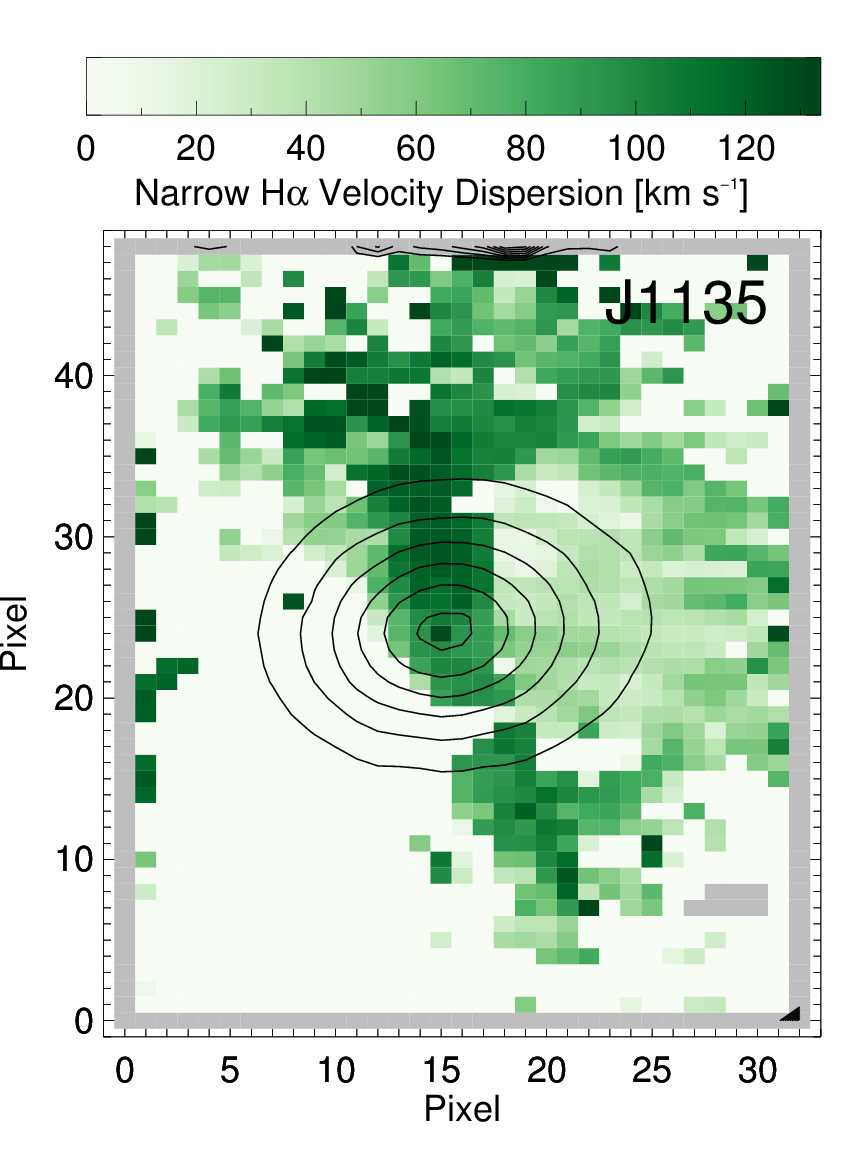}}
\raisebox{-0.5\height}{\includegraphics[width=0.16\textwidth,angle=0,trim={50 60 20 50},clip]{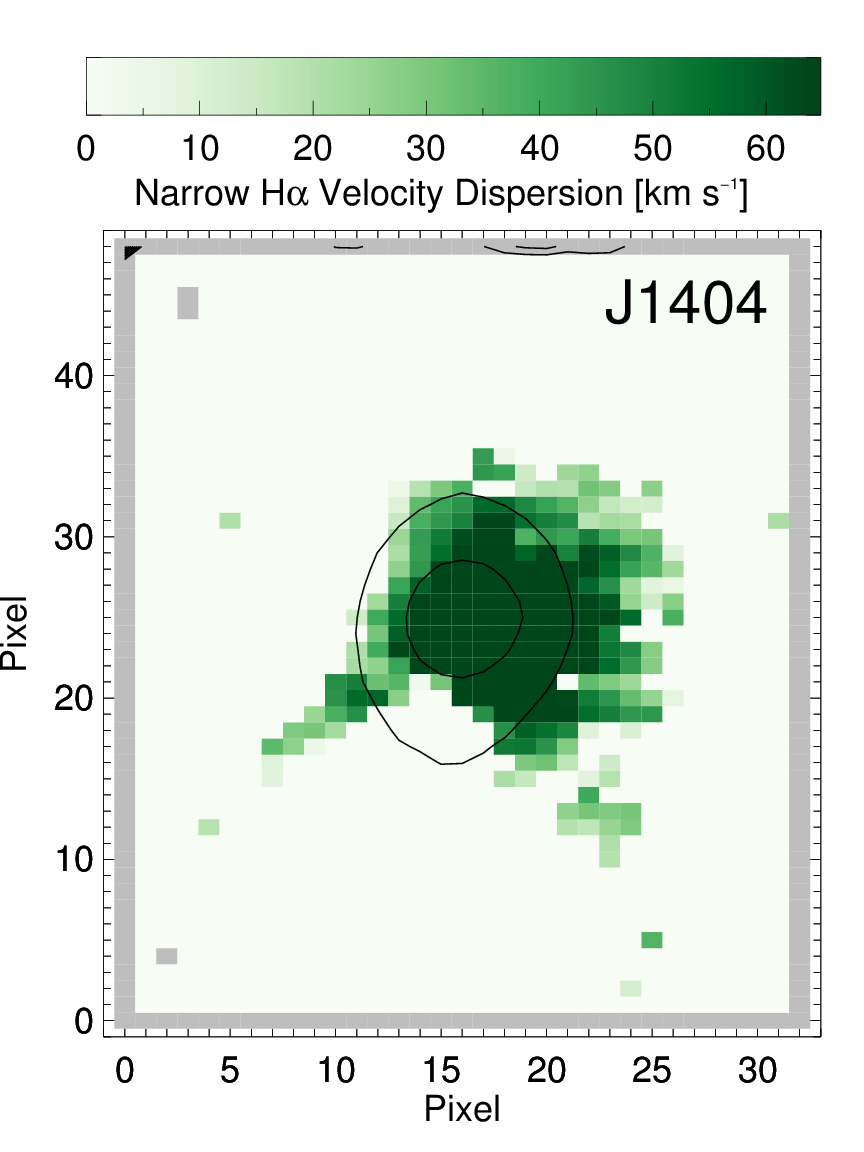}}
\raisebox{-0.5\height}{\includegraphics[width=0.16\textwidth,angle=0,trim={50 60 20 50},clip]{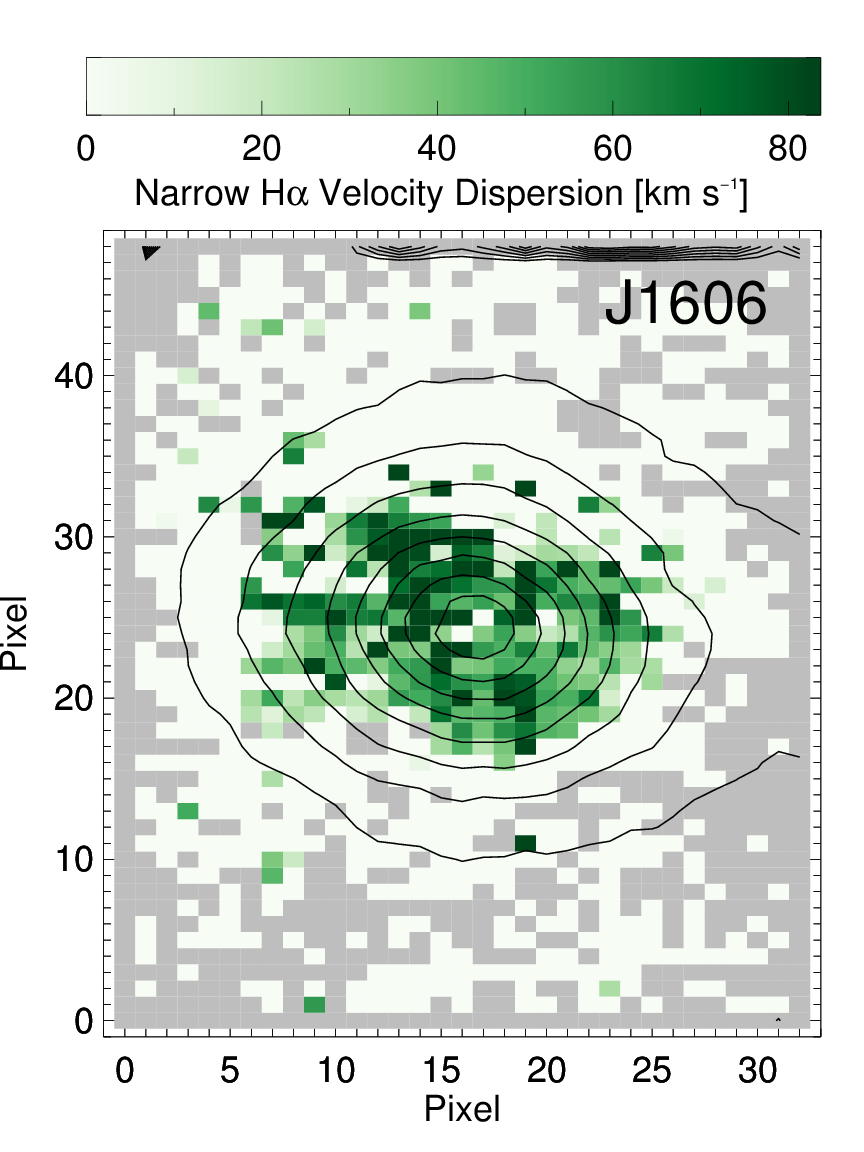}}
\raisebox{-0.5\height}{\includegraphics[width=0.16\textwidth,angle=0,trim={50 60 20 50},clip]{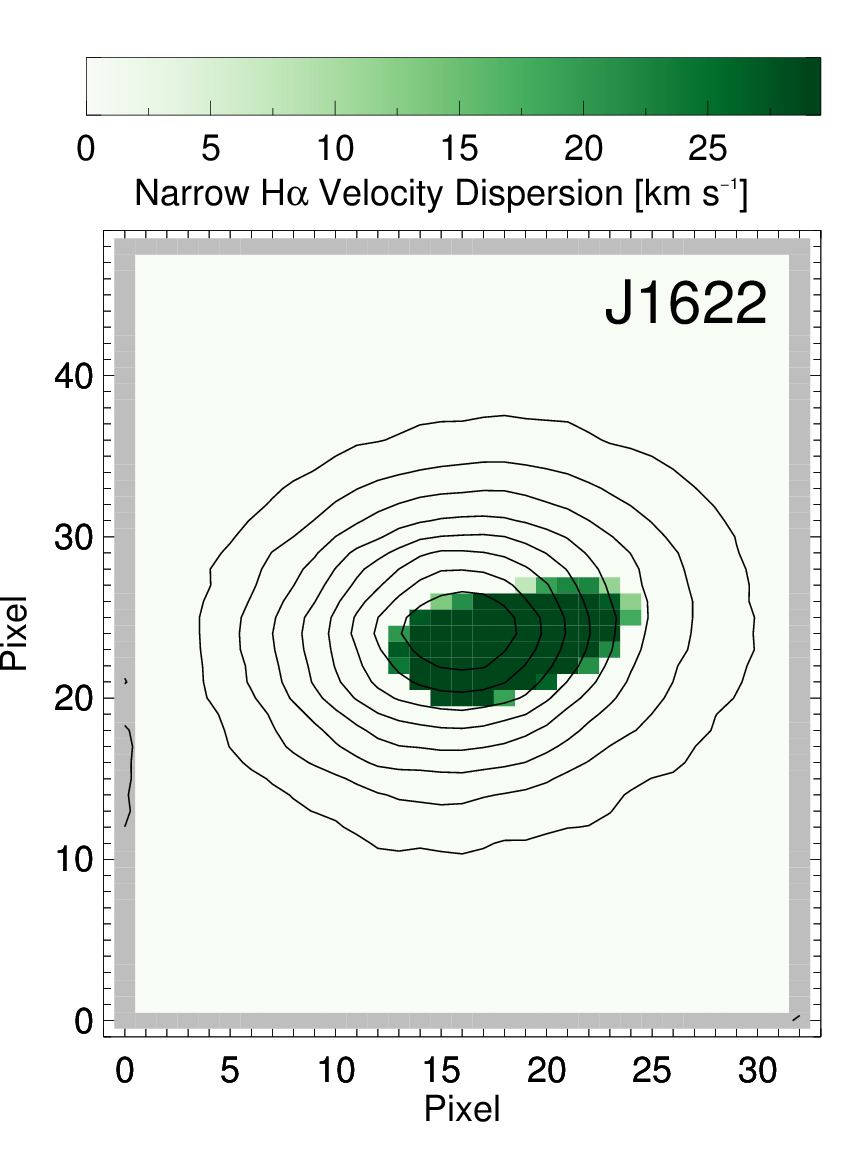}}
\raisebox{-0.5\height}{\includegraphics[width=0.16\textwidth,angle=0,trim={50 60 20 50},clip]{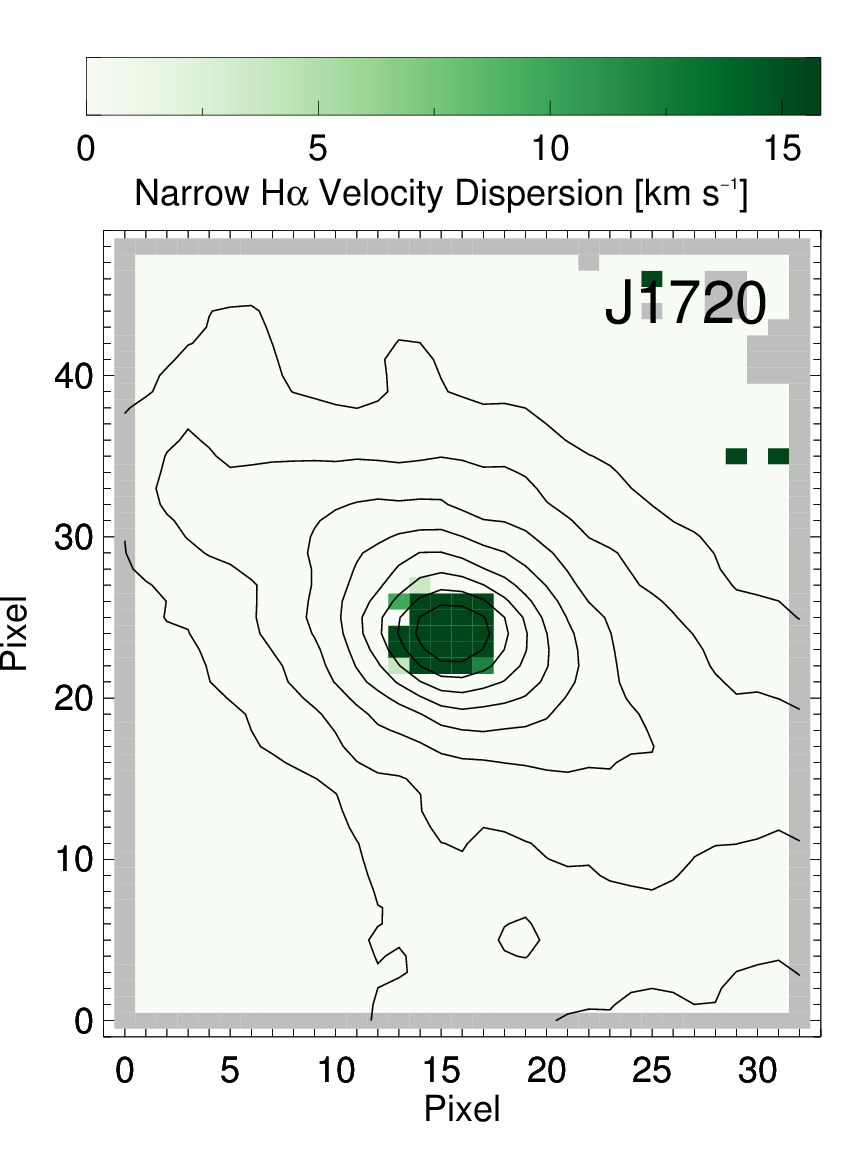}}\\
\raisebox{-0.5\height}{\includegraphics[width=0.16\textwidth,angle=0,trim={50 60 20 50},clip]{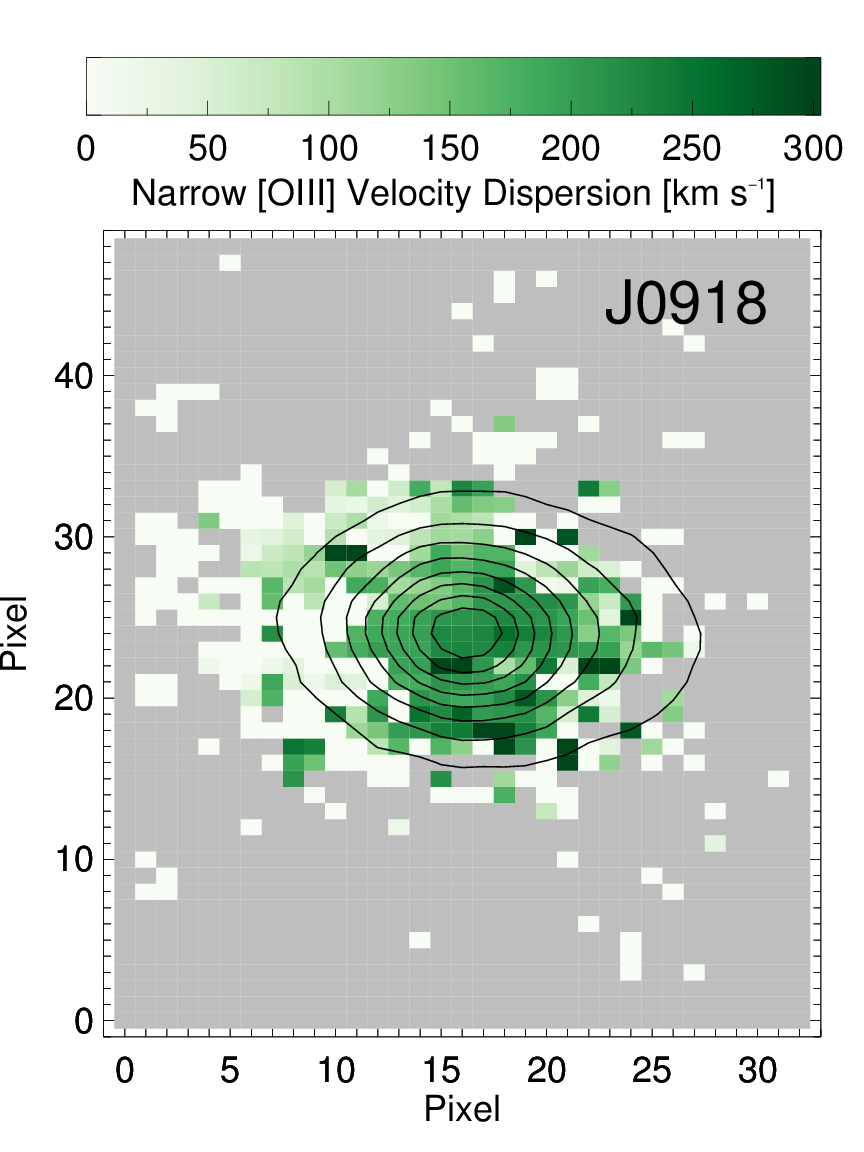}}
\raisebox{-0.5\height}{\includegraphics[width=0.16\textwidth,angle=0,trim={50 60 20 50},clip]{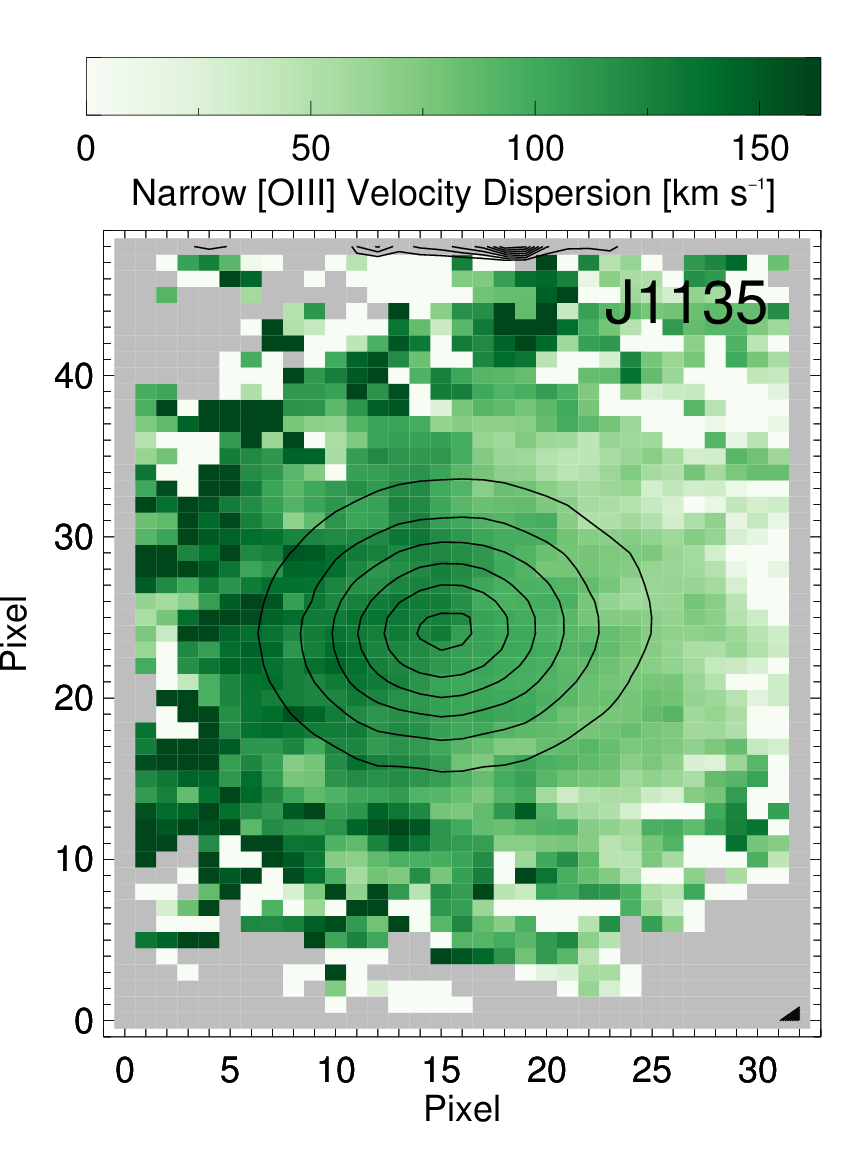}}
\raisebox{-0.5\height}{\includegraphics[width=0.16\textwidth,angle=0,trim={50 60 20 50},clip]{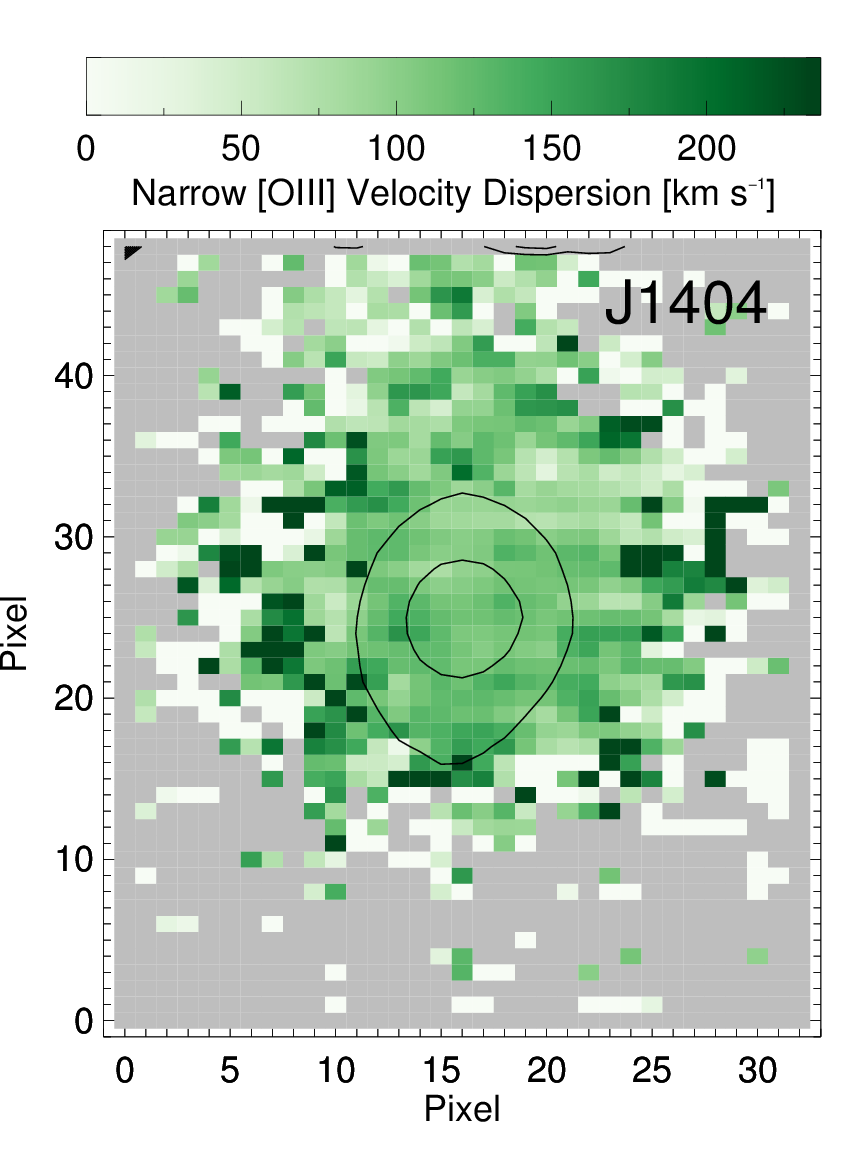}}
\raisebox{-0.5\height}{\includegraphics[width=0.16\textwidth,angle=0,trim={50 60 20 50},clip]{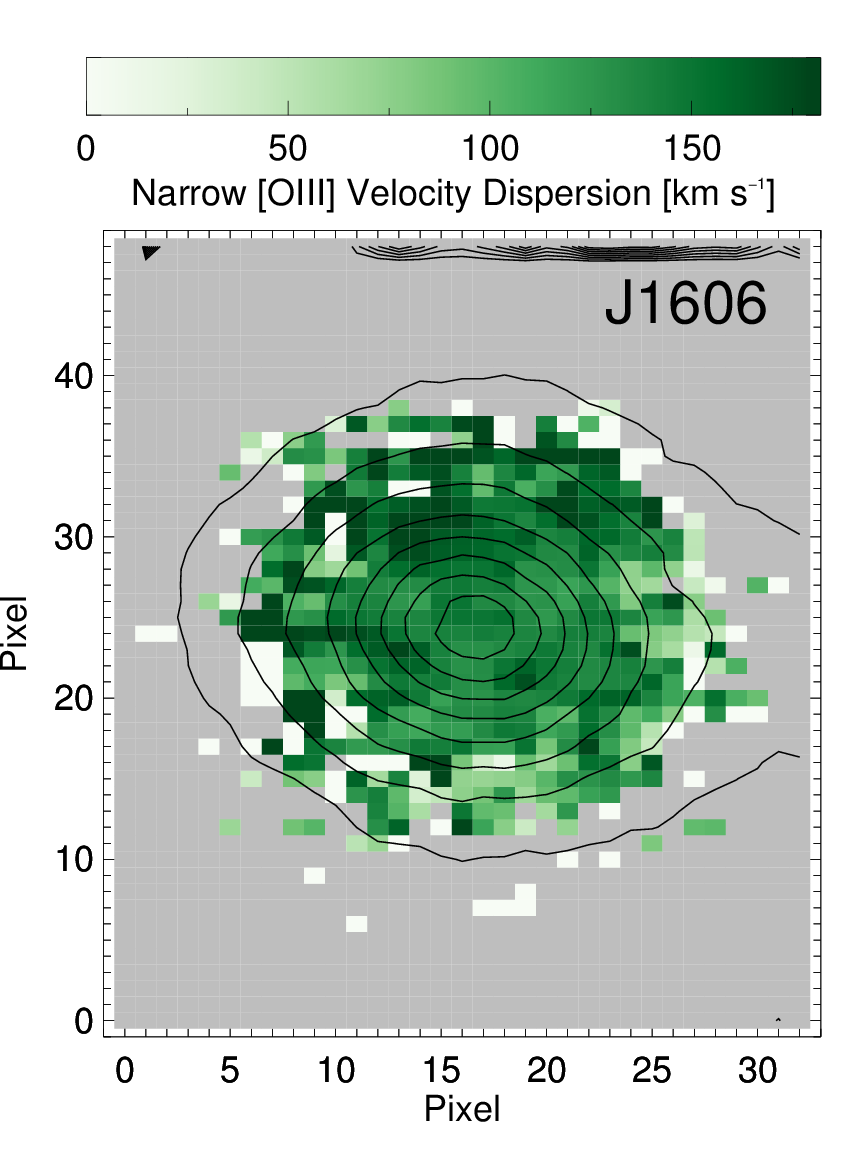}}
\raisebox{-0.5\height}{\includegraphics[width=0.16\textwidth,angle=0,trim={50 60 20 50},clip]{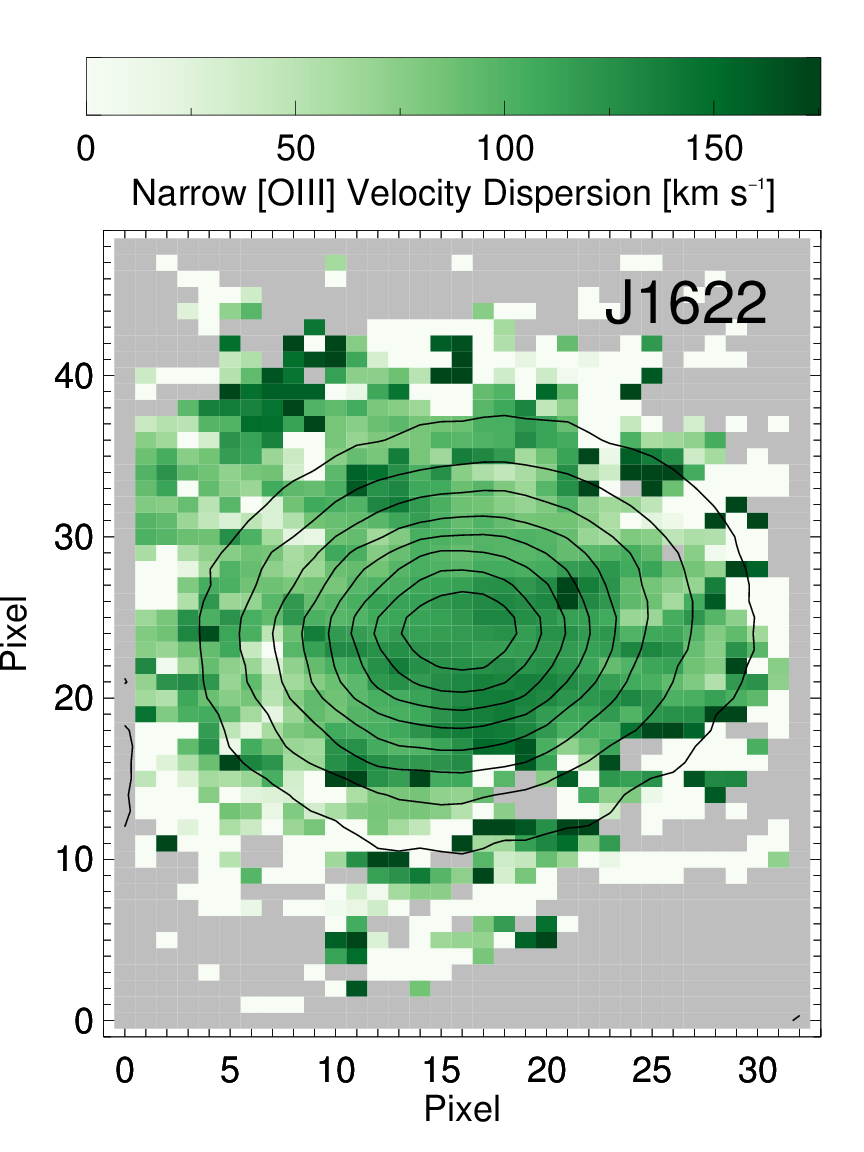}}
\raisebox{-0.5\height}{\includegraphics[width=0.16\textwidth,angle=0,trim={50 60 20 50},clip]{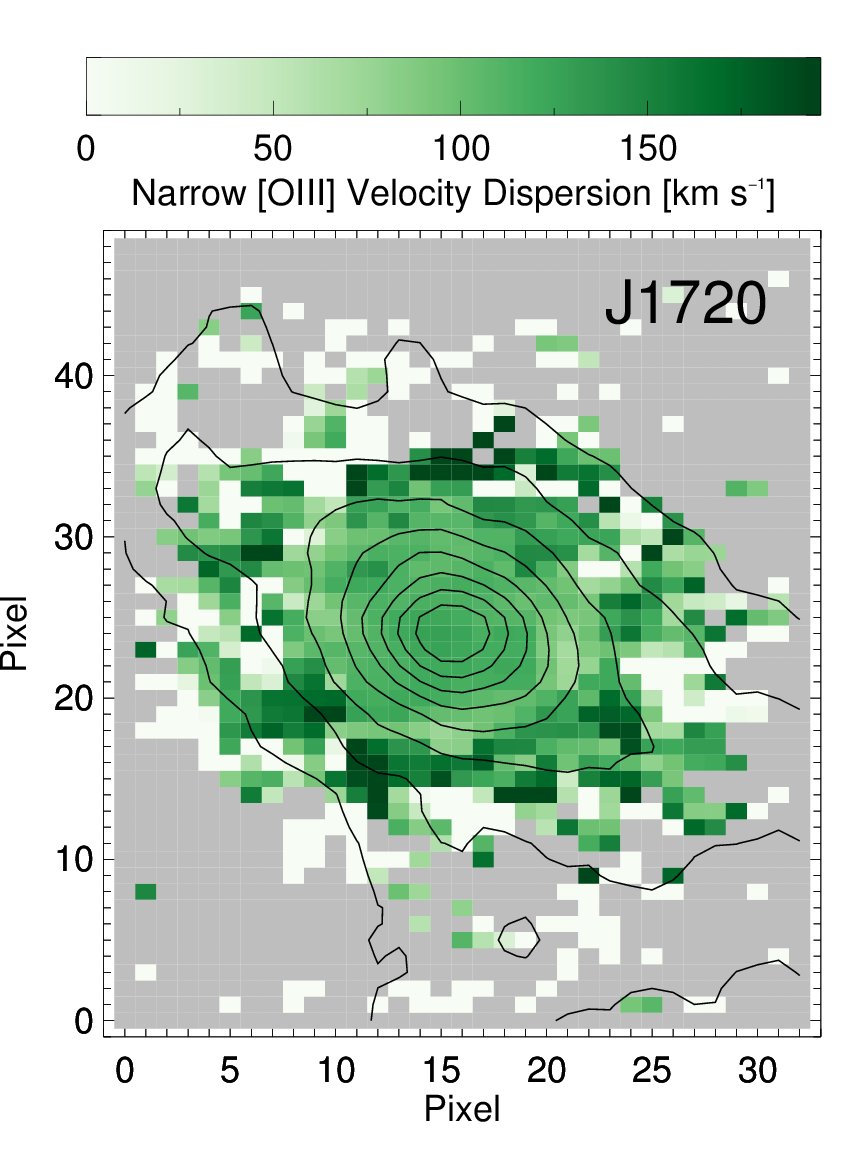}}\\
\vspace{5pt}\hrule\vspace{5pt}
{\Large Broad component}\\
\raisebox{-0.5\height}{\includegraphics[width=0.16\textwidth,angle=0,trim={50 60 20 50},clip]{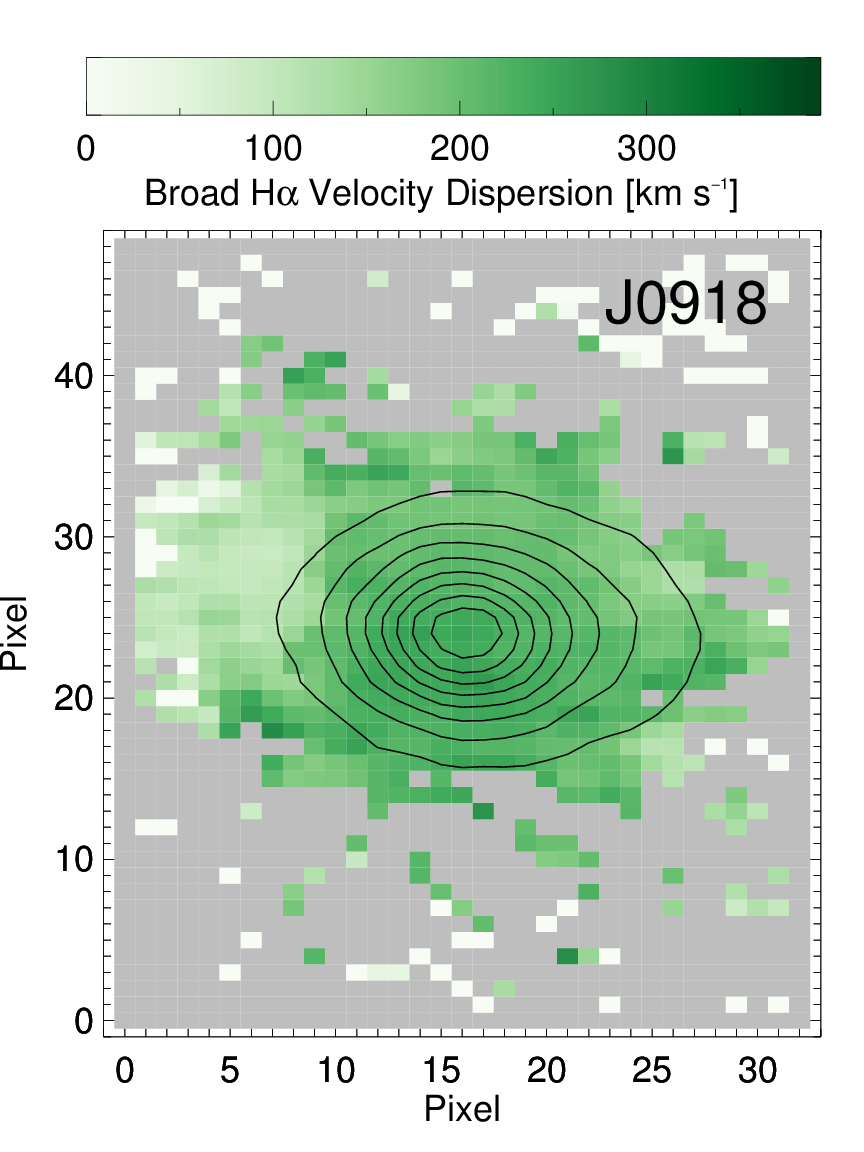}}
\raisebox{-0.5\height}{\includegraphics[width=0.16\textwidth,angle=0,trim={50 60 20 50},clip]{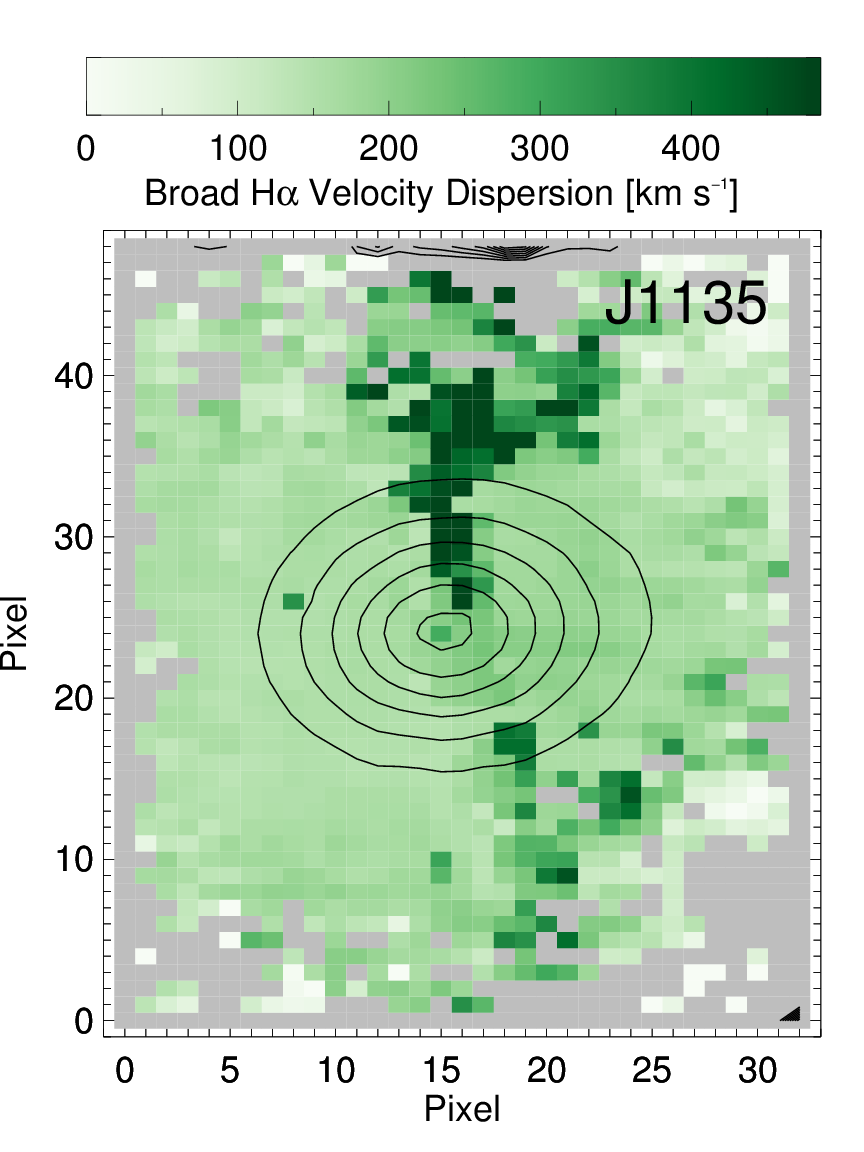}}
\raisebox{-0.5\height}{\includegraphics[width=0.16\textwidth,angle=0,trim={50 60 20 50},clip]{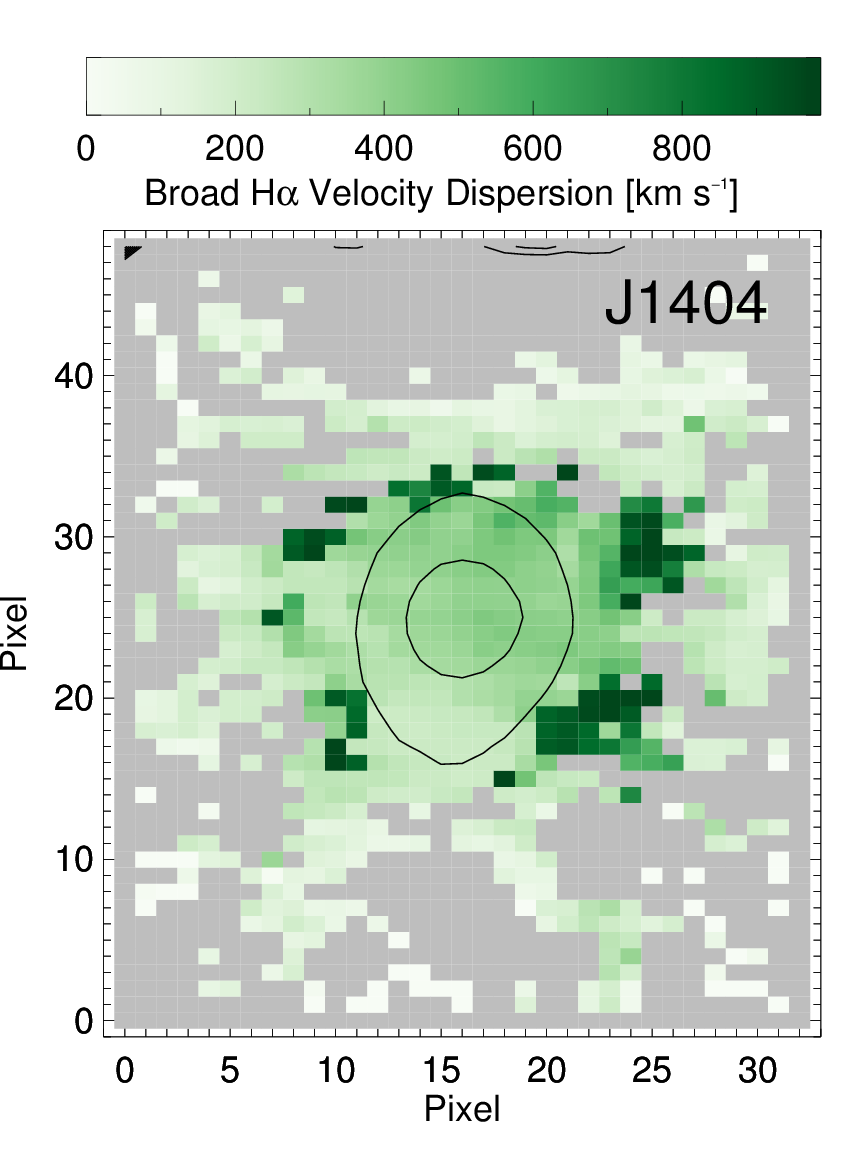}}
\raisebox{-0.5\height}{\includegraphics[width=0.16\textwidth,angle=0,trim={50 60 20 50},clip]{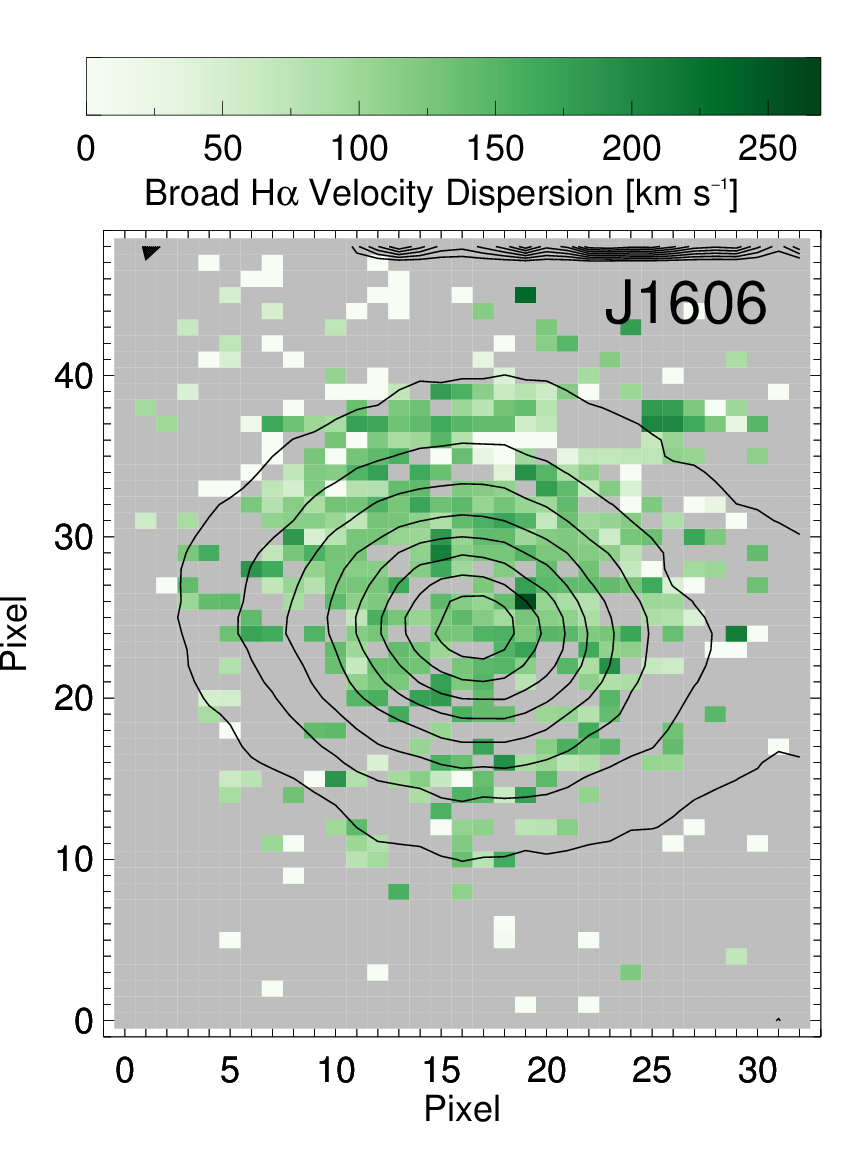}}
\raisebox{-0.5\height}{\includegraphics[width=0.16\textwidth,angle=0,trim={50 60 20 50},clip]{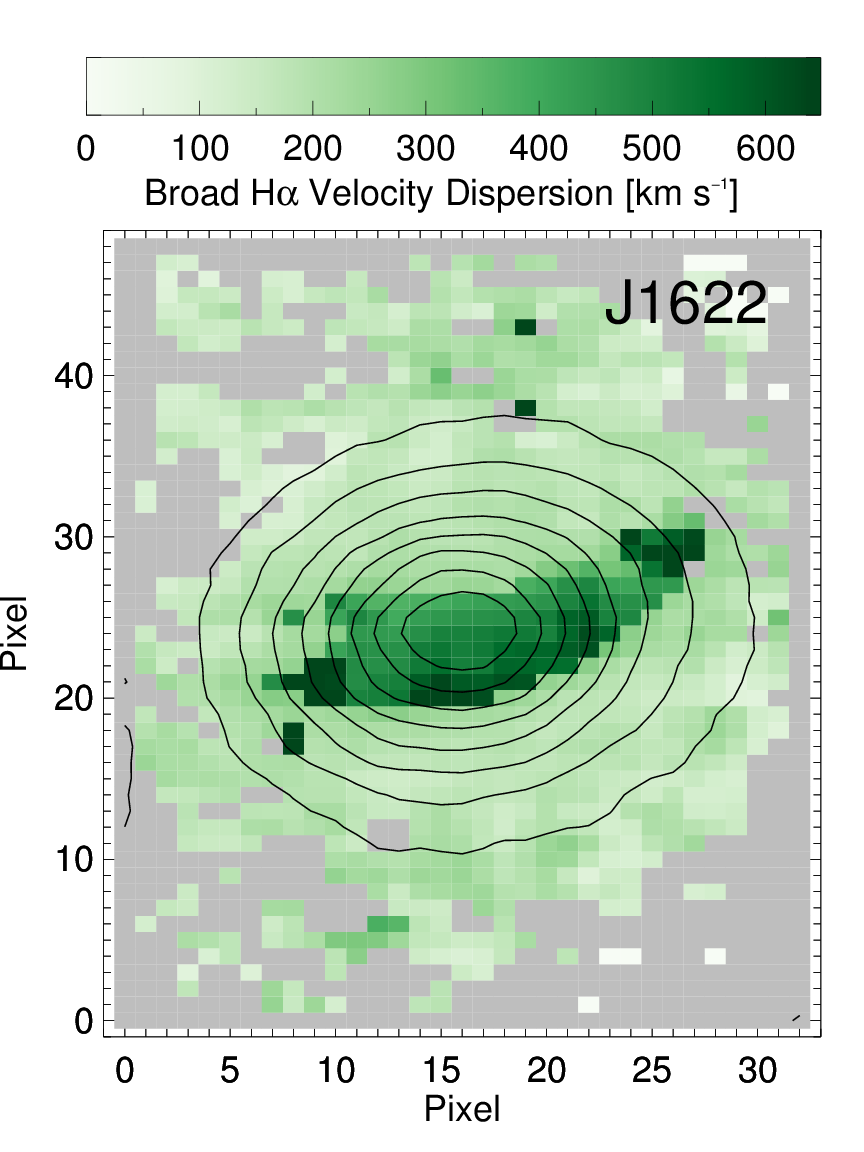}}
\raisebox{-0.5\height}{\includegraphics[width=0.16\textwidth,angle=0,trim={50 60 20 50},clip]{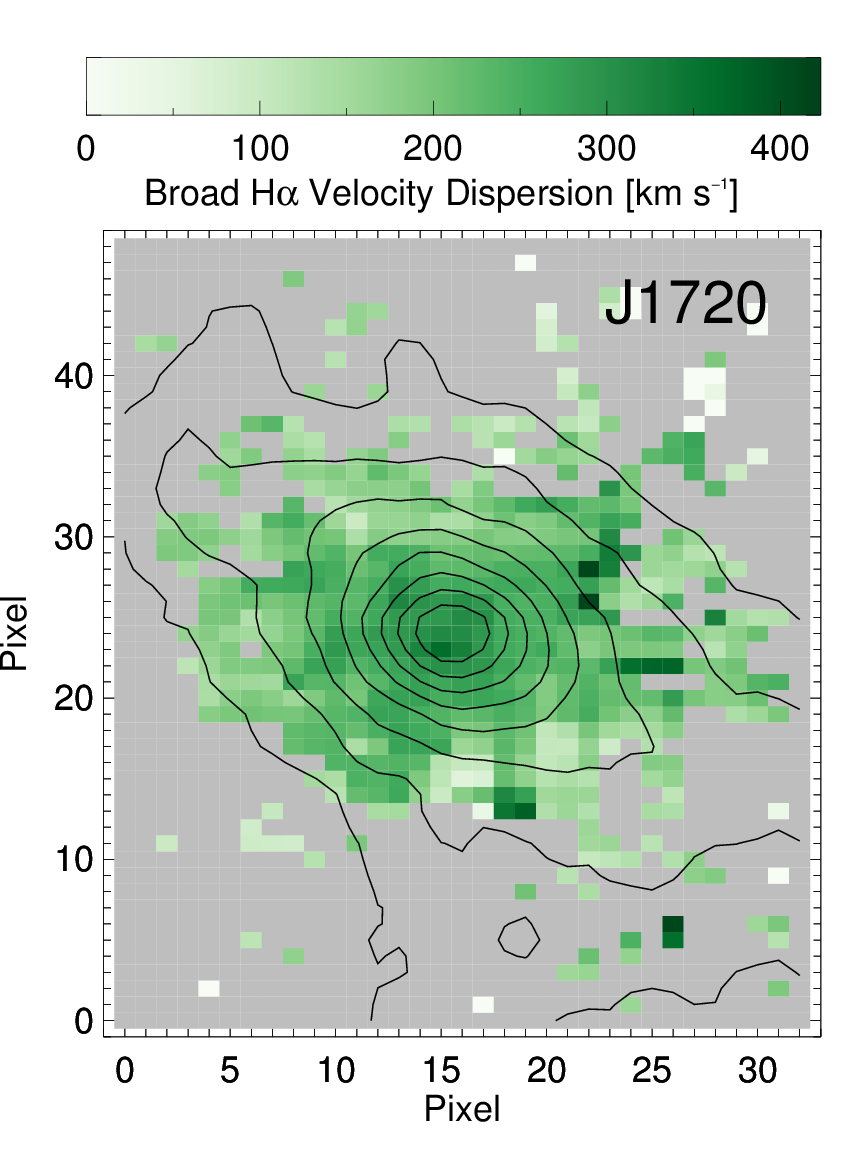}}\\
\raisebox{-0.5\height}{\includegraphics[width=0.16\textwidth,angle=0,trim={50 60 20 50},clip]{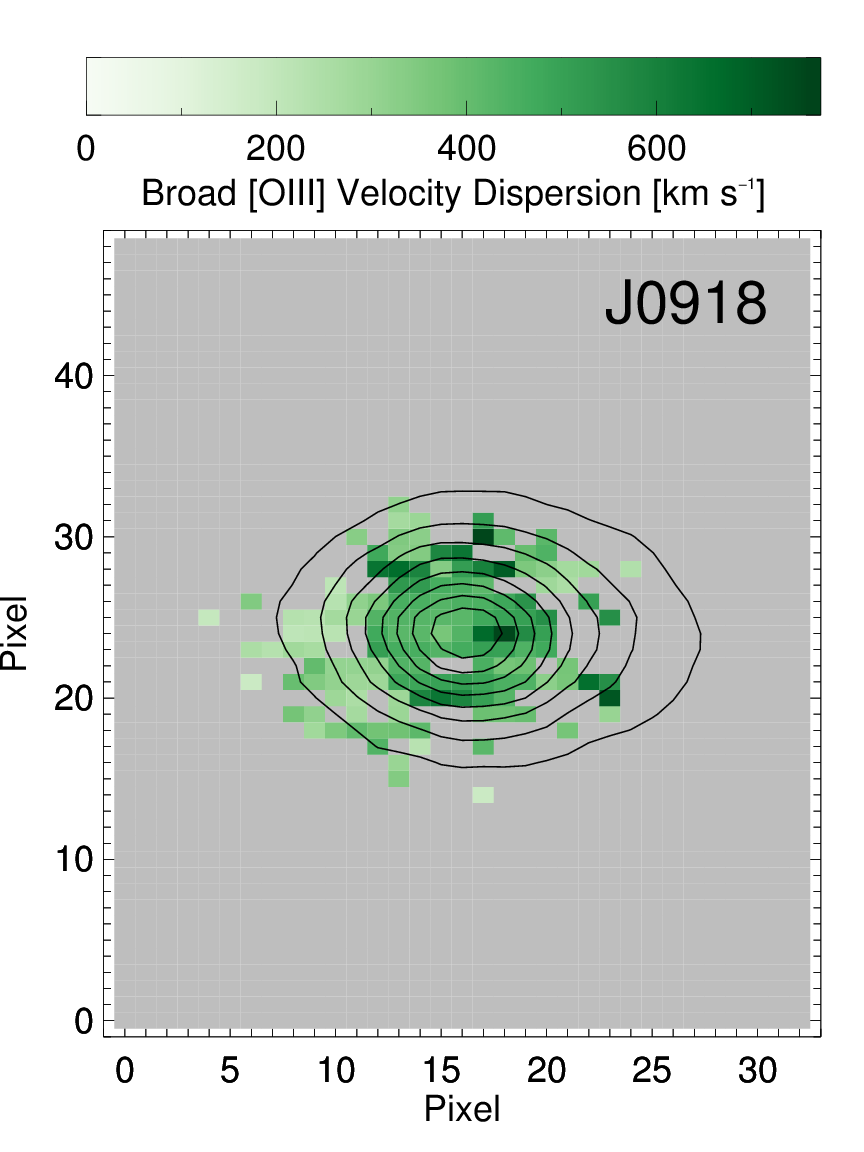}}
\raisebox{-0.5\height}{\includegraphics[width=0.16\textwidth,angle=0,trim={50 60 20 50},clip]{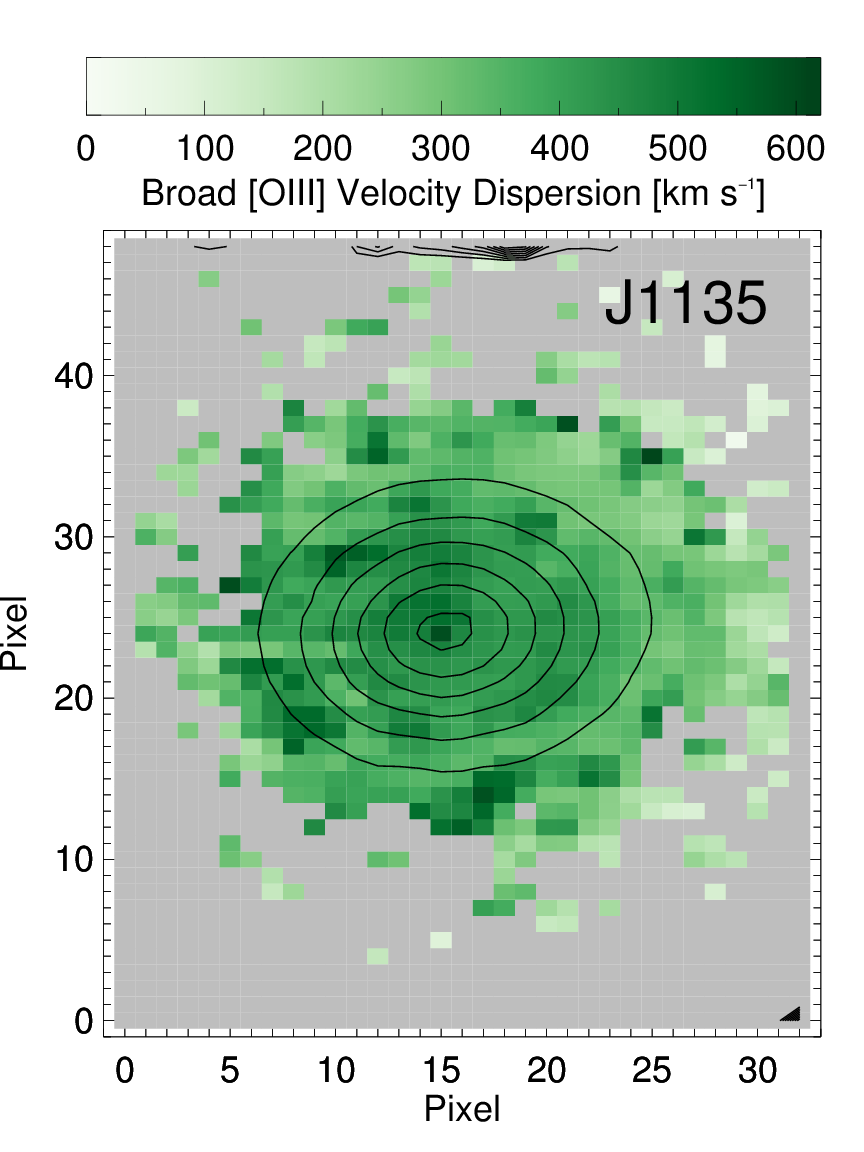}}
\raisebox{-0.5\height}{\includegraphics[width=0.16\textwidth,angle=0,trim={50 60 20 50},clip]{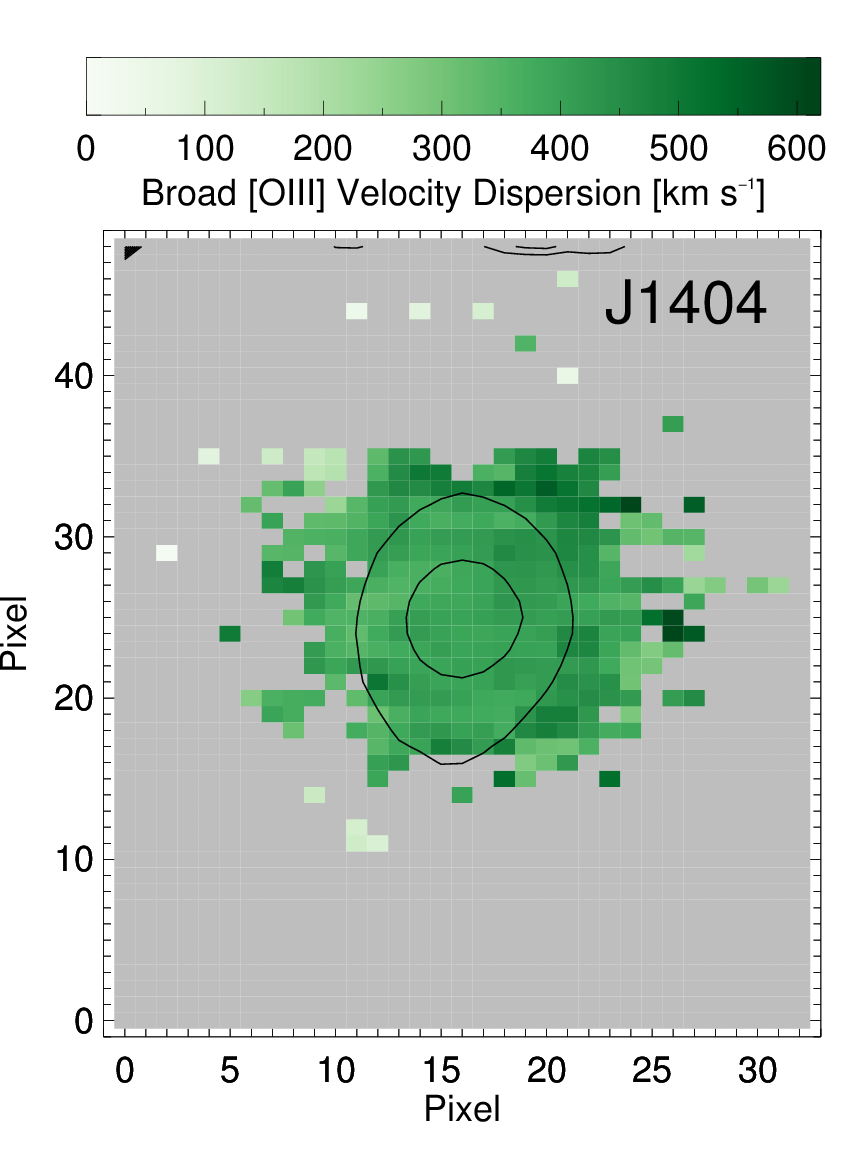}}
\raisebox{-0.5\height}{\includegraphics[width=0.16\textwidth,angle=0,trim={50 60 20 50},clip]{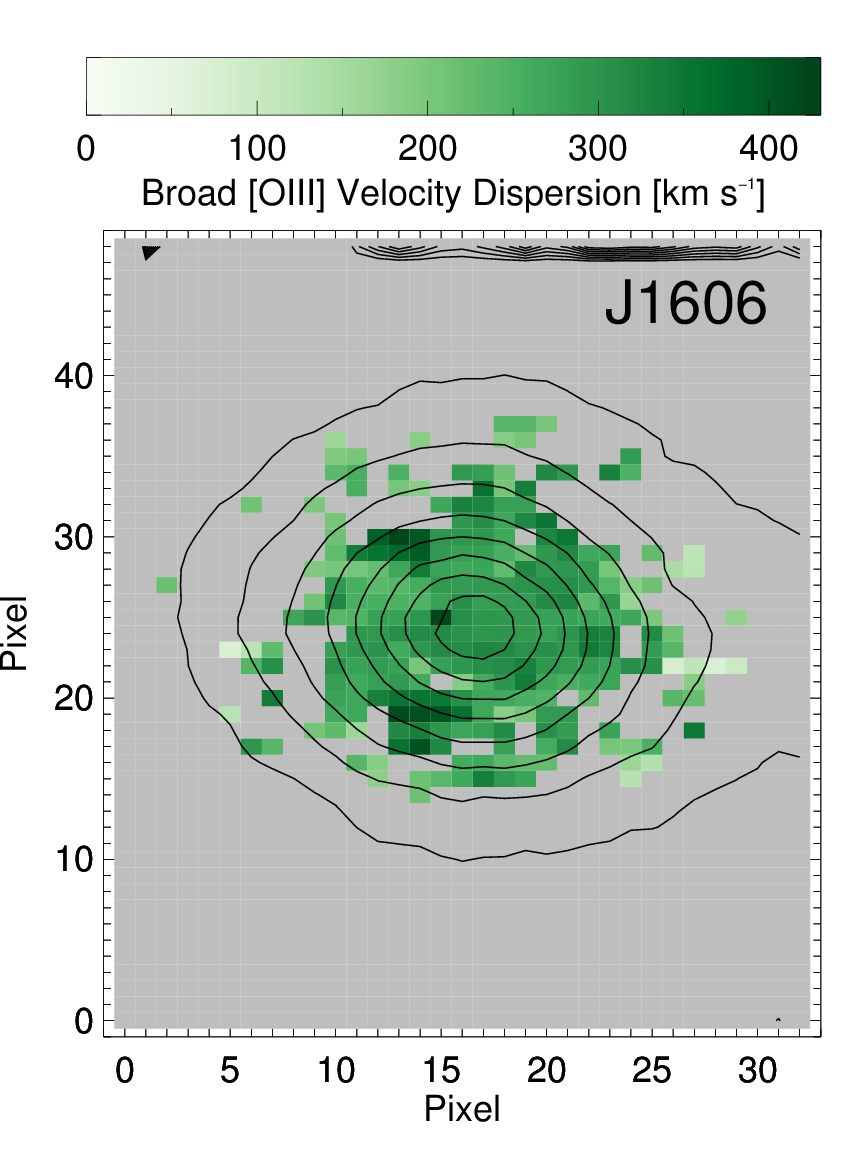}}
\raisebox{-0.5\height}{\includegraphics[width=0.16\textwidth,angle=0,trim={50 60 5 50},clip]{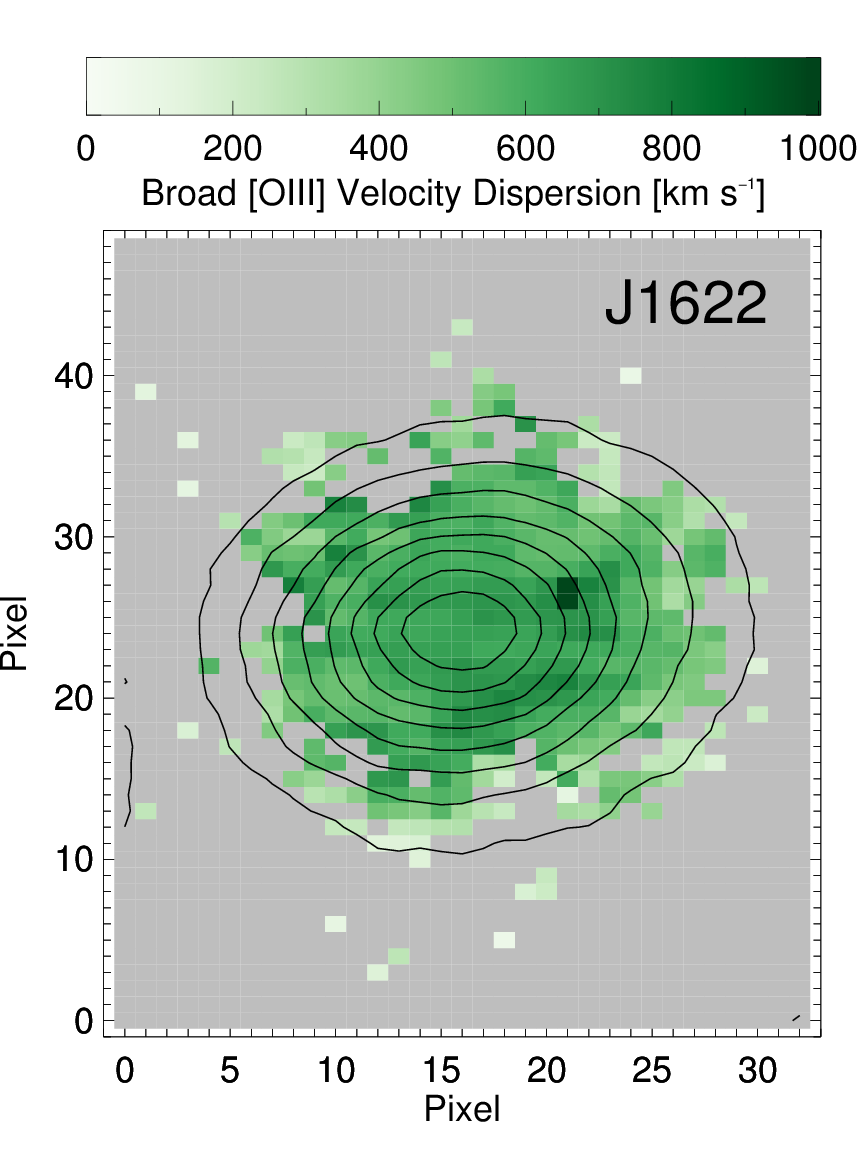}}
\raisebox{-0.5\height}{\includegraphics[width=0.16\textwidth,angle=0,trim={50 60 20 50},clip]{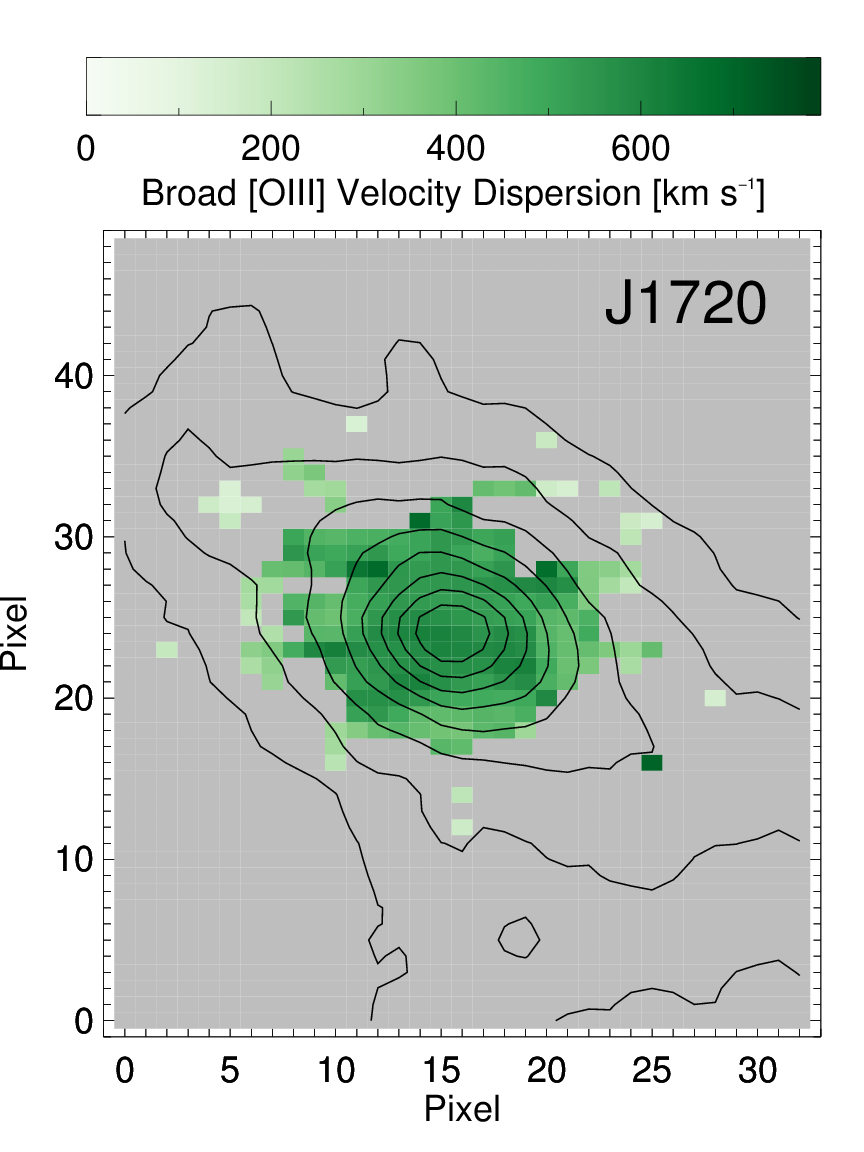}}
\caption{Velocity dispersion maps of the narrow H$\alpha$ and \mbox{[O\,\textsc{iii}]} (top) and the broad H$\alpha$ and \mbox{[O\,\textsc{iii}]} components (bottom). Contours are the same as in Fig. \ref{fig:all_flux}. Grey spaxels indicate non-detection of the emission lines (i.e., S/N$<$3).}
\label{fig:all_kins_sigma}
\end{center}
\end{figure*}

The spectrally decomposed broad and narrow components in \mbox{[O\,\textsc{iii}]} and H$\alpha$ emission lines are shown in Figs. \ref{fig:all_kins_dv} and \ref{fig:all_kins_sigma}. We first focus on the velocity of the broad and narrow components of H$\alpha$  and \mbox{[O\,\textsc{iii}]} (Fig. \ref{fig:all_kins_dv}). While the broad and narrow decomposition seems \emph{a priori} arbitrary, we show in the following that there does seem to be a physical distinction between the  two. 
The narrow H$\alpha$ and stellar velocity maps agree very well, both in terms of maximum velocities and the rotation structure. This implies that the narrow component of the H$\alpha$ emission represents the systemic velocity and follows the stellar rotation.

For the narrow \mbox{[O\,\textsc{iii}]}, four sources (J0918, J1135, J1404, and J1606) show mostly blueshifted (negative velocity) emission while 2 sources (J1622 and J1720) show redshifted emission. 
\mbox{[O\,\textsc{iii}]} seems to lack rotation signatures and appears to be kinematically decoupled from the H$\alpha$ emitting gas. For the cases where narrow \mbox{[O\,\textsc{iii}]} is blueshifted, the velocities appear mostly constant over the central part with some hints of increase with radial distances (see Section 4.4 for details). Velocity values nonetheless are consistent with the maximum absolute stellar velocities, ranging up to a few hundred km s$^{-1}$.

The velocity maps of the broad components show markedly different kinematics. For H$\alpha$ (third row of Fig. \ref{fig:all_kins_dv}), all sources show blueshifted emission at their centers with velocities that in some cases exceed the maximum stellar velocities. At outer parts, H$\alpha$ follows the stellar rotation or in some cases is undetected (i.e., the H$\alpha$ line can be fitted with a single Gaussian).
For two of the sources (J1135 and J1622) we observe elongated features near the AGN core with blueshifted emission. J1622 in addition shows a front of redshift broad \mbox{[O\,\textsc{iii}]} emission at the edge of the elongated feature of \mbox{[O\,\textsc{iii}]} blueshifted emission. The nature of this feature will be investigated further in the following paper.

The broad \mbox{[O\,\textsc{iii}]} emission (bottom row of Fig. \ref{fig:all_kins_dv}) shows the most extreme kinematics among the different components. Five sources show blueshifted broad \mbox{[O\,\textsc{iii}]} emission with velocities that are up to 3 times the maximum stellar velocities. One source (J1622) shows redshifted broad \mbox{[O\,\textsc{iii}]} emission with an implied positive gradient with radial distance, unlike the remaining 5 sources that generally show decreasing radial trends. We will investigate this radial gradients in detail in Section \ref{sec:radial}. In all cases the broad \mbox{[O\,\textsc{iii}]} velocity distribution appears very symmetric and concentrated around the AGN. The broad \mbox{[O\,\textsc{iii}]} and broad H$\alpha$ velocity maps show some degree of correlation (in terms of the velocity sign) but the broad H$\alpha$ is more extended and shows on average lower absolute velocities.

\subsubsection{Ionized gas velocity dispersion}

Next, we present the velocity dispersion maps for the H$\alpha$ and \mbox{[O\,\textsc{iii}]}, respectively for narrow and broad components in Fig. \ref{fig:all_kins_sigma}. The narrow components of both H$\alpha$ and \mbox{[O\,\textsc{iii}]} show $\sigma$ of up to a couple of hundreds km s$^{-1}$ consistent with $\sigma_{*}$. This corroborates our initial assertion that the narrow emission components are mostly dominated by the gravitational potential of the host galaxy rather than the non-gravitational influence of the AGN, as demonstrated statistically by \citet{Woo2015b}. 

Narrow H$\alpha$ shows on average smaller velocity dispersion values than narrow \mbox{[O\,\textsc{iii}]} with most sources showing unresolved  narrow H$\alpha$ emission ($\sigma_{\mathrm{H\alpha}}<\sigma_{\mathrm{instr}}$) beyond the galaxy center. Notable exceptions are J1135 and J1622, which show elongated features of elevated dispersion, tracing similar features seen in the velocity maps of Fig. \ref{fig:all_kins_dv}. The narrow \mbox{[O\,\textsc{iii}]} component on the other hand appears resolved for most spaxels with S/N$>3$ and shows consistently symmetric distribution. This may additionally indicate that the narrow \mbox{[O\,\textsc{iii}]} component is affected by the outflow. Thus, the \mbox{[O\,\textsc{iii}]} broad/narrow decomposition does not cleanly separate the two kinematic components.

The broad H$\alpha$ and \mbox{[O\,\textsc{iii}]} emission components are by definition broader than the narrow components and show values of up to 5 times the $\sigma_{*}$ derived from the SDSS spectra. Broad H$\alpha$ shows velocity dispersion values of up to 800 km s$^{-1}$. The velocity dispersion maps for broad H$\alpha$ also reveal interesting features of elevated dispersion: elongated features for J1135 and J1622 (Fig. \ref{fig:all_kins_dv}) and a ring-like structure for J1404. The broad \mbox{[O\,\textsc{iii}]} appears symmetric and concentrated. Velocity dispersion values for \mbox{[O\,\textsc{iii}]} are on average higher than H$\alpha$. 
The discrepancy between the homogeneity of the [OIII] emission kinematics and the kinematic structures seen in H$\alpha$ emission may be related to PSF smearing effects, given our seeing-limited observations ($\sim0\farcs5-0\farcs8$). Alternatively, these high velocity dispersion H$\alpha$ features may be unrelated to the AGN-driven outflow but appear co-spatial due to projection effects. 

\begin{figure*}[tbp]
\begin{center}
\includegraphics[width=0.32\textwidth,angle=0,trim={0 30 5 25},clip]{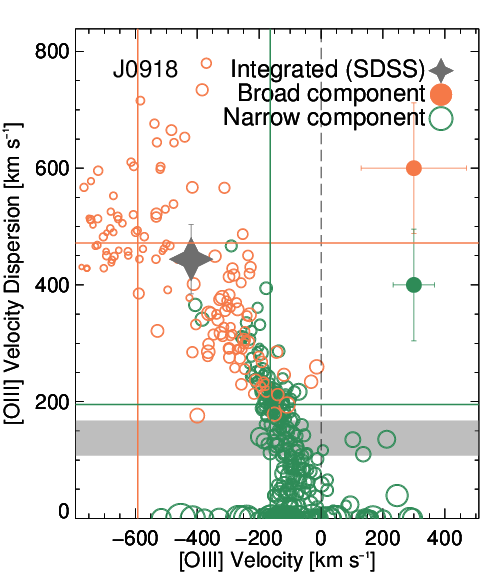}
\includegraphics[width=0.32\textwidth,angle=0,trim={0 30 5 25},clip]{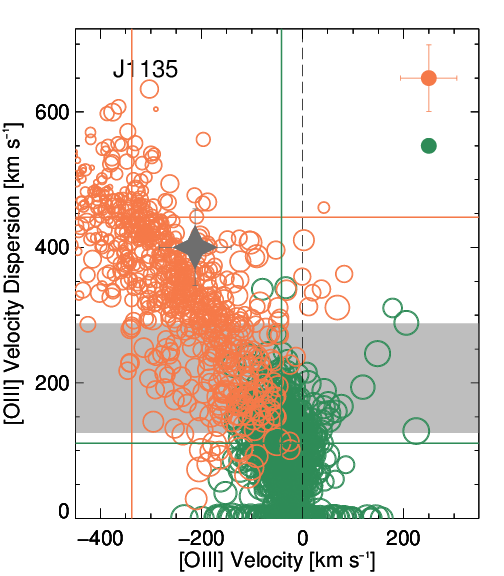}
\includegraphics[width=0.32\textwidth,angle=0,trim={0 30 5 25},clip]{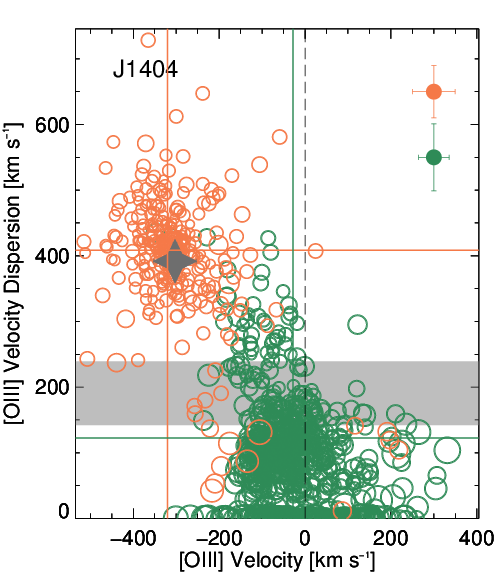}\\
\includegraphics[width=0.32\textwidth,angle=0,trim={0 0 5 25},clip]{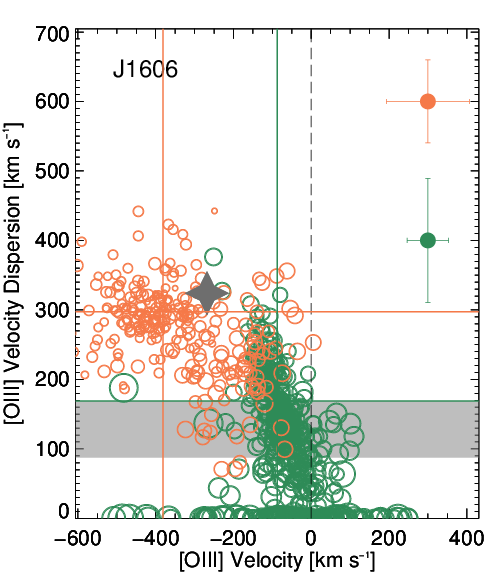}
\includegraphics[width=0.32\textwidth,angle=0,trim={0 0 5 25},clip]{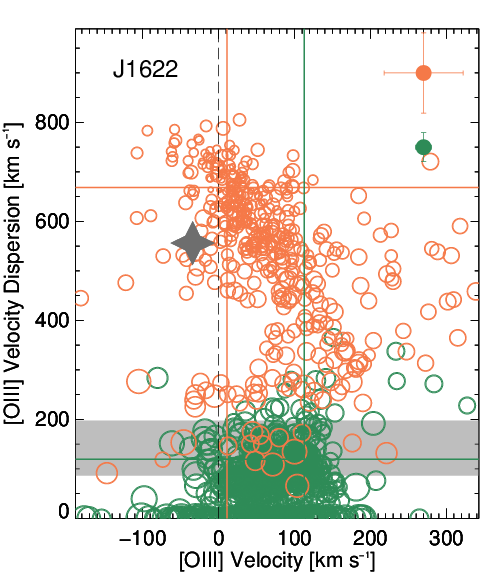}
\includegraphics[width=0.32\textwidth,angle=0,trim={0 0 5 25},clip]{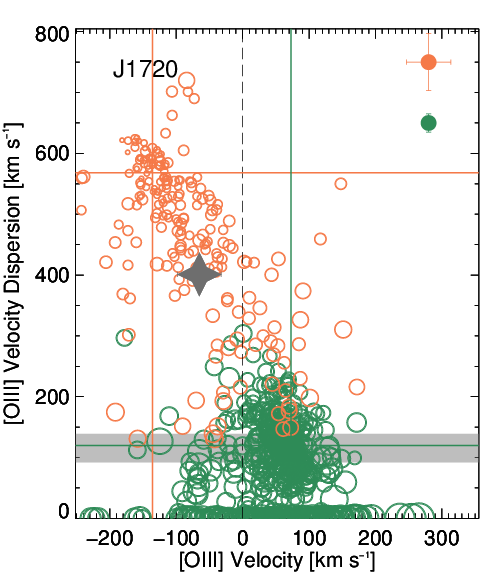}
\caption{VVD diagram of the broad (orange) and narrow (green) components of \mbox{[O\,\textsc{iii}]}, measured in individual spaxels,
along with the flux-weighted mean values (orange and green lines). The radial distance from the center (peak of the continuum) is denoted with increasing symbol sizes. For comparison, the kinematic measurements based on the SDSS spectra (large gray stars) and
the $\sigma_{*}$ value adopted from SDSS (gray-shaded area) are also overplotted while the vertical dashed line 
indicates the systemic velocity. 
The orange and green filled symbols at the top right corner show flux-weighted mean measurement uncertainties.}
\label{fig:all_vvd_oiii}
\end{center}
\end{figure*}

We note here that for J1404 the ring-like structure in broad H$\alpha$ emission corresponds to the morphology of the broad \mbox{[O\,\textsc{iii}]} emission, in that the former seems to spatially coincide with the edge of the latter. As we discuss in the Appendix, these high dispersion features may mark transitions between the outflow-dominated and star-formation dominated regions around the nucleus.

A first-order qualitative interpretation of the observed kinematics is that we are observing a superposition of two kinematically distinct components. The first one, corresponding to the narrow H$\alpha$ (and to lesser degree narrow \mbox{[O\,\textsc{iii}]}), is related to the gravitational potential of the galaxy as evidenced by the kinematic similarities with the stellar component. The second one, which is mainly traced by the broad \mbox{[O\,\textsc{iii}]}, shows extreme projected velocities and velocity dispersions (which should be indicative of the bulk motion of the ionized gas) and is concentrated around the AGN core.

{H$\beta$ shows similar kinematic behavior as H$\alpha$ and} \OIII{. A detailed analysis however is strongly limited by the low H$\beta$ fluxes, usually more than 10 times fainter than} \OIII\ {in the central parts of these AGN (this will be further explored in Paper II of this series). Nonetheless, for the central most regions we find that H$\beta$ is asymmetric (see Fig. \ref{fig:examplefit}) and requires a combination of a narrow and a broad component to be fit. Like H$\alpha$, H$\beta$ shows on average smaller velocity and velocity dispersion than} \OIII.

{We note that for both the velocity and velocity dispersion maps, "hot" spaxels exist where the spectral fit of one or both emission lines fail. This can be due to any combination of several confounding factors including low S/N, peculiar spectral shape, and problematic continuum fitting and subtraction. This leads to spaxels, most often in the outer part of the IFU, with extreme velocity and velocity dispersion values. These "hot" spaxels are usually isolated and have typically larger (factor of 2 to 3) uncertainties than their neighboring spaxels. They should not be confused with spatial conglomerations of several ($>10$) spaxels that show consistent deviations from the overall kinematic distribution (e.g., the high velocity, high velocity dispersion feature seen in J1622). The latter are signatures of areas of interest that may trace outflow-driven kinematics, shocked regions, or otherwise regions with different physical conditions.}

\subsection{Velocity-velocity dispersion diagram}

In this section we focus on the velocity versus velocity dispersion (VVD) diagram to investigate the kinematics of the ionized gas with respect to the stellar kinematics and the radial distance to the AGN. Figure \ref{fig:all_vvd_oiii} presents the VVD diagram for the broad and narrow components of \mbox{[O\,\textsc{iii}]} (shown respectively with orange and green symbols), with a clear distinction between them. The broad \mbox{[O\,\textsc{iii}]} components are predominantly found in the upper left corner of the VVD diagram, showing large velocity dispersion and significant offsets from the systemic velocity. In contrast, the narrow \mbox{[O\,\textsc{iii}]} components are distributed fairly evenly around the systemic velocity and straddle the $\sigma_{*}$ range.

We find a negative radial trend of velocity and velocity dispersion for the broad \mbox{[O\,\textsc{iii}]},
while no radial dependence of velocity and velocity dispersion is seen for the narrow component\footnote{Radial distance is denoted by the size of the symbols in Fig. \ref{fig:all_vvd_oiii}.}.
The broad \mbox{[O\,\textsc{iii}]} component shows a significant range of both velocities and dispersions
while at large radial distances these values overlap with those of the narrow \mbox{[O\,\textsc{iii}]} component. 
The negative radial trend of the broad \mbox{[O\,\textsc{iii}]} component indicates non-gravitational effects such as
outflows, whose velocity decreases radially. Thus, velocity and velocity dispersion values eventually 
become comparable to the systemic velocity and $\sigma_{*}$, respectively.
In contrast, the narrow \mbox{[O\,\textsc{iii}]} component in the outer spaxels of the FoV shows a large range of velocities that presumably reflects the rotation. This is supported by the relatively low velocity dispersions of the narrow \mbox{[O\,\textsc{iii}]} component in these spaxels.

\begin{figure*}[hbpt]
\begin{center}
\includegraphics[width=0.33\textwidth,angle=0]{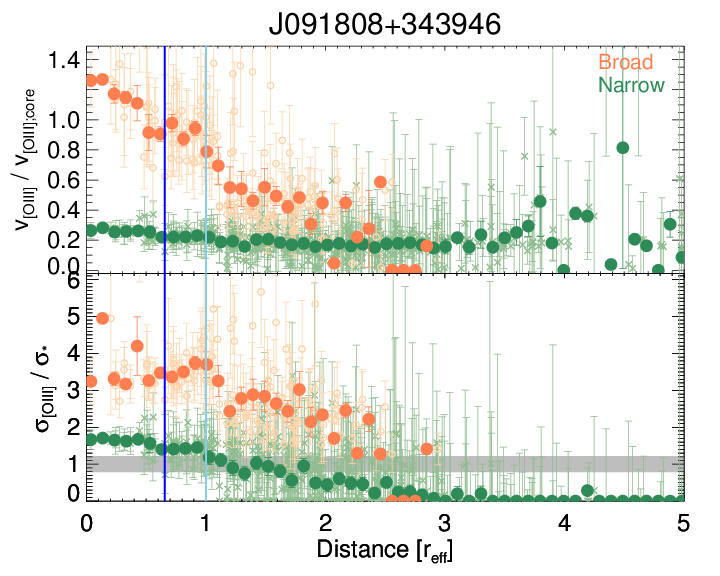}
\includegraphics[width=0.33\textwidth,angle=0]{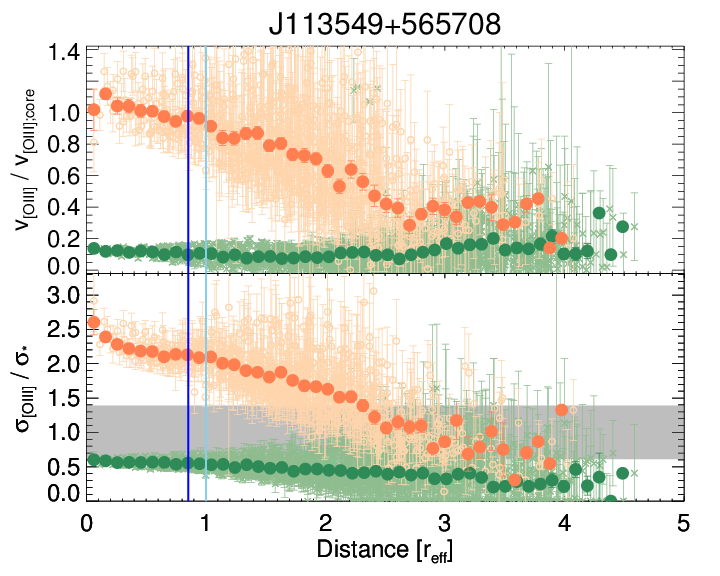}
\includegraphics[width=0.33\textwidth,angle=0]{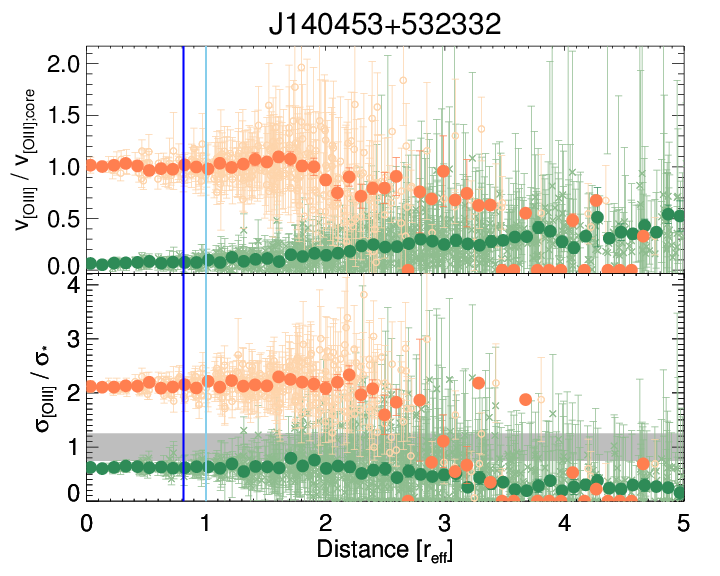}
\includegraphics[width=0.33\textwidth,angle=0]{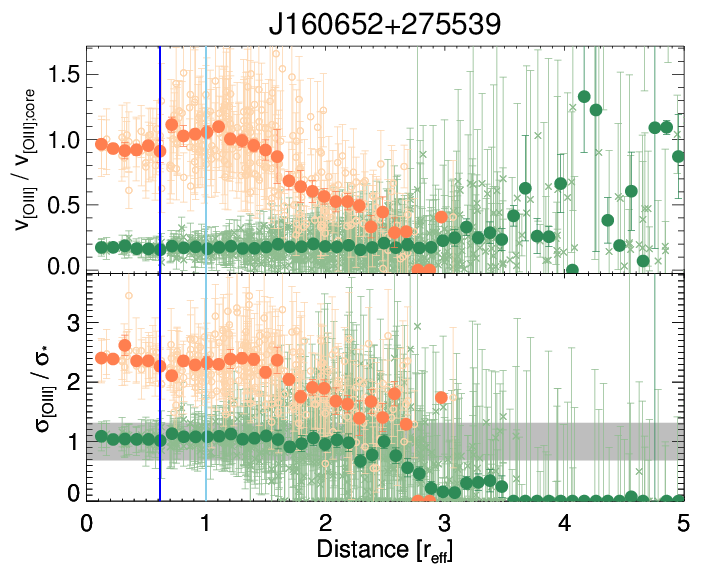}
\includegraphics[width=0.33\textwidth,angle=0]{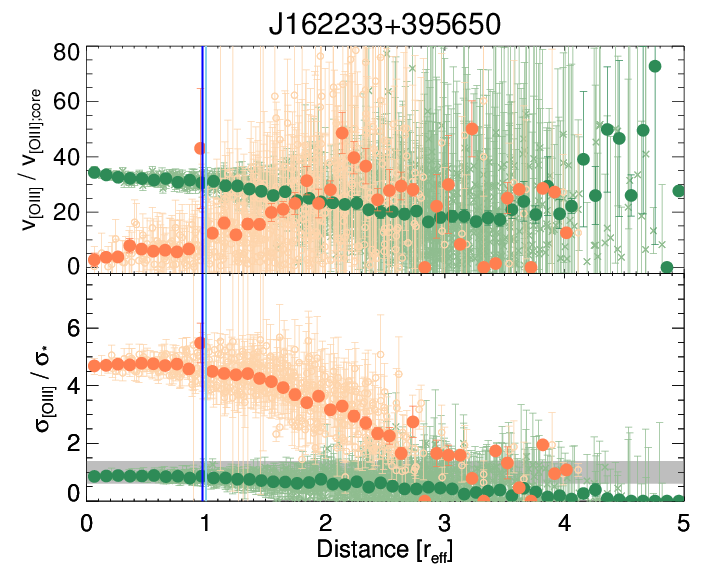}
\includegraphics[width=0.33\textwidth,angle=0]{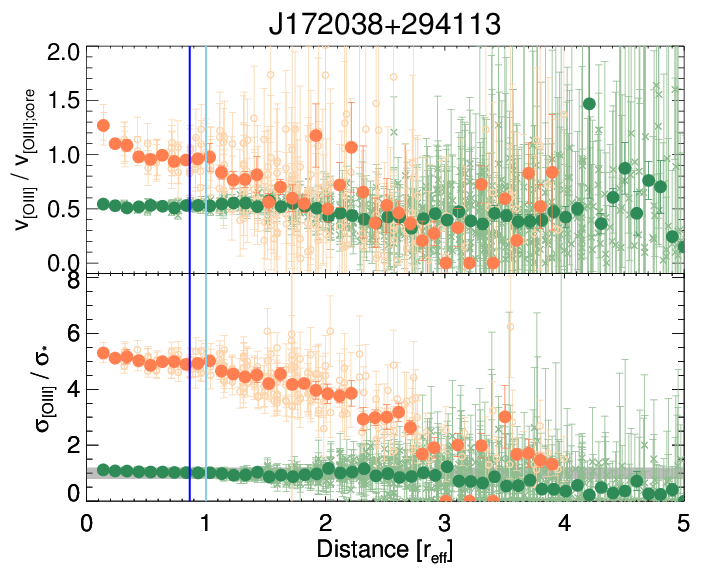}
\caption{Radial profiles of the normalized (see text for details) absolute \mbox{[O\,\textsc{iii}]} velocity (upper panel) and velocity dispersion (lower panels) of each object. The radial distance is expressed in units of r$_{\mathrm{eff}}$, calculated from the radial flux profile of the broad \mbox{[O\,\textsc{iii}]} component. The profiles of broad (orange circles) and narrow (green x's) components are shown separately,
along with the mean values and their standard errors as a function of distance (large darker symbols). Gray-shaded stripes in the lower panels denote a normalized dispersion of one ($\sigma_{\mathrm{\mbox{[O\,\textsc{iii}]}}}=\sigma_{*}$) with the stripe width equal to six times the $\sigma_{*}$ measurement uncertainty ($\pm 3\delta\sigma_{*}$), based on SDSS.}
\label{fig:radial_prof}
\end{center}
\end{figure*}

J1622 is the only source in our sample that shows consistently redshifted emission for both broad and narrow \mbox{[O\,\textsc{iii}]} components (Fig. \ref{fig:all_kins_dv}). Figure \ref{fig:all_vvd_oiii} shows that most broad components (orange symbols) have positive 
velocity (right of the vertical dashed line). However, the velocities of the broad components are consistent with those of the narrow components,
implying that the kinematics may be dominantly affected by the gravitational potential. 

In Fig. \ref{fig:all_vvd_oiii}, we also plot the velocity and velocity dispersion of \mbox{[O\,\textsc{iii}]} (large gray stars) measured from the spatially integrated SDSS. Compared to the SDSS values, we find that while the velocity dispersion of the broad \mbox{[O\,\textsc{iii}]} component is mostly reflected in the spatially integrated SDSS \mbox{[O\,\textsc{iii}]} profiles, the velocity is often underestimated, especially for the most extreme kinematics of the central part of the AGN by up to 100\%. These differences become clear by comparison with the flux-weighted mean velocity and velocity dispersion of the broad \mbox{[O\,\textsc{iii}]} components (shown with vertical and horizontal lines in Fig. \ref{fig:all_vvd_oiii}).

In the case of H$\alpha$, the VVD diagrams reveal similar patterns although with less extreme kinematics as was already shown in Figs. \ref{fig:all_kins_dv} and \ref{fig:all_kins_sigma}. Thus, we do not present the detailed plots. In brief, the broad H$\alpha$ component is found less offset from the narrow component, with the latter showing very clear bell-shaped rotational patterns in the VVD diagram such that the central spaxels show high velocity dispersion but velocities consistent with the systemic velocity at the top of the bell and outer spaxels at the bottom of the bell having low velocity dispersion and high blueshifts and redshifts, reflecting rotation.

In summary, the VVD diagram for both \mbox{[O\,\textsc{iii}]} and H$\alpha$ shows a clearly asymmetric distribution in velocities, indicating that both broad components do not follow the rotation due to a non-gravitational effect. This strongly supports a physical distinction between narrow and broad components, the latter reflecting the kinematics of outflows.

\subsection{Radial profiles of the outflow}
\label{sec:radial}

We present the radial profiles of the outflows, as these can help us estimate the projected size of the outflow-dominated region. This is one of the fundamental parameters for calculating the mass and energy of the gas entrained in the outflow. In Fig. \ref{fig:radial_prof} we show the normalized absolute velocities and velocity dispersions as a function of the effective radius, r$_{\mathrm{eff}}$, the radius within which half of the total broad \mbox{[O\,\textsc{iii}]} emission flux is contained. This definition is motivated by our finding that the broad \mbox{[O\,\textsc{iii}]} component exhibits the most extreme kinematics in the spatially resolved maps and does not suffer from gravitational effects (e.g., stellar rotation).
Velocities in Fig. \ref{fig:radial_prof} are normalized by the median velocity of the broad \mbox{[O\,\textsc{iii}]} component within one r$_{\mathrm{eff}}$ from the nucleus in order to show the relative spatial gradient. Velocity dispersions are normalized by $\sigma_{*}$ to present the non-gravitational effect. Both these normalizations allow us to identify the linear scales within which \mbox{[O\,\textsc{iii}]} shows "extreme" kinematics with respect to the gravitational potential of the host galaxy.

\begin{figure}[t]
\begin{center}
\includegraphics[width=0.45\textwidth,angle=0,trim={30 30 5 25},clip]{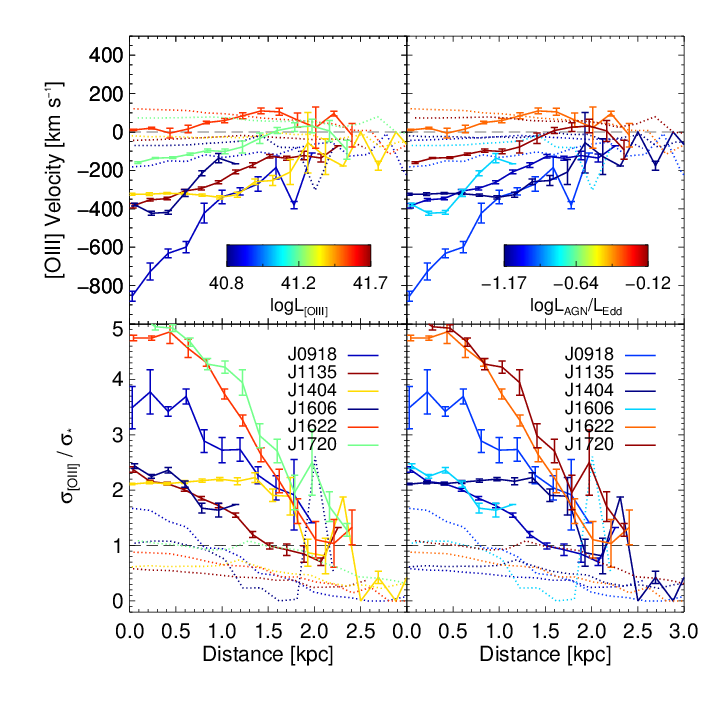}
\caption{Radial profiles of the mean velocity (top) and the mean velocity dispersion normalized by $\sigma_{*}$ (bottom) of the
broad (solid line) and narrow components of \mbox{[O\,\textsc{iii}]} (dotted line), respectively.
The color denotes the dust-uncorrected \mbox{[O\,\textsc{iii}]} luminosity (left panels) or Eddington ratio (right panels). 
Plotted uncertainties are the standard error of the mean.}
\label{fig:radial_prof_OIII}
\end{center}
\end{figure}

First, we observe a distinction between the kinematics of the broad and narrow components of \mbox{[O\,\textsc{iii}]}. Both the velocity and velocity dispersion of the broad \mbox{[O\,\textsc{iii}]} component decrease with radial distance (except for J1622), while those of the narrow \mbox{[O\,\textsc{iii}]} component show an almost flat distribution. The velocity and velocity dispersion of the broad \mbox{[O\,\textsc{iii}]} component converge to the values of the narrow \mbox{[O\,\textsc{iii}]} component at $\sim2-3r_{\mathrm{eff}}$.
Second, the extreme kinematics of the broad \mbox{[O\,\textsc{iii}]} component are prevalent within radial distances of $\sim2-3r_{\mathrm{eff}}$, beyond which the decomposition and differentiation of the two components becomes challenging due to low S/N. Third, there is an initial {plateau} of the normalized velocity dispersion for the broad \mbox{[O\,\textsc{iii}]} component, which is then followed by a fall-off starting at around $\sim1.5-2r_{\mathrm{eff}}$. 
Since we expect that the velocity dispersion of the stars and gas caused by the gravitational potential decreases with radial distance, the fact that we observe {roughly constant velocity dispersion} out to $\sim2$ r$_{\mathrm{eff}}$ implies that the outflow is actually experiencing an initial acceleration before decelerating and eventually converging to stellar kinematics (e.g., \citealt{Crenshaw2000,Fischer2013}). 
It is interesting to note that while the velocity dispersion values are different among the sample, all of them show similar radial profiles 
as a function of their r$_{\mathrm{eff}}$ as well as the decrease of the outflows to the gravitational kinematics. 

Lastly, neither the velocity nor velocity dispersion show an increasing trend with \mbox{[O\,\textsc{iii}]} luminosity.
For example, J1606 (lowest \mbox{[O\,\textsc{iii}]} luminosity in our sample) shows larger normalized velocity dispersion than J1135 (highest \mbox{[O\,\textsc{iii}]} luminosity). 
This implies that the absolute scale of the observed kinematics is mostly influenced by the inclination of the outflow to our line of sight, as well as the outflow opening angle.

In Fig. \ref{fig:radial_prof_OIII} we investigate the mean radial trend of the velocity and normalized velocity dispersion of \mbox{[O\,\textsc{iii}]} with respect to the
AGN luminosity and Eddington ratio. We use black hole mass estimates based on SDSS $\sigma_{*}$ and AGN bolometric luminosities based on the dust-uncorrected [OIII] luminosity to derive the Eddington ratios (see \citealt{Bae2014} for details). Given the significant scatter of the M$_{\mathrm{BH}}$-$\sigma_{*}$ relation and the uncertainties involved in the bolometric correction, these Eddington ratios are highly uncertain. 

A mean zero velocity in Fig. \ref{fig:radial_prof_OIII} can be a result of the canceling out of blueshifted and redshifted emission of rotating gas. On the other hand, mean negative velocities imply the presence of a non-gravitational component. Similarly, normalized velocity dispersion values significantly above one (above the gray-shaded area) indicate a non-gravitational kinematic component. Conversely, mean normalized velocity dispersions around or below one are indicative of gas following the gravitational potential of the galaxy. The $\sigma_{*}$ used for the normalization comes from the SDSS and therefore reflects the flux-weighted stellar kinematics of the nuclear stellar component. This can lead to normalized gas velocity dispersions below one, especially for the outer parts of the FoV.

We see no obvious trend between L$_{\mbox{[O\,\textsc{iii}]}}$ and the \mbox{[O\,\textsc{iii}]} kinematics, with the caveat that the range of the \mbox{[O\,\textsc{iii}]} luminosity is narrow ($\sim1$ dex). In contrast, we detect a weak trend that more extreme kinematics is observed in higher Eddington ratio 
objects. Similarly, there is some indication that sources with higher Eddington ratio also show more extended regions, with extreme \mbox{[O\,\textsc{iii}]} velocity dispersion (i.e., $\sigma_{\mathrm{\mbox{[O\,\textsc{iii}]}}}$ $>$ $\sigma_*$). 
The correlation of the [OIII] velocity dispersion with the AGN Eddington ratio is in agreement with previous studies (e.g., \citealt{Woo2015b}) and indicates that the velocity dispersion is more representative of the bulk motion of outflows as it is not affected by projection effects (e.g., \citealt{Liu2013,Harrison2014}). Nonetheless, we note that both the inclination angle and the opening angle of the outflows can be different among the 6 objects, which can dilute any intrinsic correlations with the \mbox{[O\,\textsc{iii}]} luminosity and the Eddington ratio.

In the case of H$\alpha$, the velocity and velocity dispersion of the broad component also shows a negative gradient with radial distance. We find that the decrease is more rapid, with the velocity values converging to stellar kinematics within 1-2r$_{\mathrm{eff}}^{H\alpha}$. 
We obtain the relation r$_{\mathrm{eff}}^{H\alpha}=(1.7\pm0.3)\times$r$_{\mathrm{eff}}^{\mbox{[O\,\textsc{iii}]}}$, reflecting the significantly more spatially extended nature of H$\alpha$ emission. However, the linear size of the region within which broad [OIII] and broad H$\alpha$ show non-gravitational kinematics (1-2r$_{\mathrm{eff}}^{H\alpha}$ and 2-3r$_{\mathrm{eff}}^{\mbox{[O\,\textsc{iii}]}}$, respectively) are consistent with each other. We caution that broad H$\alpha$ can also originate in star forming regions (as seen for J1720, for which broad H$\alpha$ is detected at the base of its spiral arms) and therefore we cannot rely solely on the broad H$\alpha$ flux distribution to derive the properties of the outflow. 

\begin{figure}[t]
\begin{center}
\includegraphics[width=0.45\textwidth,angle=0,trim={30 30 5 25},clip]{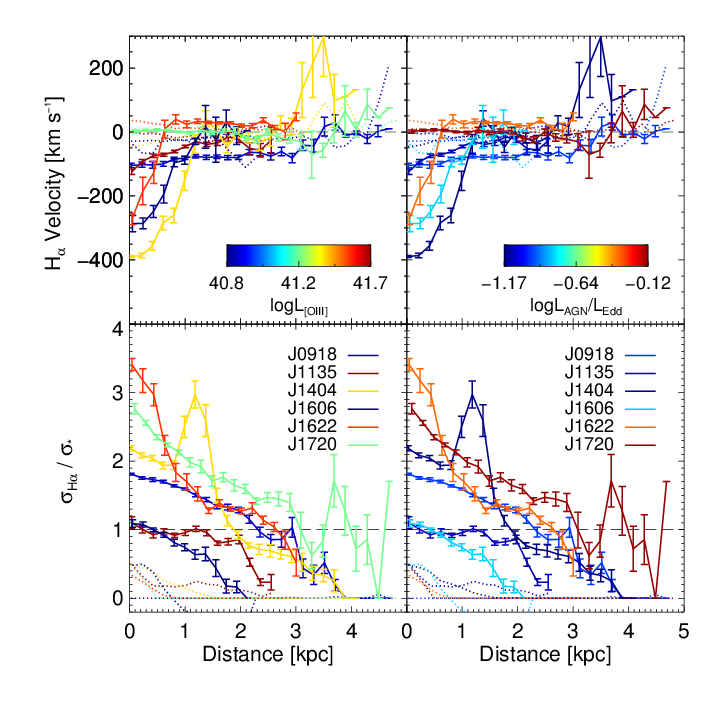}
\caption{Radial profiles of the mean velocity (top) and the mean velocity dispersion normalized by $\sigma_{*}$ (bottom) of the
broad (solid line) and narrow components of H$\alpha$ (dotted line), respectively.
The color denotes the dust-uncorrected \mbox{[O\,\textsc{iii}]} luminosity (left panels) or Eddington ratio (right panels). 
Plotted uncertainties are the standard error of the mean.}
\label{fig:radial_prof_Ha}
\end{center}
\end{figure}

In Fig. \ref{fig:radial_prof_Ha} we show the averaged kinematics radial profiles for H$\alpha$ broad and narrow components, similar to Fig. \ref{fig:radial_prof_OIII}. Qualitatively, we see the same negative trends with radial distance but with key differences compared to the \mbox{[O\,\textsc{iii}]} radial profiles. First, both the absolute velocities and normalized velocity dispersions reach much smaller values than \mbox{[O\,\textsc{iii}]}. Second, both velocity and velocity dispersion profiles show different shapes below and above $\sim1.5$ kpc. Since both narrow and part of the broad H$\alpha$ components follow the stellar rotation, this leads to a dilution of the nuclear outflow component (seen in the central kpc of the H$\alpha$ profiles). Beyond $\sim1.5$ kpc, the velocity profiles are consistent with zero velocity (as discussed above), while the velocity dispersion profiles either show flatter slopes (e.g., J1135, J1622, J1729) or even an inverse gradient for the case of J1404. Finally, unlike for Fig. \ref{fig:radial_prof_OIII}, we find no appreciable trends with either \mbox{[O\,\textsc{iii}]} luminosity or Eddington ratio. This reinforces are previous assessment that even the broad component of the H$\alpha$ emission is a poor tracer of these outflows.

\section{Physical properties of the AGN outflows}
\label{sec:physical}

\subsection{Outflow sizes}
The effective radius r$_{\mathrm{eff}}$ measured from the broad \mbox{[O\,\textsc{iii}]} component ranges from 0.4 to 0.7 kpc, while that of the narrow \mbox{[O\,\textsc{iii}]} component is up to 1.2 kpc. The broad \mbox{[O\,\textsc{iii}]} component is marginally resolved (FWHM$_{\mathrm{\mbox{[O\,\textsc{iii}]};b}}\lesssim$ seeing), with the exception of J1622. A direct comparison of the r$_{\mathrm{eff}}$ of either the broad or narrow \mbox{[O\,\textsc{iii}]} components with both L$_{\mbox{[O\,\textsc{iii}]}}$ and Eddington ratio does not show significant correlation within the limited dynamic range. 

In the case of H$\alpha$, the emission region is more extended with r$_{\mathrm{eff}}$ between 1.8 and 2.1 kpc.  
The ionized emission of H$\alpha$ is detected out to the edges of the IFU FoV,  while the \mbox{[O\,\textsc{iii}]} emission is more concentrated but still significantly detected out to $\sim3-4\arcsec$, which corresponds to a linear scale of $\sim7.4$ kpc, for the highest redshift source in the sample. 
However, we emphasize that if the maximum size of the emission line detection is taken as the size of the outflow region, one would significantly overestimate the volume of the ionized gas that is affected by the AGN-driven outflow. 
This is clearly seen in Fig. \ref{fig:radial_prof_OIII}, where at distances $>2$ kpc, the gas kinematics is consistent with the gravitational stellar kinematics. We therefore argue that a more representative outflow size can be calculated by taking into account both the kinematics and the flux distribution of broad \mbox{[O\,\textsc{iii}]}. Based on Fig. \ref{fig:radial_prof_OIII}, for each source we define the region within which the broad [OIII] kinematics show evidence for non-gravitational motions (i.e., absolute velocity $>0$ and normalized velocity dispersion $>1$). This "kinematic" size is found to be between 1.3 to 2.1 kpc (or 2 to 3 times r$_{\mathrm{eff}}^{\mbox{[O\,\textsc{iii}]}}$).
Alternatively, one can calculate the flux-weighted radius of the narrow line emitting gas (e.g., \citealt{Husemann2013,Husemann2014}), which however
does not take into account the difference between non-gravitational and gravitational kinematic components. Hence, the flux-weighted size can only be considered as an upper limit. 

\begin{figure}[bpt]
\begin{center}
\includegraphics[width=0.45\textwidth,angle=0,trim={15 50 30 60},clip]{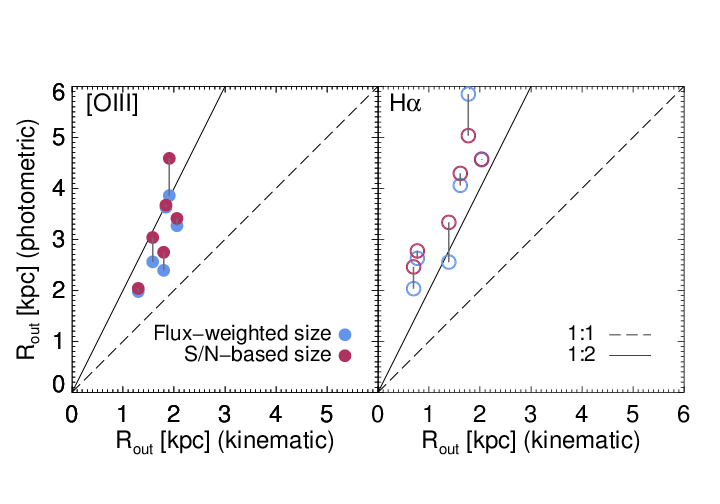}
\caption{Comparison of the kinematically measured size with the photometrically measured size of the \mbox{[O\,\textsc{iii}]} emission (left) 
and H$\alpha$ (right). We plot flux-weighted sizes (blue) and S/N-based sizes (magenta). 
The dashed and solid lines indicate the 1:1  and 1:2 ratio between them, respectively. 
The kinematically measured sizes are based on the broad components, while the photometrically measured sizes are based on the  
total emission line profiles.}
\label{fig:size_comp}
\end{center}
\end{figure}

In Fig. \ref{fig:size_comp} we compare the measured outflow sizes using the three different methods discussed: (1) based on the kinematics of the broad \mbox{[O\,\textsc{iii}]} and H$\alpha$ emitting gas 
(2) based on the flux-weighted distance (following \citealt{Husemann2013}, and (3) based on the \mbox{[O\,\textsc{iii}]} detection with S/N $>$ 5. 
Note that for this exercise we exclude spaxels classified as HII region on their emission line flux ratios, but include composite and LINER regions (more detailed analysis will be presented in the following paper).
Based on the flux-weighted calculation, we obtain the size of the broad \mbox{[O\,\textsc{iii}]} as 1.5--3 kpc, and the size of 
the total \mbox{[O\,\textsc{iii}]} emission line or the total H$\alpha$ emission line as up to 3.7 and 4.1 kpc, respectively, significantly larger than the kinematically estimated size. 
We find that both photometric methods overestimate the size of the region within which non-gravitational outflow kinematics is observed by up to a factor of 2 for \mbox{[O\,\textsc{iii}]}.
On face value, this in effect leads to an overestimation of the outflow lifetimes (which are usually calculated as the size of the outflow divided by its bulk velocity) and consequently to an underestimation of the energy and mass outflow rates. We will explore this effect more in detail in the second paper of this series.
In Table \ref{tab:properties} we present the flux-weighted mean properties within the  r$_{\mathrm{eff}}$, of each target. 

\begin{deluxetable*}{c c c c c c c c c c c}
\tabletypesize{\footnotesize}
\tablecolumns{12}
\tablewidth{0pt}
\tablecaption{Measurement of basic properties of the \mbox{[O\,\textsc{iii}]} and H$\alpha$ emission regions. \label{tab:properties}}
\tablehead{\colhead{ID}	&	\colhead{$r_{\mathrm{eff}}$}	&	\colhead{Comp}	&	\colhead{L$_{\mbox{[O\,\textsc{iii}]}}$}	&	\colhead{v$_{\mbox{[O\,\textsc{iii}]}}$}	&	\colhead{$\sigma_{\mbox{[O\,\textsc{iii}]}}$}	&	\colhead{L$_{H\alpha}$}	&	\colhead{v$_{H\alpha}$}	&	\colhead{$\sigma_{H\alpha}$}	&	\colhead{\mbox{[O\,\textsc{iii}]}}	& \colhead{H$\alpha$}	\\
 \colhead{ } & \colhead{[kpc]} & 	\colhead{ } & \colhead{[erg s$^{-1}$]} & \multicolumn{2}{c}{[km s$^{-1}$]} & \colhead{[erg s$^{-1}$]} & \multicolumn{2}{c}{[km s$^{-1}$]} & \multicolumn{2}{c}{Kinematics} }
\startdata
\multirow{3}{*}{J0918}     & \multirow{3}{*}{     0.74    } &     T &    40.08     &     -380    $\pm$      50     &      436    $\pm$      66     &     40.77     &      -53    $\pm$      23     &      146    $\pm$      22    &  \\
 &  &     B &    39.69     &     -673    $\pm$     130     &      499    $\pm$     119     &     40.38     &      -90    $\pm$      43     &      229    $\pm$      32    &  Blue & Blue\\
 &  &     N &    39.85     &     -178    $\pm$      30     &      229    $\pm$      39     &     38.44     &      -27    $\pm$      14     &        0    $\pm$       0    & Blue & I \\
\multirow{3}{*}{J1135}     & \multirow{3}{*}{     0.60    } &     T &    41.56     &     -162    $\pm$      20     &      328    $\pm$      28     &     41.69     &      -57    $\pm$      87     &      185    $\pm$      84    &  \\
 &  &     B &    41.13     &     -354    $\pm$      50     &      458    $\pm$      40     &     41.47     &      -85    $\pm$     115     &      200    $\pm$     120    &  Blue & Blue\\
 &  &     N &    41.35     &      -42    $\pm$      10     &      114    $\pm$      14     &     39.85     &      -10    $\pm$     148     &       80    $\pm$      44    &  Blue & Rot\\
\multirow{3}{*}{J1404}     & \multirow{3}{*}{     0.69    } &     T &    40.81     &     -229    $\pm$      10     &      368    $\pm$      13     &     41.03     &     -154    $\pm$     147     &      376    $\pm$     207    &  \\
 &  &     B &    40.66     &     -319    $\pm$      20     &      404    $\pm$      21     &     40.75     &     -247    $\pm$     250     &      474    $\pm$     243    & Blue & Blue\\
 &  &     N &    40.30     &      -24    $\pm$      10     &      121    $\pm$      15     &     39.14     &      -31    $\pm$      98     &       46    $\pm$      31    &  Blue/Rot & Rot\\
\multirow{3}{*}{J1606}     & \multirow{3}{*}{     0.43    } &     T &    40.68     &     -205    $\pm$      20     &      271    $\pm$      18     &     41.37     &     -148    $\pm$     326     &     1139    $\pm$     680    &  \\
 &  &     B &    40.30     &     -379    $\pm$      70     &      303    $\pm$      33     &     40.55     &      -93    $\pm$     250     &       68    $\pm$      68    &  Blue & Blue\\
 &  &     N &    40.45     &      -73    $\pm$      20     &      140    $\pm$      15     &     39.37     &      -42    $\pm$     176     &       51    $\pm$      53    &  Blue & Rot\\
\multirow{3}{*}{J1622}     & \multirow{3}{*}{     0.63    } &     T &    41.13     &       54    $\pm$      30     &      519    $\pm$      68     &     41.59     &        0    $\pm$      53     &      202    $\pm$     105    &  \\
 &  &     B &    40.90     &        4    $\pm$     130     &      683    $\pm$     157     &     41.03     &      -37    $\pm$     136     &      353    $\pm$     153    &  Blue/I	&	Red\\
 &  &     N &    40.74     &      118    $\pm$      10     &      124    $\pm$      13     &     42.44     &       18    $\pm$      53     &       13    $\pm$      12    & Red & Rot\\
\multirow{3}{*}{J1720}     & \multirow{3}{*}{     0.64    } &     T &    40.49     &      -38    $\pm$      10     &      430    $\pm$      19     &     40.48     &        4    $\pm$      23     &      215    $\pm$      62    &  \\
 &  &     B &    40.20     &     -144    $\pm$      20     &      582    $\pm$      29     &     40.20     &        4    $\pm$      40     &      287    $\pm$      69    &  Blue & I\\
 &  &     N &    40.18     &       73    $\pm$       0     &      122    $\pm$       5     &     38.53     &        5    $\pm$      27     &       18    $\pm$      10    &  Red/Rot & Rot\\
\enddata
\tablecomments{
Col. 1:  target ID, Col. 2: the effective radius of the broad \mbox{[O\,\textsc{iii}]} component, Col. 3: line component (Total, Broad, Narrow), Col. 4: dust-uncorrected \mbox{[O\,\textsc{iii}]} luminosity, Col. 5: \mbox{[O\,\textsc{iii}]} velocity, Col. 6: \mbox{[O\,\textsc{iii}]} velocity dispersion, Col. 7: H$\alpha$ luminosity, Col. 8: H$\alpha$ velocity, Col. 9: H$\alpha$ velocity dispersion, Col. 10 and 11: visual classification of the \mbox{[O\,\textsc{iii}]} and H$\alpha$ kinematics. Kinematics classifications are: Blue - predominantly blueshifted emission, Red - predominantly redshifted emission, Rot - Ionized gas following stellar rotation, I - Irregular features. 
Note that kinematic measurements are flux-weighted within one effective radius, r$_{\mathrm{eff}}$.}
\end{deluxetable*}

\subsection{Outflow sizes and AGN luminosity}
\label{sec:size}

\begin{figure}[bpt]
\begin{center}
\includegraphics[width=0.45\textwidth,angle=0,trim={5 30 15 10},clip]{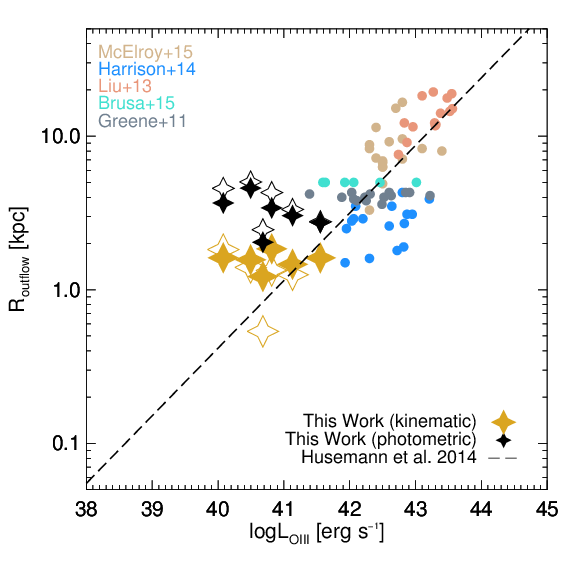}
\caption{Outflow size versus \mbox{[O\,\textsc{iii}]} luminosity, for our sample (stars) and the literature samples (circles). 
Our kinematically measured sizes are derived based on \mbox{[O\,\textsc{iii}]} (filled symbols) or H$\alpha$ emission (open symbols). 
The dashed line indicates the size-luminosity relation based on Type 1 AGNs by \citet{Husemann2014} . For our sample we plot both kinematically (gold) and photometrically measured sizes (black).}
\label{fig:lit_size_AGN}
\end{center}
\end{figure}

In this section, we compare the size of the outflows with the AGN \mbox{[O\,\textsc{iii}]} luminosity. In order to increase the dynamical range, we combine our sample with samples taken from the literature. In total, we compile objects from 5 literature studies, including 3 samples based on IFU observations (\citealt{Liu2013,Harrison2014,McElroy2015}) and 2 samples based on spatially resolved long-slit spectroscopy (\citealt{Greene2011,Brusa2015}). The combined sample covers more than 4 orders of magnitude in \mbox{[O\,\textsc{iii}]} luminosity, with our sample being at the lower end of AGN \mbox{[O\,\textsc{iii}]} luminosities. Note that the size of the outflows is defined differently among the different studies. Despite this limitations, we find a well-defined size-luminosity relation (Fig. \ref{fig:lit_size_AGN}). Instead of using AGN bolometric luminosities, we choose to use dust-uncorrected \mbox{[O\,\textsc{iii}]} luminosity to avoid uncertainties related to the bolometric correction and extinction correction and for direct comparison with other literature studies.

We get a Kendall's $\tau$ correlation coefficient of 0.46 with very high significance (p$<0.0001$). We observe, however, that the implied correlation appears to have a varying slope, with AGN luminosities $\lesssim10^{42.5}$ erg s$^{-1}$ showing a flatter slope compared to their more luminous counterparts. 

As shown in Fig. \ref{fig:size_comp}, for most of the 
literature samples the actual size of the outflow may be overestimated since it is mostly based on the ionized gas flux distribution, 
without taking into account the actual spatially resolved kinematics of \mbox{[O\,\textsc{iii}]}. Furthermore, most previous studies 
did not characterize the spatially resolved stellar kinematics, which are important in understanding the effect of the gravitational effect on the gas kinematics.
The latter is of importance for higher luminosity AGNs hosted by more massive galaxies (see the comparison between photometric and kinematic size in Fig. \ref{fig:size_comp}).

In Fig. \ref{fig:lit_size_AGN} we also plot the size-luminosity relation derived by \citet{Husemann2014} based on a sample of luminous Type 1 AGNs (dashed line; Eq. 2 from that paper). We find outflow measurements from Type 2 AGNs to be largely consistent with the relation. A linear regression fit to the combined sample gives a slope of $0.30\pm0.04$\footnote{We note that in absence of robust uncertainties and a largely inhomogeneous dataset of outflow sizes, the slope uncertainty quoted only reflects the statistical scatter of the size measurements.}. This is marginally flatter ($\sim2\sigma$) than the slope of $0.44\pm0.06$ derived by \citet{Husemann2014} but consistent within the uncertainties with the slope of $0.33\pm0.04$ derived by \citet{Schmitt2003} for a sample of local Seyfert galaxies observed with HST narrow-band imaging.

\section{Discussion}
\label{sec:discussion}

In this section we discuss the implications of our results concerning the geometry of AGN-driven outflows and the importance of decomposed kinematics in accurately constraining the size of the outflows. The morphology and stellar kinematics of our sample show rotating disks that play a role in determining the kinematics of the NLR.  
However, an additional kinematic component is clearly present, leading to high velocity dispersion and velocity shift from the systemic velocity. We find that most objects (5 out of 6) in the sample show negative velocity, which is consistent with previous IFU studies (e.g., \citealt{Liu2013,Rupke2013,Harrison2014,Carniani2015}).
A higher fraction of blueshifted \mbox{[O\,\textsc{iii}]} than redshifted \mbox{[O\,\textsc{iii}]} among high luminosity AGNs
is also found by statistical studies of type 2 AGNs (e.g., \citealt{Bae2014, Woo2015b}).

The separation of the outflow component from the gravitational component is demonstrated in the VVD diagram (Fig. \ref{fig:all_vvd_oiii}), which is a helpful tool in visualizing the different kinematic components. Compared to the systemic velocity and $\sigma_{*}$, we find that \mbox{[O\,\textsc{iii}]} in the central spaxels predominantly 
shows higher velocity dispersion and high velocity shift, indicating an outflow-dominated region, while the gravity-dominated region is 
separate in the lower velocity dispersion regime. The outflow may be experiencing an initial acceleration, as indicated by the velocity dispersion radial profiles for \mbox{[O\,\textsc{iii}]} (Fig. \ref{fig:radial_prof_OIII}), but decelerates beyond a few r$_{\mathrm{eff}}$. Together with the dependence on the radial distance in the VVD diagram, these suggest that 
the outflow kinematics weakens radially.

We expect some degree of dust extinction for all of our targets, as evidenced by the presence of a rotation-dominated stellar disk in at least five of them. Furthermore, with velocity dispersions from 2 to up to 5 times larger than the $\sigma_{*}$, we find strong evidence for either a wide-angled or quasi-spherical geometry for these outflows. The combination of obscuration and the opening angle of the outflow can, to some degree, dictate the observed kinematics of these outflows. In a separate paper (Bae et al. 2015, in prep.) we will use 3-D biconical outflow models to explore the position of a source on the VVD diagram.

Determining the size of AGN-driven outflows is a thorny subject that has been approached in many different ways in the literature. Previously, several studies have relied on purely flux-based measurements using, e.g., \mbox{[O\,\textsc{iii}]} to define the outflow region size (e.g., \citealt{Nesvadba2006,Bennert2006,Greene2011,Liu2013,Rupke2013,Harrison2014}). Alternatively, some simplified consideration of the kinematics was also included in the calculation of the outflow size (e.g., measuring the size of only the broad Gaussian component of \mbox{[O\,\textsc{iii}]}, \citealt{Rodriguez2013,Brusa2015,McElroy2015,Carniani2015}). 

However, the size we are interested in is that of the region where the ionized gas kinematics is dominated by the AGN-driven outflow. As we showed in Fig. \ref{fig:radial_prof}, even for the broad component of \mbox{[O\,\textsc{iii}]}, which exhibits the most extreme kinematics, both the velocity and velocity dispersion radially decrease and become consistent with stellar kinematics
beyond $\sim$2-3 r$_{\mathrm{eft}}^{\mathrm{\mbox{[O\,\textsc{iii}]}}}$.
Thus, if the size of the region, where the broad component of \mbox{[O\,\textsc{iii}]} (or worse the total \mbox{[O\,\textsc{iii}]}) is detected,
is taken as the outflow size, this would lead to an overestimation by on average 30\% to 50\% and up to a factor of two (Fig. \ref{fig:size_comp}). If the H$\alpha$ emission line is used instead of \mbox{[O\,\textsc{iii}]}, the outflow size would be more severely overestimated since the H$\alpha$ emission line profile can be affected by gas kinematics due to star formation and stellar motion. {It is conceivable that AGN-driven outflows extend beyond the GMOS FoV, which covers out to $\sim$4 kpc from the center. However, since both the velocity and velocity dispersion of the ionized gas drop off and converge with the stellar kinematics within the FoV as shown in Fig. 8, it is not likely that the current data is missing larger scale outflows.}

Since we find outflow sizes to be no larger than a couple of kpc, with most of the extreme kinematics confined within the central kpc of our sources, the impact of the outflows seems
to be limited. It is difficult to understand how appreciable amounts of energy can be deposited at galaxy-wide scales to quench star formation or eject gas reservoirs in the galaxy outskirts. Case in point, the Petrosian radii of our 6 galaxies (based on SDSS photometry) are found to be between 5 and 12 kpc, corresponding to  $\sim$5 to 10 times of the outflow sizes. Even assuming that our kinematic method underestimates the "sphere of influence" of AGN feedback, BPT-based sizes of the AGN photoionized regions (explored in depth in a following paper) are still approximately 3 to 6 times smaller than the Petrosian radius of their hosts.

\section{Summary and Conclusions}
\label{sec:conclusions}

We have presented a careful analysis of the kinematic properties of 6 luminous Type 2 AGNs based on the wealth of information in the IFU data and kinematically and spatially resolved measurements. We summarize our findings:
\begin{itemize}

\item The emission line profile of H$\alpha$ and to lesser extent \mbox{[O\,\textsc{iii}]} appear to be a superposition of 
two kinematic components (i.e., broad and narrow components) : a non-gravitational outflow component (high velocity shift and velocity dispersion) concentrated within
the central 1.3--2.1 kpc, and a gravitational component in the outer region that follows stellar rotation. (Fig. \ref{fig:all_kins_total}).

\item The narrow component of H$\alpha$ follows the stellar kinematics closely, while the broad components of both H$\alpha$ and \mbox{[O\,\textsc{iii}]}, and to a lesser degree the narrow component of \mbox{[O\,\textsc{iii}]}, show extreme kinematics. These include large velocity dispersion (up to 5 times $\sigma_{*}$) and blueshifted emission relative to the systemic velocity (Figs. \ref{fig:all_kins_dv} and \ref{fig:all_kins_sigma}).

\item The VVD diagram separates extreme outflow kinematics from gravitational motions. There is a decrease of velocity dispersion and velocity of \mbox{[O\,\textsc{iii}]} down to the levels that is consistent with $\sigma_{*}$ and the systemic velocity, with increasing radial distance (Fig. \ref{fig:all_vvd_oiii}).

\item The radial profiles of the velocity and velocity dispersion show negative trends with radial distance. Based on the kinematics and the flux distribution of the outflowing gas, for the first time the size of the outflow can be kinematically determined as 1.3--2.1 kpc (Figs. \ref{fig:radial_prof} and \ref{fig:radial_prof_OIII}).

\item Size estimation based on the flux (either using the flux distribution or the S/N) is severely overestimated compared to the size of the region where the outflow kinematics dominate over gravitational motion, by up a factor of 2 for \mbox{[O\,\textsc{iii}]} and 
by more than a factor of 2 for H$\alpha$ (Fig. \ref{fig:size_comp}).

\item There is a positive correlation of the size of the outflow with the AGN \mbox{[O\,\textsc{iii}]} luminosity with  (Fig. \ref{fig:lit_size_AGN}).

\end{itemize}

Our IFU observations have revealed powerful outflows with more extreme kinematics (velocity and velocity dispersion) than what was implied by their spatially integrated SDSS spectra. This by itself underlines that spatially integrated spectroscopic studies of AGNs may be underestimating the incidence of AGN-driven outflows and the kinematics of the gas entrained in them.

We have shown  that identifying and characterizing outflows is not straightforward due to the confounding factors related to physically distinct but co-spatial kinematic components (e.g., stellar rotation, star formation-driven outflows). Despite the clear presence of strong outflows in these Type 2 AGNs, the actual impact in terms of feedback to their host galaxies is unclear, particularly given their relatively small size. Despite of the small sample size, as the selected AGNs are extremely rare objects in the local universe due to their strong outflow
signatures, the potency of AGN-driven outflows as negative feedback agents seems questionable. 

In the following paper we will explore the ionization mechanisms at play in these sources and the driving mechanism behind the extreme gas kinematics observed. We will also touch upon the star formation properties and their spatial relation to the outflow kinematically-dominated region. 

\acknowledgments{We thank the anonymous referee for their helpful comments. This research was supported by the National Research Foundation of Korea (NRF) grant funded by the Korea government (MEST) (No. 2010-0027910). This work was supported by K-GMT Science Program funded through the Project BIG3: "From Big Bang to Big Data with Big Eyes" operated by Korea Astronomy and Space Science Institute (KASI). Based on observations obtained at the Gemini Observatory processed using the Gemini IRAF package, which is operated by the Association of Universities for Research in Astronomy, Inc., under a cooperative agreement with the NSF on behalf of the Gemini partnership: the National Science Foundation (United States), the National Research Council (Canada), CONICYT (Chile), the Australian Research Council (Australia), Minist\'{e}rio da Ci\^{e}ncia, Tecnologia e Inova\c{c}\~{a}o (Brazil) and Ministerio de Ciencia, Tecnolog\'{i}a e Innovaci\'{o}n Productiva (Argentina). This research has made use of NASA's Astrophysics Data System Bibliographic Services. For this research, we have made extensive use of the TOPCAT software (\citealt{Taylor2005}), which is part of the suite of Virtual Observatory tools. This research has made use of the NASA/IPAC Extragalactic Database (NED), which is operated by the Jet Propulsion Laboratory, California Institute of Technology, under contract with the National Aeronautics and Space Administration.}

\bibliographystyle{aa}
\bibliography{bibtex}

\begin{appendix}
\label{sec:appendix}
Here we provide comments specific to individual sources.
\vspace{10pt}
\section{J091808+343946}
\label{sec:J0918}

{J0918 is at redshift z=0.0973 (FoV$\sim8.5\times6$ kpc, resolution $\sim0.85$ kpc). It is a face-on disk with a b/a (major-to-minor axis) ratio of 0.94. The source  appears elongated in the West-East direction in the IFU continuum flux map, however, this is a seeing effect, as the spectrophotometric star observed during the observing runs shows similar elongation in its IFU flux map. }\mbox{[O\,\textsc{iii}]} {is significantly detected (S/N$>$3) only in the central $\sim1.5\arcsec\,=\,2.5$ kpc, and appears spatially unresolved. On the contrary, H$\alpha$ is much more extended, significantly detected out to the edge of the FoV.}

There is little evidence for rotation in the stellar velocity map of this source. The broad \mbox{[O\,\textsc{iii}]} component is very compact and appears blueshifted with velocities up to more than $\sim600$ km s$^{-1}$. The velocity dispersion of the broad component reaches values up to $\sim700$ km s$^{-1}$ ($\sim 4$$\sigma_{*}$). The broad H$\alpha$ component is blueshifted in the central region with some hints for redshifted emission at the lower right edge of the IFU, but with velocities consistent with the stellar component. 

\section{J113549+565708}
\label{sec:J1135}

{J1135 is at z=0.0515 (FoV$\sim5.8\times3.4$ kpc, resolution $\sim0.77$ kpc). It is the most luminous galaxy in our sample, with an $r$-band magnitude of 14.7 AB mag and an }\mbox{[O\,\textsc{iii}]} {luminosity of $10^{43.1}$ erg s$^{-1}$.
It is a face-on disk (b/a=0.82) with two spiral arms and hints of morphological distortions indicative of a recent ($\sim$Gyr) interaction. A structure observed in the North-West of the galaxy may be a remnant of that interaction. Both }\mbox{[O\,\textsc{iii}]} {and H$\alpha$ are detected out to the edge of the IFU FoV. Interestingly, while H$\alpha$ is slightly more extended than }\mbox{[O\,\textsc{iii}]}{, the two show very similar spatial distributions.}

The stellar velocity map of J1135 shows evidence for a rotation pattern in the North-West to South-East direction. The broad \mbox{[O\,\textsc{iii}]} component appears blueshifted, reaching velocities of $-300$ km s$^{-1}$. The \mbox{[O\,\textsc{iii}]} broad component dispersion is $\sim600$ km s$^{-1}$, 3 times higher than $\sigma_{*}$ ($\sim200$ km s$^{-1}$). The H$\alpha$ broad component shows a spatial superposition of two components: a rotation component that follows the stellar rotation and an elongated structure in the North-South direction with strong blueshifted emission ($\gtrsim200$ km s$^{-1}$). The same feature is seen in the broad H$\alpha$ dispersion map. The maximum dispersion is $\sim400$ km s$^{-1}$, two times the $\sigma_{*}$ and larger than the broad \mbox{[O\,\textsc{iii}]} component dispersion.

\section{J140453+532332}
\label{sec:J1404}

{J1404 is at z=0.0813 (FoV$\sim7.3\times5.1$ kpc, resolution $\sim0.88$ kpc). It has a b/a axis ratio of 0.38 and appears as a highly inclined disk galaxy elongated in the North-South direction. The }\mbox{[O\,\textsc{iii}]} {emission is concentrated with a somewhat asymmetric shape, although faint }\mbox{[O\,\textsc{iii}]} {emission is detected out to the edge of the FoV. In comparison, the H$\alpha$ flux map shows very extended emission, significantly detected out to the edge of the FoV.}

The stellar velocity map of J1404 reveals a rotation pattern in a North to South direction. The broad \mbox{[O\,\textsc{iii}]} component shows blueshifted emission. The broad \mbox{[O\,\textsc{iii}]} velocity dispersion takes large values $\sim400$ km s$^{-1}$ or $\sim2$$\sigma_{*}$. The broad H$\alpha$ component shows a strong blueshifted core, consistent in shape and average velocity with the \mbox{[O\,\textsc{iii}]} broad component. Gas at the outer part of the FoV appears to follow the stellar rotation. The velocity dispersion of the broad H$\alpha$ shows values consistent with the \mbox{[O\,\textsc{iii}]} broad component, with a ring of elevated velocity dispersion that is also seen in the narrow \mbox{[O\,\textsc{iii}]} component. Beyond the ring the velocity dispersion drops sharply to values consistent with $\sigma_{*}$. This is a strong indication that we are observing the terminal region of an outflow driven from the core of the galaxy.

\section{J160652+275539}
\label{sec:J1606}

{J1606 is at z=0.0461 (FoV$\sim4.4\times3$ kpc, resolution $\sim0.44$ kpc), is the closest galaxy in the sample. It has a b/a axis ratio of 0.88 and appears as an almost face-on disk with hints of spiral structure in its SDSS image. The galaxy appears elongated in the West-East direction (beyond the seeing effect). No }\mbox{[O\,\textsc{iii}]} {emission is detected beyond $\sim1\arcsec-1.5\arcsec$ from the nucleus.}

J1606 shows rotation in a North to South direction. The broad \mbox{[O\,\textsc{iii}]} emission component appears blueshifted across the \mbox{[O\,\textsc{iii}]}-detected spaxels, with velocities reaching values above $-400$ km s$^{-1}$. The velocity dispersion of the broad \mbox{[O\,\textsc{iii}]} is $\sim2.5$ times the $\sigma_{*}$. For both the velocity and velocity dispersion of the broad \mbox{[O\,\textsc{iii}]} component, there is evidence for a drop-off beyond $\sim1\arcsec$ distances from the nucleus, approaching the stellar properties. The broad H$\alpha$ emission component appears very blueshifted in the inner part of the IFU, spatially coinciding with the blueshifted broad \mbox{[O\,\textsc{iii}]} emission. There is evidence for rotation along the stellar disk in the outer parts of the broad H$\alpha$ velocity map. The velocity dispersion of the broad H$\alpha$ component is the highest near the nucleus with values of $\sim200$ km s$^{-1}$.

\section{J162233+395650}
\label{sec:J1622}

{J1622 is at z=0.0631 (FoV$\sim5.8\times4.1$ kpc, resolution $\sim0.58$ kpc). It has a b/a axis ratio of 0.76 and appears as a face-on disk. The galaxy appears slightly elongated in the SouthWest-NorthEast direction. The H$\alpha$ emission is significantly detected across the full FoV, showing a similar elongation as the continuum. }\mbox{[O\,\textsc{iii}]} {is centrally concentrated ($<\sim2\arcsec$) with no obvious elongation.}

The stellar velocity map of J1622 shows evidence for a rotational pattern in a South-West to North-East direction. The \mbox{[O\,\textsc{iii}]} line is very concentrated and appears redshifted. There is an indication for a ring of elevated velocity around $\sim0.8\arcsec$ from the continuum center. The velocities are comparable to the stellar maximum velocity, $\sim200$ km s$^{-1}$. The broad \mbox{[O\,\textsc{iii}]} component velocity dispersion in contrast shows values up to 800 km s$^{-1}$ or $\sim6$$\sigma_{*}$. The broad H$\alpha$ emission appears blueshifted in the central part of the IFU, forming a pronounced elongated feature that crosses to the South of the nucleus in a South-East to North-West direction. Right to the South of this elongated feature, velocities are found to be positive, with maximum velocities of $\sim200$ km s$^{-1}$. The region of redshifted H$\alpha$ emission seems to trace the elongated blueshifted H$\alpha$ feature and extends out to roughly the scale of the redshifted broad \mbox{[O\,\textsc{iii}]} emission. The outer part of the broad H$\alpha$ appears consistent with the stellar velocity map. The velocity dispersion of broad H$\alpha$ shows the same spatially elongated feature as the velocity map, with velocity dispersion values of $\sim600$ km s$^{-1}$.

The very large velocity dispersion for both \mbox{[O\,\textsc{iii}]} and H$\alpha$ emission lines implies a very large opening angle, while the absence of significantly blueshifted emission indicates that there is no obscuring screen and we are observing the outflow with a large viewing angle. An alternative scenario would require a significant overlap (either physical or through projection) between AGN- and star formation-ionized gas. This would lead to a smoothing out of any kinematically distinct components linked to the AGN, especially considering that the massive stars producing the stellar winds should be uniformly distributed across the stellar disk and around the AGN in the nucleus.

\section{J172038+294112}
\label{sec:J1720}

{J1720 is at redshift $z=0.0995$ (FoV$\sim8.8\times6.2$ kpc, resolution $\sim1$ kpc) is the farthest galaxy in our sample. It is a face-on disk galaxy with two spiral arms and a b/a axis ratio of 0.95. The continuum IFU flux map shows a core with a NorthWest-SouthEast elongation and emission from the base of the spiral arms. The }\mbox{[O\,\textsc{iii}]} {emission from J1720 is centrally concentrated with no obvious asymmetries (detected out to $\sim2$\arcsec). In contrast, the H$\alpha$ map shows very extended emission with both a bright central component and emission tracing the spiral arms in the NorthWest and SouthEast corners of the FoV.}

The stellar velocity map of J1720 shows rotation in the North-East to South-West direction. The broad \mbox{[O\,\textsc{iii}]} component shows blueshifted nuclear emission with maximum velocities of $\sim200$ km s$^{-1}$. The velocity dispersion of the broad \mbox{[O\,\textsc{iii}]} component is very high (up to $700$ km s$^{-1}$ or $\sim5\sigma_{*}$) with indications that the maximum velocity dispersion is observed at $\sim0.5\arcsec$ from the nucleus. The broad H$\alpha$ component appears blueshifted around the galaxy core, with velocities consistent with the broad \mbox{[O\,\textsc{iii}]} component. The spiral arms also show broad H$\alpha$ emission but with velocities following the stellar velocities. The velocity dispersion of the broad H$\alpha$ component is up to 5$\sigma_{*}$ near the galaxy nucleus.




\end{appendix}
\end{document}